\documentclass{aa}  
\usepackage{graphicx}
\usepackage{color}
\usepackage{txfonts}
\usepackage{hyperref}
\usepackage{lscape}
\usepackage{natbib}
\usepackage{amsmath}
\usepackage{amsfonts}
\usepackage{wasysym}
\usepackage{graphics}
\usepackage{times}
\usepackage{parskip}
\usepackage{pdflscape}
\usepackage{geometry}
\usepackage{marginnote}
\usepackage{multicol}
\usepackage{soul}

\usepackage{natbibspacing}
    \renewcommand{\dbltopfraction}{0.9} % fit big float above 2-col. text
\newfont{\nlx}{cmssdc10 scaled 900}
\newfont{\tlx}{cmssdc10 scaled 600}
\newfont{\mfont}{cmssdc10 scaled 810}
\definecolor{myblue1}{rgb}{0.0,0.604,0.831} 
\definecolor{myblue2}{rgb}{0.0,0.49,0.6745}
\definecolor{myblue3}{rgb}{0.0156,0.4078,0.9921}
\definecolor{myblue4}{rgb}{0.0,0.44,0.87}
\definecolor{myred1}{rgb}{0.529,0.019,0.017}
\definecolor{mycyan}{rgb}{0.63921569,0.0,0.48235294}

\newcommand{\brem}[1]{\textcolor{black}{\nlx #1}}
\newcommand{\mbrem}[1]{\textcolor{black}{\mfont #1}}

\newcommand{\PutLabel}[3]{\put(#1,#2){#3}}
\newfont{\hvss}{cmssdc10 scaled 1540}
\hypersetup{
    pdftoolbar=true,				% show Acrobat's toolbar? 
    pdfmenubar=true,				% show Acrobat's menu?
    pdffitwindow=false,			% window fit to page when opened
    pdfstartview={FitH},			% fits the width of the page to the window
    pdftitle={title},				% title
    pdfauthor={Author},			% author
    pdfsubject={Subject},			% subject of the document
    pdfcreator={Author},			% creator of the document
    pdfproducer={Author},			% producer of the document
    pdfkeywords={Population \& Evolutionary} ,        % list of keywords
    pdfnewwindow=true,			% links in new window
    colorlinks=true,				% false: boxed links; true: colored links
    linkcolor=cyan,				% color of internal links (change box color with linkbordercolor)
    citecolor=myblue4,				% color of links to bibliography
    filecolor=cyan,				% color of file links
    urlcolor=cyan				% color of external links
}
\def\starlight{\sc Starlight\rm}
\def\RY{${\cal RY}$}
\def\mbh{${\cal M}_{\bullet}$}

\def\reff{$R_{\rm eff}$}

\def\ha{H$\alpha$}

\def\msun{$\mathrm{M}_{\odot}$}
\def\zsun{$\mathrm{Z}_{\odot}$}

\def\D4000{$D_{4000}$}

\def\rr{$R^{\star}$}
\def\rbulge{$R_{\rm B}$}
\newcommand{\sbb}{mag/$\sq\arcsec$}
\newcommand{\dmb}{$<\!\!\!\delta\mu_{9{\rm G}}\!\!\!>$}
\def\mstar{${\cal M}_{\star}$}

\def\sstar{$\Sigma_{\star}$}

\def\flb{$\delta{\rm L_B}$}

\def\tmass{$\langle \log t_\star \rangle_{{\cal M}}$}

\def\ewha{EW(H$\alpha$)}

\def\tcut{$t_{\mathrm{cut}}$}
\def\mbulge{$M_{\mathrm{B}}$}
\def\ml{\small ${\cal M/L}$\normalsize}
\def\mmb{$M_{\rm B}$}

\def\dio{$\delta_{\rm io}$}

\def\mueff{$\mu_{\rm eff}$}
\def\SFQ{\mbrem{SFQ}}

\def\oD{\brem{oD}}
\def\iD{\brem{iD}}

\def\isan{\brem{isan}}
\def\kmsec{km/s}

\def\tmass{$t_{\star,\cal M}$}
\def\sfha{\mfont SFH1\rm}
\def\sfhb{\mfont SFH2\rm}
\def\sfhc{\mfont SFH3\rm}

\def\tln2ha{$\log$([N\,{\sc ii}]${\scriptstyle 6584}$/H$\alpha$)}
\def\tlo3hb{$\log$([O\,{\sc iii}]${\scriptstyle 5007}$/H$\beta$)}
\newcommand{\tref}[1]{\textcolor{myblue4}{#1}}
\def\bid{\brem{B/iD}}
\def\tmig{$\tau_{\rm m}$}
\def\pegase{{\sc Pegas\'e}}
\newcommand\btab[5]{\begin{table*}[#1]\label{#3}{\parbox{#4}{\caption{#2}}\rule[-0.5ex]{0cm}{0.5ex} }
\begin{tabular*}{#4}{#5} \label{#3} }
\begin{document} 
% ==============================================================================================================
\title{Inside-out star formation quenching and the need for a revision 
of bulge-disk decomposition concepts for spiral galaxies}
\titlerunning{Inside-out star formation quenching and resulting biases in galaxy bulge-disk decomposition studies}
\authorrunning{Papaderos et al.}
% ===============================================================================================================
   \author{
          Polychronis Papaderos      % (Πολυχρὸνης Παπαδερὸς)
          \inst{\ref{IA-FCiencias},\ref{ULisbon}}   
          \and
          Iris Breda
         \inst{\ref{IA-FCiencias},\ref{ULisbon},\ref{IA-CAUP},\ref{IAA-CSIC}}          
          \and
          Andrew Humphrey
          \inst{\ref{IA-CAUP}}
          \and          
          Jean Michel Gomes
          \inst{\ref{IA-CAUP}}          
          \and 
          Bodo L. Ziegler
          \inst{\ref{UWien}}          
         \and 
         Cirino Pappalardo 
          \inst{\ref{IA-FCiencias},\ref{ULisbon}}            
          }        

\institute{Instituto de Astrofísica e Ciências do Espaço, Universidade de Lisboa, OAL, Tapada da Ajuda, PT1349-018 Lisboa, Portugal 
\label{IA-FCiencias}
\and 
Departamento de Física, Faculdade de Ciências da Universidade de Lisboa, Edifício C8, Campo Grande, PT1749-016 Lisboa, Portugal \label{ULisbon}
\and 
Instituto de Astrof\'{i}sica e Ci\^{e}ncias do Espaço - Centro de Astrof\'isica da Universidade do Porto, Rua das Estrelas, 4150-762 Porto, Portugal \label{IA-CAUP} 
\and 
 Instituto de Astrof\'isica de Andaluc\'ia (CSIC), Glorieta de la Astronom\'ia s/n, 18008 Granada, Spain
 \label{IAA-CSIC}
\and 
 Department of Astrophysics, University of Vienna, T\"urkenschanzstr. 17, A-1180 Vienna, Austria
 \label{UWien}
\\
             \email{papaderos@astro.up.pt}
             }
\date{Received ??; accepted ??}

\abstract
%Context
{Our knowledge about the photometric and structural properties of bulges in late-type galaxies (LTGs) is founded upon image decomposition 
into a S\'ersic model for the central luminosity excess of the bulge and an exponential model for the more extended underlying disk.
We argue that the standard practice of adopting an exponential model for the disk all the way to its center is inadequate because 
it implicitly neglects the fact of star formation (SF) quenching in the centers of LTGs. 
Extrapolating the fit to the observable star-forming zone of the disk (outside the bulge) inwardly
overestimates the true surface brightness of the disk in its SF-quenched central zone (beneath the bulge).
We refer to this effect as \dio.
Using predictions from evolutionary synthesis models and applying {\sc RemoveYoung}, a tool that allows  
the suppression of stellar populations younger 
than an adjustable age cutoff from integral field spectroscopy data, we estimate the \dio\ in the centers of massive SF-quenched LTGs to be up to $\sim$2.5 (0.7) $B$ ($K$) mag.
The primary consequence of the neglect of \dio\ in bulge-disk decomposition studies is the oversubtraction of the disk underneath the bulge, leading to a systematic underestimation of the true luminosity of the latter. 
Secondary biases impact the structural characterization (e.g., S\'ersic exponent $\eta$ and effective radius) and color gradients 
of bulges, and might include the erroneous classification of LTGs with a moderately faint bulge as bulgeless disks.
Framed in the picture of galaxy downsizing and inside-out SF quenching, \dio\ is expected to differentially impact galaxies 
across redshift and stellar mass \mstar,
thus leading to systematic and complex biases in the scatter and slope of various galaxy scaling relations.
We conjecture that correction for the \dio\ effect will lead to a down-bending of the bulge versus supermassive black hole relation 
for galaxies below log(\mstar/\msun)$\sim$10.7. 
A decreasing \mbh/\mstar\ ratio with decreasing \mstar\ would help to consistently explain the scarcity and weakness of accretion-powered nuclear activity in low-mass spiral galaxies.
Finally, it is pointed out that a well-detectable \dio\ ($>$2 $r$ mag) can emerge early on through inward migration of star-forming clumps from the disk in combination with a strong contrast of emission-line equivalent widths between the quenched proto-bulge and its star-forming periphery. Spatially resolved studies of \dio\ with the JWST, ELT, and Euclid could therefore offer key insights into the chronology and physical drivers of SF quenching in the early phase of galaxy assembly.
}

% 5 {} token are mandatory
\maketitle
%________________________________________________________________

\vspace*{0.2cm}
\section{Introduction \label{intro}}

An explicit assumption in bulge-disk decomposition studies of late-type galaxies (LTGs) is that the exponential model for the visible part of the disk (i.e., outside the bulge) is valid all the way to the galaxy center (below the bulge). Subtraction of that model from the galaxy allows extracting the central luminosity excess of the bulge, the fitting of which with a \citet{Sersic63} function yields the total magnitude, the S\'ersic index $\eta$, effective radius \reff, and the effective surface brightness $\mu_{\rm eff}$ of the bulge. 

These quantities are fundamental to our understanding of the nature and structural characteristics of galaxy bulges, and to their empirical subdivision into classical bulges and pseudobulges \citep[cf.][among others]{Gad09,FisDro10,FisDro11,FerLor14,Men17,Neumann17}. 
They also constitute the observational foundation for various galaxy scaling relations \citep[e.g.,][]{Kormendy77,FJ76,DD87}, including the correlation of the bulge versus supermassive black hole (SMBH) \citep[][see also, \tref{Kormendy \& Ho 2013} for a review]{Richstone98,FerMer00,Ho08,Simmons13}.

Regardless of whether the photometric analysis of the bulge is carried out in 1D or 2D and whether it is done sequentially, through subtraction of the disk first and subsequent fitting of the bulge, or simultaneously, through nonlinear fitting of both, the key assumption in all cases is that the intensity profile of the disk below the bulge (hereafter, inner disk, \iD) is simply the inward extension of the exponential profile of the outer disk (\oD).
% :::::::: FIGURE 1 ::::::::::::::::::::::::::::::::::::
%\setlength{\unitlength}{1mm}
\begin{center}
\begin{figure*}[h!]
\begin{picture}(86,54)
\put(0,30){\includegraphics[height=2.4cm]{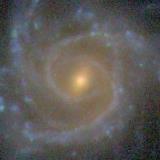}}
\put(25,30){\includegraphics[height=2.4cm]{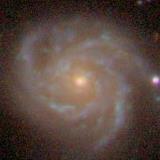}}
\put(50,30){\includegraphics[height=2.4cm]{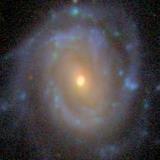}}
\put(75,30){\includegraphics[height=2.4cm]{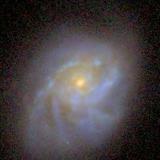}}
\put(0,5){\includegraphics[height=2.4cm]{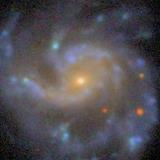}}
\put(25,5){\includegraphics[height=2.4cm]{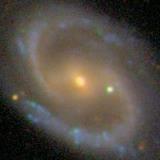}}
\put(50,5){\includegraphics[height=2.4cm]{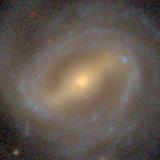}}
\put(75,5){\includegraphics[height=2.4cm]{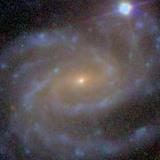}}
\put(100,0){\includegraphics[width=8.7cm]{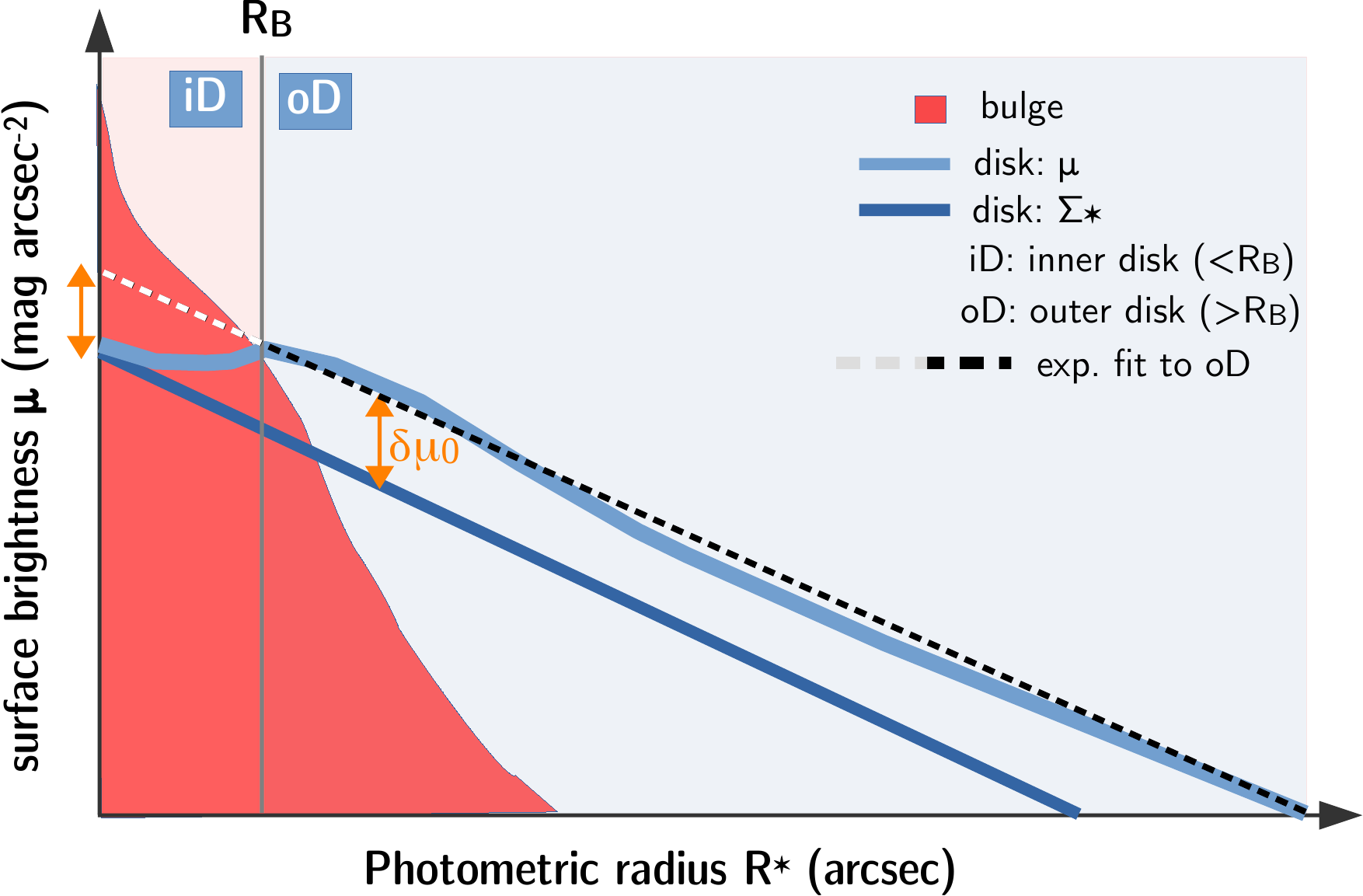}}
\end{picture}
\caption{\brem{left:} SDSS true-color thumbnails of galaxies from the sample analyzed by \citet{BP18}, illustrating the typical morphology of Milky Way-sized LTGs in the local Universe, 
with their inner reddish zone and their surrounding blue star-forming disk.
From top left to bottom-right: \object{NGC 0309}, \object{NGC 0234}, \object{NGC 3811}, \object{NGC 0873}, \object{UGC 4256}, 
\object{NGC 2543}, \object{NGC 0171}, \object{UGC 4308}.
\brem{b)} Schematic representation of the bulge (shaded red area) and the disk (light blue) in a face-on late-type galaxy. The surface brightness profile of the disk results from the projection of an exponential stellar surface density \sstar\ (dark blue), whereby the \ml\ ratio in the inner SF-quenched zone of the disk (\iD) within the bulge radius \rbulge\ is higher than that in the outer (\rr$>$\rbulge) star-forming zone of the disk (\oD). 
A consequence of this is that the central surface brightness $\mu_0$ of the disk is fainter by $\delta\mu_0$ mag than the value 
implied by inward extrapolation of the exponential fit to the outer disk (dashed line) and that the standard bulge-disk decomposition 
overestimates the integrated magnitude of the disk within \rbulge\ by \dio\ (mag). 
This in turn entails an oversubtraction of the disk, thus a systematic underestimation of the luminosity of the bulge.
\label{sketch}}
\end{figure*}
\end{center}      
% :::::::: FIGURE 1 ::::::::::::::::::::::::::::::::::::

\vspace*{-1.0cm}
However, it is worth contemplating what further assumptions are implicitly encapsulated in the postulate that the exponential model for the visible outer disk is valid all the way to the galaxy center. 
The first assumption is that the stellar surface density \sstar\ (\msun\ kpc$^{-2}$) of the disk follows a single exponential profile to \rr=0\arcsec, and the second assumption is that its stellar mass-to-light ratio (\ml) is spatially constant.

Here, we only focus on the second assumption. 
To a first approximation (leaving aside the age-metallicity degeneracy), a radially constant \ml\ translates into a homologous spectral energy distribution (SED), a spatially uniform star formation history (SFH) and current specific star formation rate (sSFR), as previously pointed out in \citet[][hereafter \textcolor{myblue4}{B20b}]{Breda20b}. 

This, however, stands in diametric contrast to the well-established phenomenon of star formation quenching (\SFQ) in the centers 
of massive ($L\ga L^{\star}$) late-type galaxies. A strong depression or complete cessation of star formation (SF) activity within 
the bulge radius \rbulge\ of these systems has been documented through photometry \citep[e.g.,][]{BP94,PelBal96,dJvK94} and, more recently, the centrally decreasing SF surface density, sSFR, and \ha\ equivalent width, and increasing age within \rbulge\ 
using integral field spectroscopy (IFS) data \citep{Perez13,Fang13,Gon14,CatT17,Belfiore18,Zibetti17,BP18,Quai19,Kalinova21}.

That \SFQ\ commences early on in the dense centers of galaxies above typically log(\mstar/\msun)$\sim$10.5 \citep{Strateva01} 
has observationally been established through studies of galaxies at higher redshift \citep{Tacchella15,Mosleh17}, in agreement with cosmological simulations \citep[e.g.,][]{Tacchella16}.
The most direct illustration of \SFQ\ is through visual inspection of true-color images of local LTGs with their typically 
reddish bulge and surrounding blue star-forming disk (Fig.~\ref{sketch},left).

The physical origin and timescales of inside-out \SFQ\ in galaxies, even though a fundamental subject that has motivated numerous previous investigations, is not of primary importance to our considerations next. 
Important is merely the simple argument that the mechanisms that inhibit SF in the bulge must also act toward inhibiting SF in the inner disk, given that these two stellar components are cospatial and their SF activity (if any at all) is fed by a common reservoir of gas \citep[][hereafter \tref{B20a}]{Breda20a}. 
Because partial or complete cessation of SF entails an increase in the \ml\ ratio, a consequence of \SFQ\ is that the \ml\ ratio in the inner disk (\rr$\leq$\rbulge) must be higher than than in the outer disk (\rr$>$\rbulge).
If this is so, the SF-quenched inner disk must be dimmer than the inwardly extrapolated model for the star-forming outer disk.
Therefore, the immediate implication of \SFQ\ for bulge-disk decomposition studies is the overestimation (and oversubtraction) of the disk inside \rbulge, consequently, the under-estimation of the luminosity of the bulge.

A schematic illustration of this effect (dimmer inner disk than the inwardly extrapolated exponential fit to the outer disk) 
is given in the right panel of Fig.~\ref{sketch}. For a thin face-on disk, the surface brightness profile 
(SBP; in $L_{\odot}$\ pc$^{-2}$ or \sbb; light blue) is the product of the stellar surface density \sstar\ (dark blue) by the inverse \ml.
For the sake of simplicity, \sstar\ is here assumed to follow an exponential profile all the way to \rr=0\arcsec\ (see, however, \textcolor{myblue4}{B20b} and Sect.~\ref{cd}).
If \ml\ were to be constant throughout the disk (i.e., in the case of no \SFQ), then the fit to the visible outer disk at \rr$>$\rbulge\ (dashed line) would yield an exact match to the profile of the invisible inner disk (\rr$\leq$\rbulge), thus warranting a correct determination of the central luminosity excess of the bulge (shaded red area). 
In the case of \SFQ, however, the central surface brightness $\mu_0$ of the disk is fainter by $\delta\mu_0$ mag than the value inferred from the exponential model for the visible outer disk.
As a result, standard bulge-disk decomposition entails an overprediction of the luminosity of the disk 
underneath the bulge by a factor on the order of dex($\delta\mu_0$/-2.5)$^{-1}$.
In the following, we refer to the difference between the true integrated magnitude of the disk inside \rbulge\ 
and the value predicted from the exponential disk model as \dio\ ($\geq$0 mag).

The obvious next question is whether this \dio\ effect is relevant at all.
If it is small (e.g., $\la$0.2 mag), it might rightfully be argued that it is absorbed within the error budget of bulge-disk decomposition studies for spiral and lenticular galaxies and thus can be ignored. If significantly larger, however, then it should deserve a closer examination because it might systematically impact photometric and structural studies of bulges.

The aim of this pilot study is to draw attention to \dio\ and motivate observational and theoretical work toward better understanding it. 
To this end, we provide empirical estimates of this effect and offer a concise discussion of the biases that neglecting it entails for the photometric characterization of galaxy bulges.
Section~\ref{dio} presents estimates based on simple evolutionary models and spectral population synthesis of IFS data for local LTGs and shows that \dio\ can exceed 2 mag in the optical and 0.6 mag in the near-infrared (NIR). Section~\ref{photometry} addresses the effect of the neglect of \dio\ on determinations of the luminosity and color of bulges. We also show that \dio\ can prevent detection of a moderately faint bulge embedded within a centrally SF-quenched LTG and lead to its erroneous classification as bulgeless. In Sect.~\ref{dis} we briefly comment on the mass-dependent evolution of \dio\ from the perspective of galaxy downsizing and how neglecting it might be imprinted on the slope and scatter of galaxy scaling relations, including the relation between absolute magnitude \mbulge\ of the bulge and SMBH mass \mbh. Finally, we discuss possible empirical approaches for estimating \dio\ and accounting for it in galaxy decomposition schemes. 
A summary of our conclusions is given in Sect.~\ref{summary}.

% :::::::::::::::::::::::::::::::::::::::::::::::::::::::::::::::::::::
\section{Estimates of \dio\ \label{dio}}
% :::::::::::::::::::::::::::::::::::::::::::::::::::::::::::::::::::::
The exact value of \dio\ depends on the integrated apparent magnitude of the disk inside the bulge radius \rbulge, 
\begin{equation}
-2.5 \log_{10}\left( 2 \pi \int_{0}^{R_{\rm B}} R^{\star}\, I(R^{\star})\, dR^{\star} \right) + {\rm constant}
\label{eq:magdisk}
\end{equation}
with \rr\ denoting the photometric radius (\arcsec) and $I$(\rr) the disk's intensity profile (erg s$^{-1}$ cm$^{-2}$  $\sq\arcsec$) in a given photometric band. Because the central part of the disk underneath the bulge is observationally inaccessible, a direct determination of \dio\ is not possible. Nevertheless, a rough estimate of it can be obtained from evolutionary synthesis models by assuming that the bulge and inner disk 
share a similar SFH (Sect.~\ref{ev-dio}) or through a comparison of the luminosity contribution by young stellar populations 
in the inner and outer zone of the disk. The latter task can be achieved through postprocessing of population spectral synthesis (PSS)
fits to spatially resolved IFS data with the age-slicing tool {\sc RemoveYoung}, as we detail in Sect.~\ref{ifs-dio}.\\
For the sake of brevity, we henceforth denote with "bulge" the entirety of stellar populations enclosed within \rbulge, 
that is, the bulge itself and the inner disk, with a possible contribution from a bar. 
% =================================================================================
\subsection{Estimate of \dio\ from evolutionary synthesis models \label{ev-dio}}
% =================================================================================
For an exponential stellar surface density profile \sstar(\rr), \dio\ (mag) primarily depends on the ratio $\psi$ of the mean \ml\ of the disk 
inside and outside \rbulge\ as
\begin{equation}
\delta_{\rm io}\,\,{\rm (mag)} \simeq 2.5\cdot \log(\psi) = 2.5\cdot \log\left(\frac{\cal{M/L}_{\rm iD}}{\cal{M/L}_{\rm oD}}\right).
\label{eq1}
\end{equation}
% ::::::::::::::: Figure 2 ::::::::::::::::::::::::::::::::::::::::::::::::::::::::::::::
\begin{center}
\begin{figure}[!ht] 
\begin{picture}(86,110)
\put(0,60){\includegraphics[width=8.6cm]{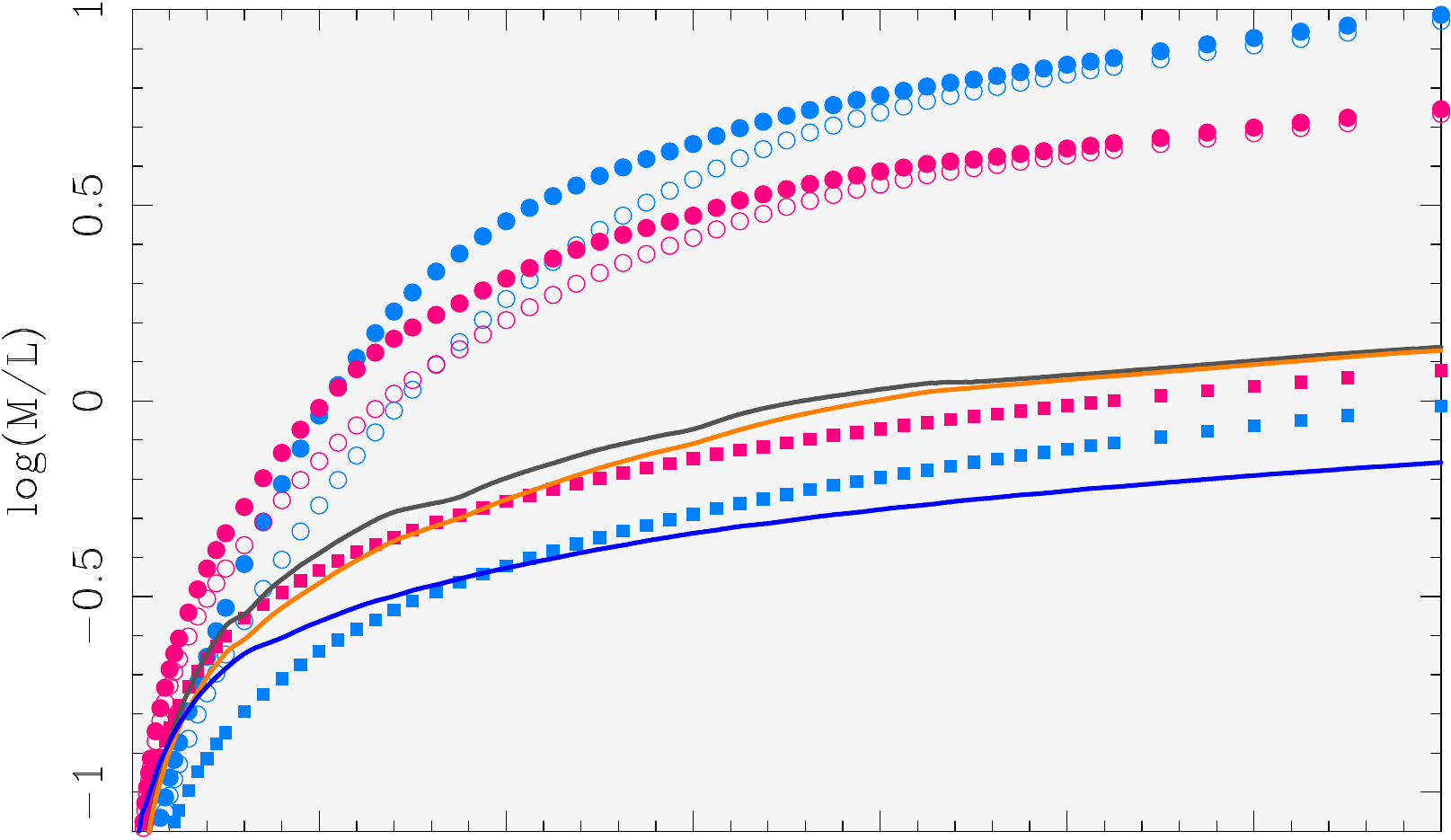}}
\put(0,0){\includegraphics[width=8.7cm]{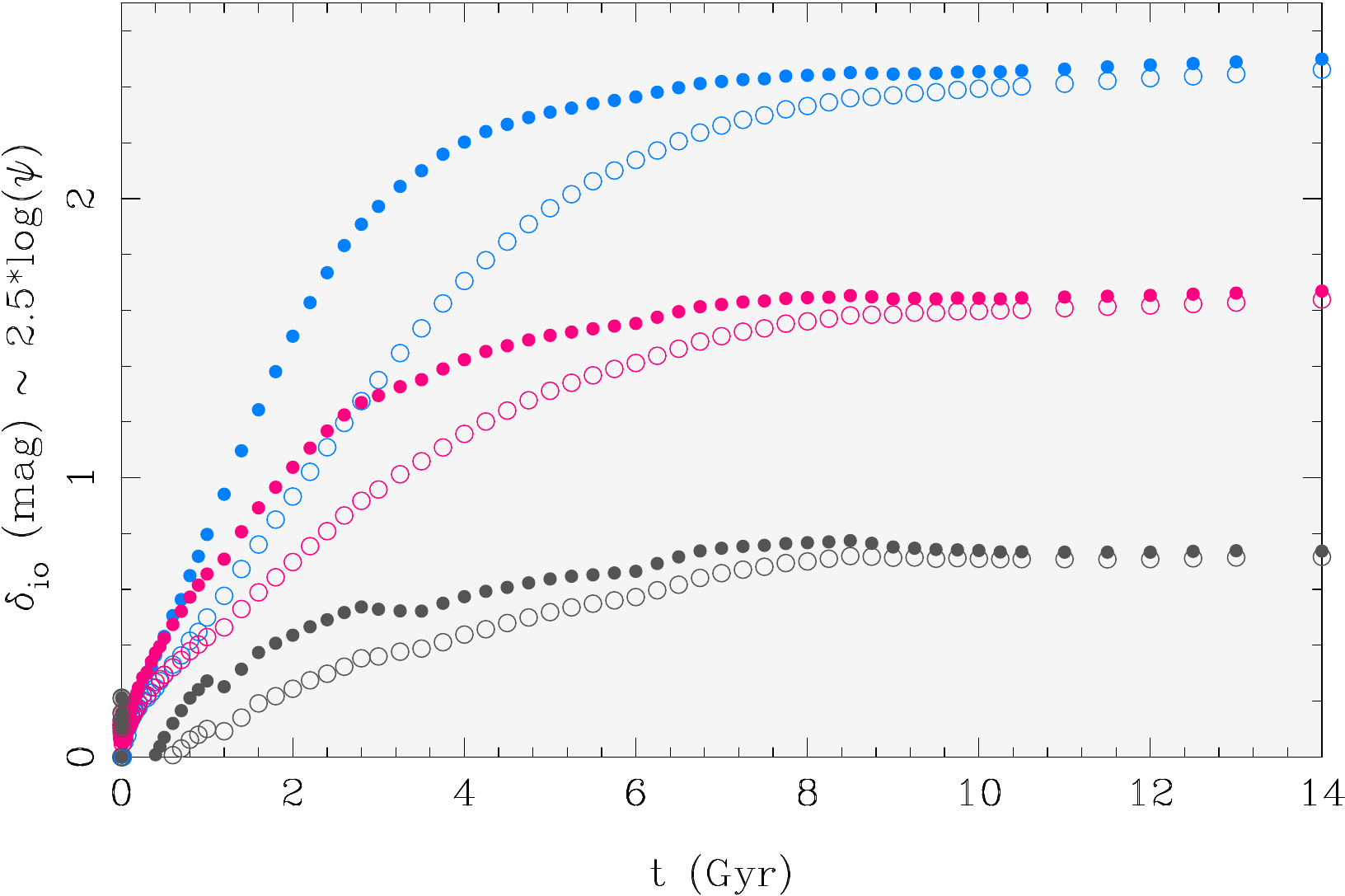}}
\end{picture}
\caption{\brem{upper panel:} Logarithm of the \ml\ ratio as a function of time, as obtained with {\sc P\'egase} assuming continuous star formation at a constant SFR (\sfha), and an exponentially decreasing SFR with an e-folding time $\tau$ of 0.5 Gyr and 1 Gyr (\sfhb\ and \sfhc, respectively). Predictions for the $B$ and $R$ band (blue and red, respectively) are shown with symbols (squares, solid circles and open circles for \sfha, \sfhb, and \sfhc, respectively) and those for the $K$ band with curves (blue, orange and black for \sfha, \sfhb\ and \sfhc, respectively).
\brem{lower panel:} Evolution of \dio, as approximated by Eq.~\ref{eq1}, when assuming \sfhb\ and \sfhc\ for the bulge (solid and open symbols, respectively) with predictions for the $B$, $R$, and $K$ band shown in blue, red and black, respectively. 
\label{mlratio}} 
\end{figure}
\end{center}      
% ::::::::::::::: Figure 2 ::::::::::::::::::::::::::::::::::::::::::::::::::::::::::::::
The ratio $\psi$ in turn encapsulates the stellar mass assembly history and metallicity in the two disk zones, 
and obviously depends on the photometric filter considered.
An estimate of \ml\ can be obtained using standard semiempirical SFH parametrizations, according to which galactic disks are characterized by a nearly-constant star formation rate (SFR) \citep[e.g.,][]{Gallagher84} whereas stellar spheroids are better described by an exponentially declining SFR with an e-folding timescale $\tau$ that scales inversely with present-day stellar mass \mstar\ \citep[][see also \textcolor{myblue4}{Poggianti et al. 1999, Gavazzi et al. 2002}]{Sandage86,GRV87}.

Using \pegase\ \citep{FRV97} we follow the photometric evolution of the outer disk (\rr$>$\rbulge) assuming continuous SF at a constant SFR and a fixed stellar metallicity \zsun/5 (hereafter \sfha). The bulge (\rr$\leq$\rbulge) is approximated by exponentially decreasing SFR models with a $\tau$ of 0.5 Gyr and 1 Gyr for solar metallicity (\sfhb\ and \sfhc, respectively). 
These two models imply a currently low SFR and a predominantly old stellar population in the centers of LTGs, in agreement with optical and NIR colors \citep{deJong96a,deJong96b,GA01} and constraints from spectral synthesis \citep{Gal05,Perez13,Zibetti17,BP18}. Furthermore, the mass-weighted age implied by the two latter SFHs (\tmass$\ga$12 Gyr) agrees well with determinations from spectral population synthesis for high-mass (log(\mstar/\msun)$>$10.5) bulges \citep[cf., e.g., Fig.~6 and Table B.1 in][]{BP18}. The \pegase\ models in Fig.~\ref{mlratio} take nebular emission into account and refer to a Salpeter initial mass function (IMF) between 0.1 and 100 \msun.

The upper panel shows the evolution of \ml\ in $B$ (Johnson) and $R$ (Cousins) (blue and red, respectively) for \sfha\ (squares), \sfhb\ (solid circles) and \sfhc\ (open circles). 
The respective values in the $K$ band are shown with the blue, orange and black curve. As expected, the \ml\ ratio evolves strongest in the $B$ band and only mildly increases with time in the NIR.

The lower panel illustrates the evolution of \dio, as approximated by Eq.~\ref{eq1}, in $B$ (blue), $R$ (red) and $K$ (black) when assuming \sfhb\ or \sfhc\ for the bulge (solid and open symbols, respectively): both SFH scenarios imply a steep increase within the first $\sim$4 Gyr of galactic evolution ($z\approx 1.43$, assuming that LTG formation began 0.5 Gyr after the Big Bang) and a leveling-off after $\sim$8 Gyr ($z\approx 0.6$) to a nearly constant \dio\ of $\sim$2.5 mag in $B$, 1.6 mag in $R$, and 0.7 mag in $K$.

The strongly simplified SFH parameterizations in Fig.~\ref{mlratio} allow the insight that \dio\ becomes significant ($>$1 $B$ mag) early on and is non-negligible in present-day LTGs. On the other hand, it should be kept in mind that the underlying working hypothesis of stellar populations within \rbulge\ sharing a similar evolutionary history, although broadly consistent with observations and supported by plausibility arguments, does most certainly not capture the complexity of the diverse and highly interlinked processes that shape the evolution of gas and stars at the centers of LTGs.

The standard scenario \citep[see, e.g.,][and references therein]{KorKen04} distinguishes between classical bulges (CBs) and pseudobulges (PBs). The first are thought to emerge quasi-monolithically early on, prior to, and independently of the disk, and the second gradually build up over the billion-year-long secular evolution of LTGs through in situ SF and inward migration of stars and star-forming clumps from the disk.
The picture of two distinct routes to bulge formation is not undisputed, however. For example, \citet[][hereafter \textcolor{myblue4}{BP18}]{BP18} found no evidence for the age bimodality implied by the standard scenario from a spectral synthesis study of a representative sample of LTGs. Their analysis shows instead that bulges span a continuous sequence in age with the latter following a tight correlation with their present-day \mstar\ and \sstar.

Nevertheless, an elementary discussion of the relevance of \dio\ from the perspective of the standard bulge formation scenario may be attempted. Whereas a disk underneath the bulge is unlikely to exist just after the dominant phase of CB formation, it is conceivable that stellar diffusion and inward migration of material from the outer disk subsequently lead to the gradual build-up of an inner disk. 
However, it is not immediately apparent why such dynamical processes per se ensure the fine-tuning required to make the 
\sstar\ profile of the inner disk an exact inward extension of the exponential \sstar\ profile of the outer disk.
Furthermore, dynamical heating of the inner disk through its billion-year-long interaction with the kinematically hotter CB might inflate 
it into a triaxial entity that is hardly separable from the bulge itself. Such considerations suggest that CBs either lack an underlying inner disk, or that their inner disk, if present, substantially deviates from the exponential \sstar\ slope of the outer disk \citep[cf. the discussion in][]{Breda20b}. This would then call for a revision of structural determinations of CBs (e.g., absolute magnitude, S\'ersic exponent, and effective radius) on the basis of an adequate image decomposition scheme that allows for a down-bending of the disk inside \rbulge. 
Because a central depletion of the disk can from the photometric point of view be regarded as equivalent to an increase in the disk \ml\ to infinity, a high \dio\ should be a generic characteristic of CB-hosting LTGs and as such an indispensable element to consider in their study.

The amplitude of \dio\ in PBs likely depends on the relative contribution of migration and in situ SF to the stellar mass growth inside their \rbulge. The first process (inward migration of stars and SF clumps from the disk) leads to negative radial age gradients with a slope that inversely scales with the average migration velocity \textcolor{myblue4}{(B20a)} and can amplify preexisting \SFQ\ patterns at the centers of LTGs (cf. \textcolor{myblue4}{BP18} for a discussion)\footnote{This is due to the 'stellar mass filtering effect' discussed in \citet[][see also \textcolor{myblue4}{Papaderos \& \"Ostlin 2012}]{P02}, namely the depopulation of the high-mass end of the IMF with a main-sequence lifetime shorter than the migration timescale. As these authors remark, neglect of this effect can mimick a spatially varying IMF or lead to a systematic overestimation of age when color maps are used to age-date stellar populations.}. Quite importantly, because inward migration results in a higher \ml\ in the bulge than in the disk, it also naturally acts toward enhancing \dio\ in PB-hosting LTGs.

It follows from these considerations that the presence of \dio\ is consistent with the standard bulge formation scenario. On the other hand, without tight observational and theoretical constraints it is difficult to establish whether an inner disk invariably exists in both 
CBs and PBs, and, if so, whether its SFH is similar to that of the bulge (and bar). Addressing these questions not only requires further observational work (e.g., correction of bulge color gradients for a possible underlying inner disk, or a combined 
spectrophotometric and kinematical bulge-disk decomposition) but also the understanding of the influence of active galactic nuclei (AGN) on the SFH both within \rbulge\ and over the entire galaxy. This subject encapsulates several poorly understood aspects, such as the directionality and chronology of negative AGN feedback, and its possible coupling with galaxy mass \mstar. For instance, \textcolor{myblue4}{BP18} (their Fig. 6f) concluded from diagnostic emission-line ratios after \citet[][hereafter BPT]{BPT81} that gas excitation in bulges of lower-mass LTGs (log(\mstar/\msun)$\la$10) is dominated by photoionization by OB stars whereas more massive ones (log(\mstar/\msun)$\ga$10.5) generally fall on the locus of Composites, LINERs and Seyferts. This empirical insight, also supported by previous work \citep[e.g.,][]{Strateva01,KH13}, led these authors to conjecture that the SMBH-to-bulge mass ratio \mbh/\mbulge\ (or, alternatively, the Eddington ratio or SMBH spin parameter) increase with \mstar, with the transition between SF- and AGN-dominated gas excitation occurring at around 10$\la$log(\mstar/\msun)$\la$10.5 and a \sstar$\simeq 10^9$ \msun/kpc$^2$. 
Circumstantial support for this hypothesis comes from the observed inversion of radial stellar age gradients in bulges from zero and positive values in sub-L$^{\star}$ LTGs to predominantly negative values above log(\mstar/\msun)$\sim$10.5 \textcolor{myblue4}{(B20a)}.
\begin{center}
\begin{figure*} %[h!]
\begin{picture}(200,38)
\put(0,0){\includegraphics[height=3.8cm]{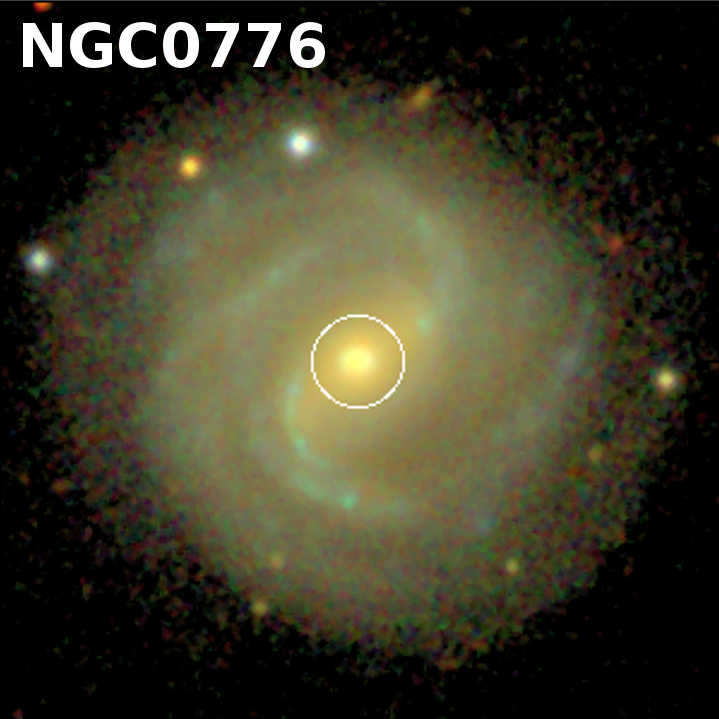}}
\put(40,0){\includegraphics[height=3.8cm]{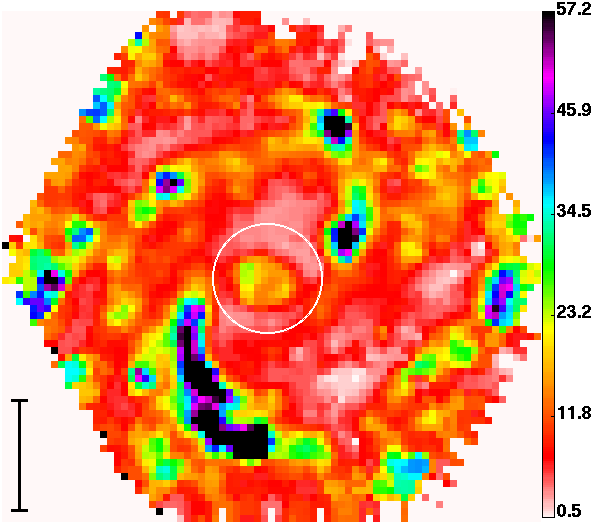}}
\put(86,-2){\includegraphics[height=4cm]{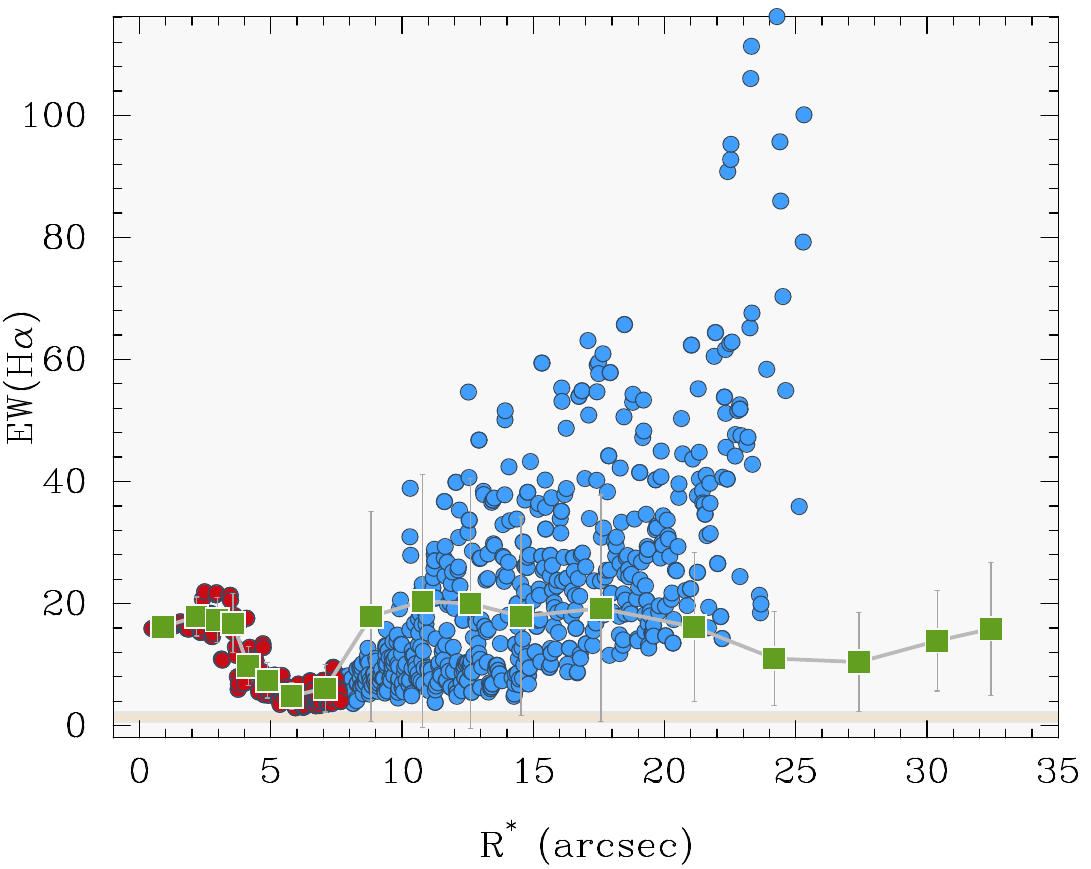}}
\put(138,-2){\includegraphics[height=4cm]{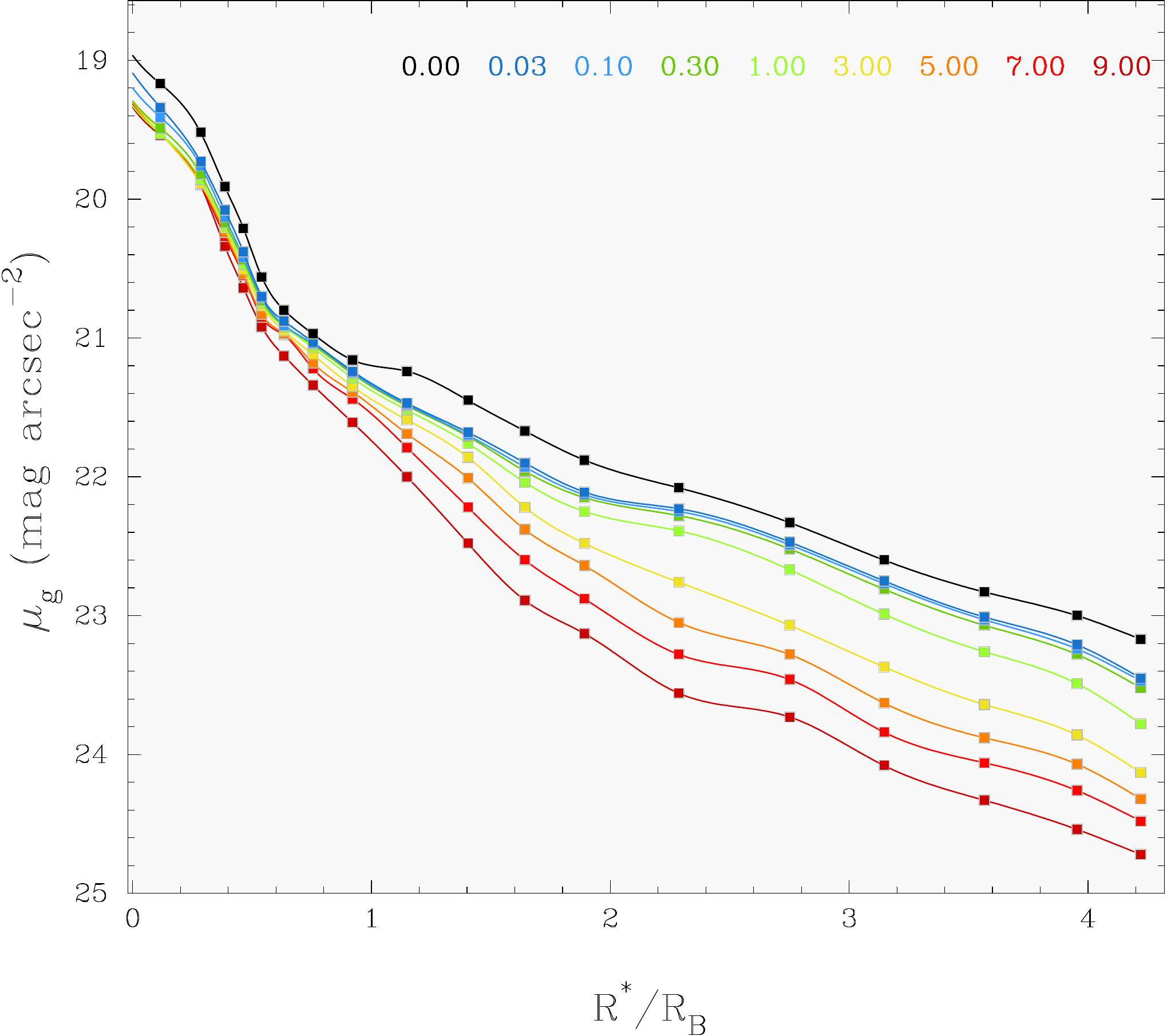}}
\PutLabel{2}{2}{\hvss \textcolor{white}{a}}
\PutLabel{42}{32}{\hvss b}
\PutLabel{94}{32}{\hvss c}
\PutLabel{145}{5}{\hvss d}
\end{picture}
\caption{\brem{a)} True-color SDSS image composite of the LTG \object{NGC 0776} (D=65.5 Mpc; log(\mstar/\msun)=11.1). The circle depicts the isophotal diameter of the bulge \citep[15\farcs3;][]{B19};
\brem{b)} \ewha\ map of the galaxy displayed in the range between 0.5 \AA\ and 57 \AA. The vertical bar corresponds to a projected scale of 5 kpc; \brem{c)} Radial \ewha\ profile, with dots showing determinations for single spaxels inside and outside \rbulge\
(red and blue, respectively) and squares corresponding to the mean \ewha\ within irregular isophotal annuli (\isan) adapted to the morphology of the emission-line-free stellar continuum (see \tref{BP18} for details). Error bars show the standard deviation about the mean of single-spaxel determinations within each \isan. The shaded gray horizontal stripe between 0.5 and 2.4 \AA\ marks the range in \ewha\ that can be accounted for by photoionization by the evolved ($>$100 Myr) post-AGB stellar component \citep[e.g.,][]{Cid11,GP16-ETGs}. \brem{d)} $g$-band SBPs of the galaxy after the suppression of stellar populations with ages younger than 0.03, 0.1, 0.3, 1, 3, 5, 7, and 9 Gyr (cf. the color-coding at the top right) 
from CALIFA IFS data. The profile for \tcut=0 (black) corresponds to the SBP that would be obtained when nebular line 
emission alone were removed.}
\label{ry1}
\end{figure*}
\end{center} 
% ================================================================================================
\subsection{Observational estimate of \dio\ from population spectral synthesis \label{ifs-dio}}
% ================================================================================================
A second estimate on \dio\ can be obtained from a comparison of the luminosity contribution of young stellar populations inside and outside the bulge. This is possible through spectral modeling of spatially resolved IFS data in conjunction with postprocessing of population vectors
(the best-fitting combination of spectral templates obtained from PSS) with the tool {\sc RemoveYoung} \citep[\RY;][]{GP16-RY}.
This stratigraphy (age slicing) technique consists of removing simple stellar population (SSP) spectra younger than an adjustable age cutoff \tcut\ from the population vector, of the subsequent reconstruction of the SED of the residual older stellar component, and, finally, of the computation of its apparent magnitude in various bands through convolution of its SED with (currently 25) filter transmission curves. An additional output from \RY\ is the stellar mass \mstar\ as a function of \tcut. The determination of the magnitude and mass of stellar populations within different age intervals is implicit to the concept of this tool and has been implemented in the IFS processing pipeline {\sc Porto3D} \citep{P13,GP16-ETGs}.

{\sc RemoveYoung} was for the first time systematically applied to IFS data for a representative sample of 135 LTGs spanning a stellar mass 8.9$\la$log(\mstar/\msun)$\la$11.5 by BP18.
Their sample was compiled from the CALIFA survey \citep{Sanchez12-DR1,Sanchez16-DR3} which was conducted with the Potsdam Multi-Aperture Spectrometer \citep[PMAS;][]{Roth05} in its PPaK mode \citep[][]{Verh04,Kelz06} and the V500 grating, and reduced as described in \citet[][and references therein]{GB15CALIFA}. Spectral modeling was carried out in a spaxel-by-spaxel mode with {\sc Starlight} \citep{Cid05} using SSPs from \citet{BruCha03} for 38 ages between 1 Myr and 13 Gyr (see \tref{BP18} for details). 
\RY\ was applied for nine \tcut\ back to an age of 9 Gyr \citep[an empirical estimate of the average lookback time where the age resolution of {\sc Starlight} models to CALIFA data becomes lower than 1 Gyr; cf.][]{B19} with the goal of investigating the mass assembly history of LTGs in a radially resolved manner.

While the \RY-based analysis by \tref{BP18} reinforces the insight that the centers of massive LTGs assemble and quench first \citep[e.g.,][]{Perez13,Tacchella15} it offers a novel route to the study of inside-out \SFQ\ through the determination of the luminosity fraction of stellar populations of different age to the surface brightness profiles (SBPs) of galaxies. 
In particular, \tref{BP18} proposed to use the difference \dmb\ (mag) between the mean $r$-band surface brightness within the bulge radius\footnote{These authors defined \rbulge\ as the extinction-corrected isophotal radius at which a S\'ersic model to the central luminosity excess of the bulge has a surface brightness of 24 $r$ \sbb. An alternative definition of \rbulge\ involves the radius at which the surface brightness of the disk becomes equal to that of the bulge \citep[e.g.,][]{San14}.} at \tcut=0 Gyr and 9~Gyr as a diagnostic for the evolutionary and physical properties of LTG bulges\footnote{Determinations for a \tcut=0 refer to the best-fitting stellar SED to an observed spectrum, that is, they only involve a correction for nebular line emission.}. 

The rationale for this was that \dmb\ tightly correlates with \mstar, \sstar, and the mass-weighted stellar age and metallicity of LTG bulges.
Based on \dmb, they tentatively subdivided their sample into three classes: galaxies hosting \brem{iA} bulges (\dmb$<$--1.5 mag) populate the low-mass end (log(\mstar/\msun)$<$10.3) of the LTG sequence and typically show the lowest bulge-to-total (B/T) mass ratio 
and bulge-to-disk age contrast. \brem{iB} bulges reside in LTGs with log(\mstar/\msun)$\sim$10.5 and show an intermediate \dmb, while high-mass \brem{iC} bulges in LTGs with log(\mstar/\msun)$\ga$10.7 are characterized by a \dmb\ greater than --0.5 mag. This last class is characterized by the highest B/T ratio and bulge-to-disk age contrast (up to $\sim$3 Gyr). The morphological and structural properties of these three LTG classes point to a loose association with bulgeless, pseudo-bulge, and classical-bulge LTGs, , respectively. 

Consistently with the observed increase in the bulge-to-disk age contrast with increasing galaxy mass (BP18; their Fig. 7c), SBPs computed with \RY\ for different \tcut\ witness a continued growth of the disk in all LTGs, regardless of their \mstar, in contrast to the evidence for a dependence of \SFQ\ on \mstar\ within the bulge (cf. Fig.~5 in \tref{BP18}): Suppression of stellar populations younger than 9 Gyr has virtually no effect on SBPs of high-mass \brem{iC} bulges, which implies that they have completed their assembly and entered the ensuing \SFQ\ early on. 

In contrast, removal of young stellar populations leads to a strong dimming by 1--2 $r$ mag in the disk of these most massive LTGs. 
At the antipodal end of low-mass LTGs (\brem{iA} class), suppression of stellar populations of increasingly high age leads to a roughly symmetric dimming both in the bulge and in the disk, which points to a nearly homologous radial growth of \sstar. 
Therefore, the combined evidence from the analysis of LTGs with \RY\ suggests a trend for an increasing \dio\ with increasing \mstar. This is consistent with the fact that the mass fraction of stars older than 9 Gyr increases by a factor $\sim$4 within the bulge (from 21\% in \brem{iA} bulges to 85\% in \brem{iC} bulges), but only by a factor $\sim$2 in the outer disk (\tref{BP18}).

Figure~\ref{ry1} shows an example of the application of \RY\ to the \brem{iC} LTG \object{NGC 0776}. From panel \brem{b}, it is apparent that the H$\alpha$ equivalent width \ewha\ increases from the center to the periphery, as also reflected in the radial distribution of values for individual spaxels (dots in panel \brem{c}) for which the spectrum was fit with a mean percentual deviation ADEV$<$6 \citep[cf.][]{Cid05,GP16-ETGs}. 
Squares correspond to mean values within irregular isophotal annuli (\isan) adapted to the morphology of the emission-line free continuum (cf. \tref{BP18} for details).
Panel~\brem{d} shows the $g$-band SBPs after suppression of stellar populations youger than 0.03, 0.1, 0.3, 1, 3, 5, 7, and 9 Gyr, as obtained by postprocessing the spectral modeling output from \starlight\ with \RY. Removal of stellar populations for all \tcut\ has little effect on the bulge, which, taken at face value, implies that SF at the center of the galaxy has been practically extinguished since at least 9 Gyr ago. In contrast, the outer disk (\rr/\rbulge$\ga$1) has continued building up, as is apparent from the fact that suppression of stellar populations younger than 9 Gyr effects a dimming by $\sim$1.5 $g$ mag.  

% :::::::: FIGURE 3 ::::::::::::::::::::::::::::::::::::
%\setlength{\unitlength}{1mm}
% :::::::: FIGURE 4 ::::::::::::::::::::::::::::::::::::
\begin{center}
\begin{figure}
\begin{picture}(86,64)
\put(8,0){\includegraphics[width=7cm]{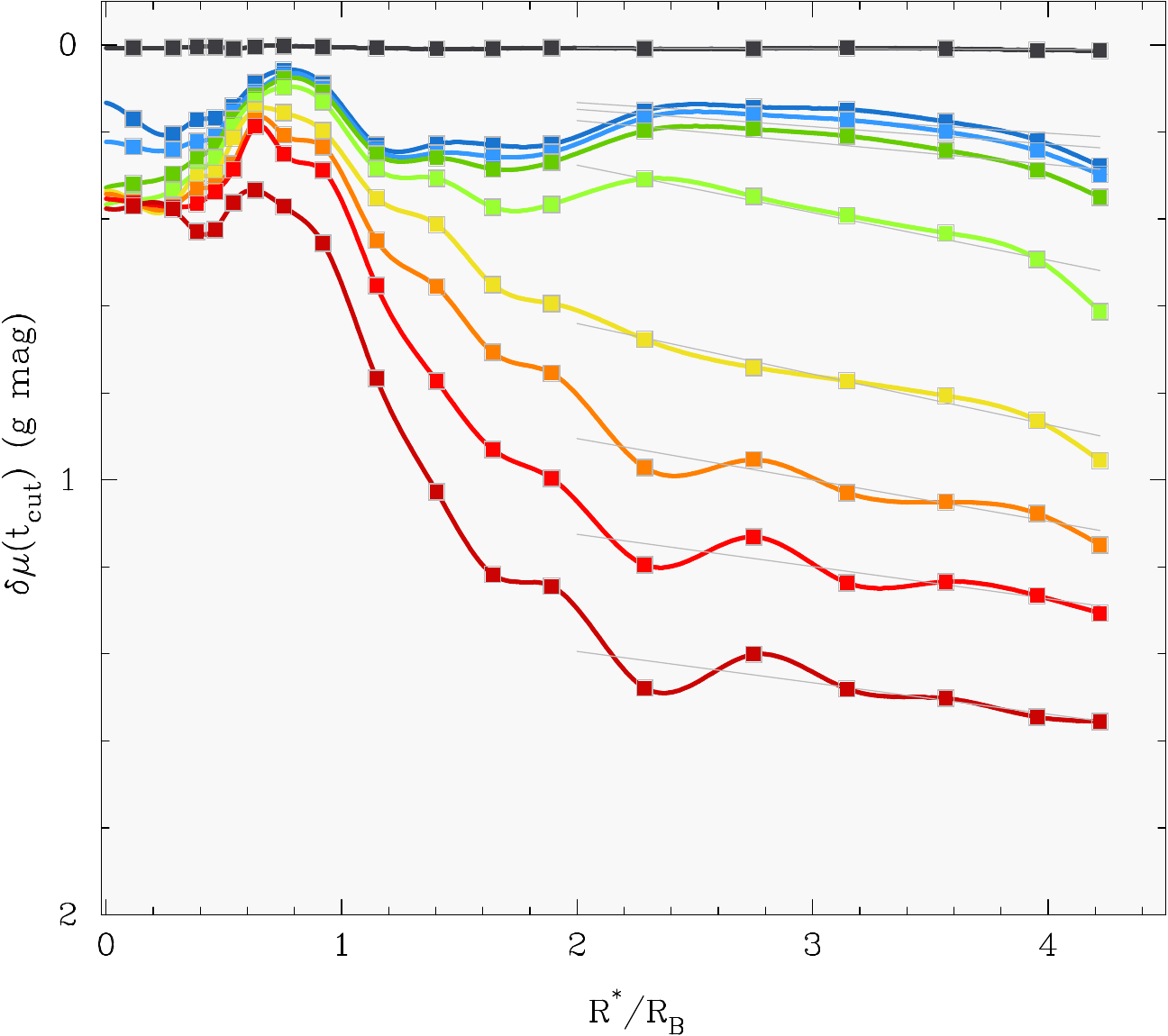}}
\end{picture}
\caption{Dimming $\delta\mu$ (mag) relative to the observed, emission-line corrected $g$-band SBP of \object{NGC 0776} (black) as a function of normalized radius \rr/\rbulge\ after suppression of stellar populations younger than a \tcut\ between 0.03 and 9 Gyr. 
The color-coding is same as in Fig.~\ref{ry1}d. Gray lines show linear fits for \rr/\rbulge$\geq$2, i.e. in a radius interval in which the contribution of the bulge and bar become negligible.} 
\label{ry2}
\end{figure}
\end{center}      
% :::::::: FIGURE 4 ::::::::::::::::::::::::::::::::::::

% :::::::: FIGURE 5 ::::::::::::::::::::::::::::::::::::
\begin{center}
\begin{figure}
\begin{picture}(86,74)
\put(0,49){\includegraphics[width=8.44cm]{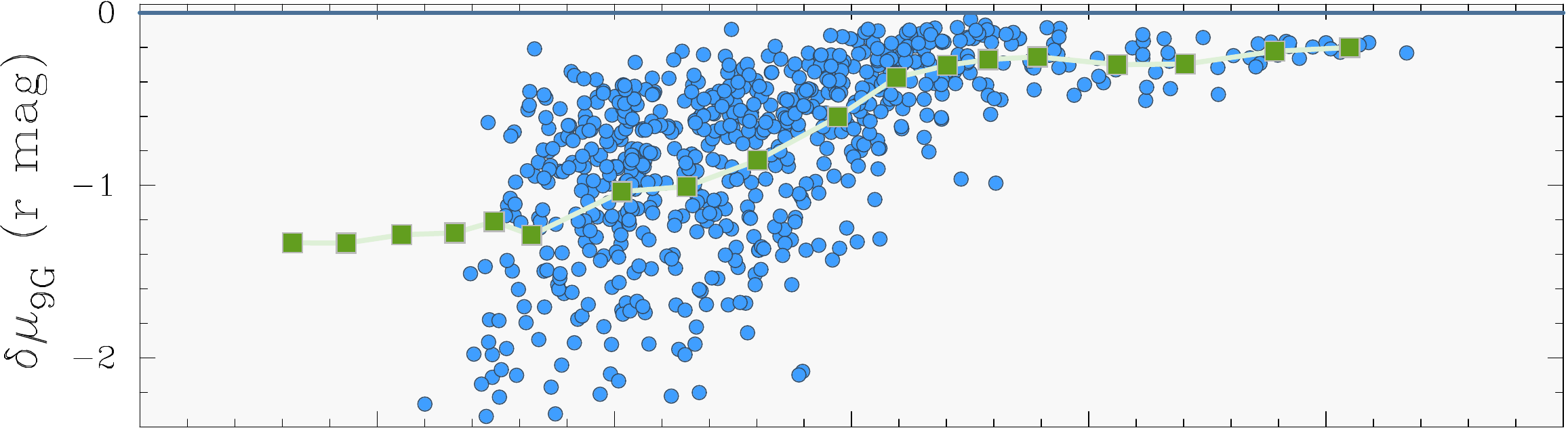}}
\put(0,0){\includegraphics[width=8.6cm]{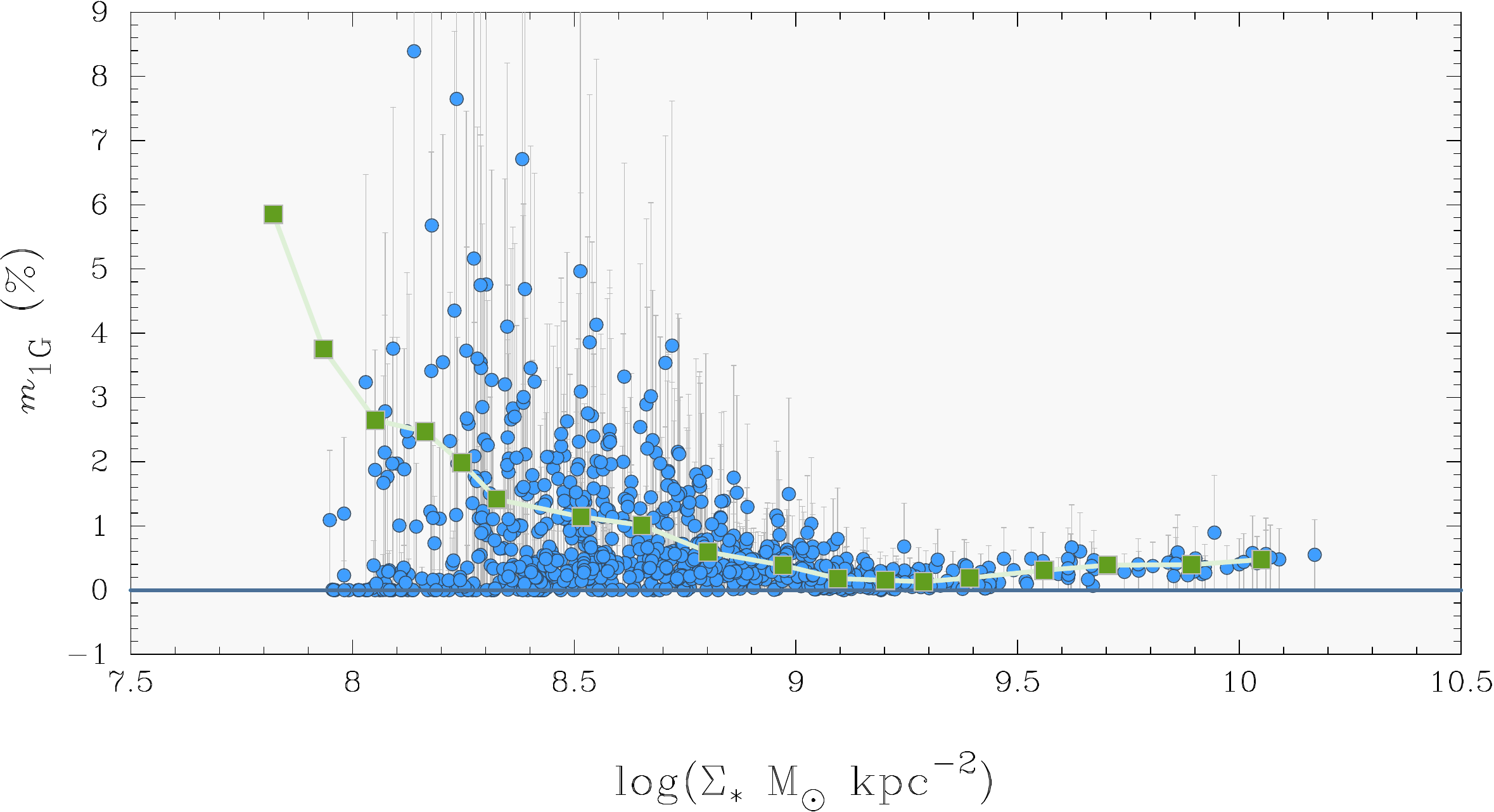}}
\PutLabel{79}{53}{\hvss a}
\PutLabel{79}{39}{\hvss b}
\end{picture}
\caption{\object{NGC 0776}: \brem{a)} Difference \dmb\ between the $r$-band magnitude of the old ($\geq$9 Gyr) and existing stellar component vs. logarithm of the stellar surface density \sstar\ (\msun\ kpc$^{-2}$). Symbols have the same meaning as in Fig.~\ref{ry1}\tref{c}. \brem{b)} Percentual mass fraction of stars formed over the past one Gyr (lower panel) vs. \sstar.} 
\label{ry3}
\end{figure}
\end{center}      
% :::::::: FIGURE 5 ::::::::::::::::::::::::::::::::::::

% :::::::: FIGURE extra 1 (illustration of the SHF) ::::
\begin{center}
\begin{figure}
\begin{picture}(86,92)
\put(4,0){\includegraphics[width=8cm]{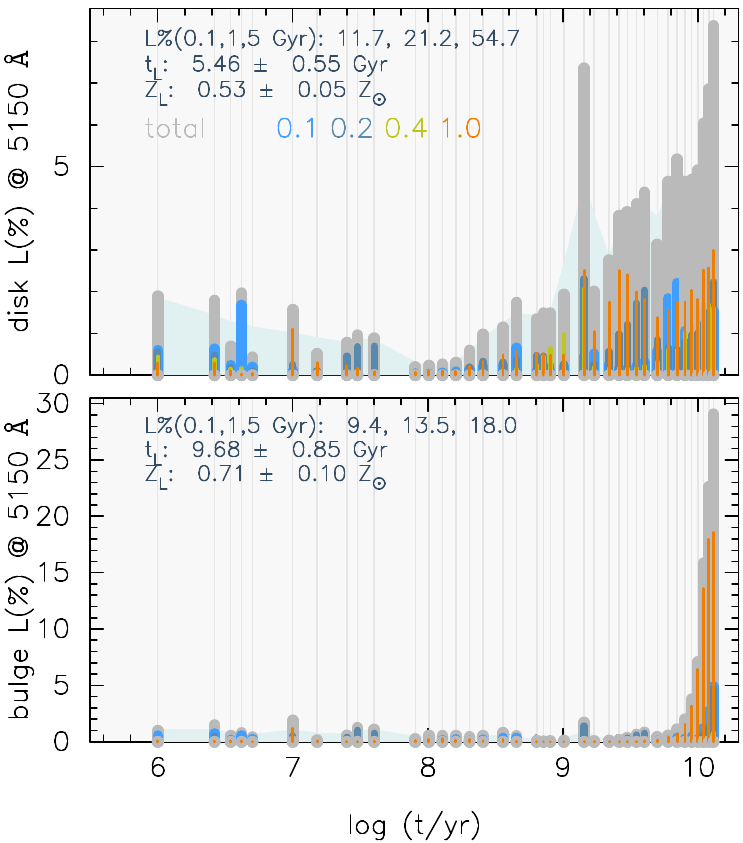}}
\end{picture}
\caption{\object{NGC 0776}: Comparison of the SFH in the bulge and disk (lower and upper panel, respectively).
Vertical bars, color-coded according to stellar metallicity (open blue, dark blue, green and orange for \zsun/10, \zsun/5, \zsun/2.5 and \zsun, respectively) show the contribution (in percent) of individual 
simple stellar populations (SSPs) to the monochromatic luminosity of the continuum at 5150 \AA.
The sum of all SSP contributions at a given age is shown in gray. Thin vertical lines mark the (38) ages available in the SSP library used (see BP18 for details).
The labels in the upper left part of each diagram inform about the luminosity fraction of stellar populations younger than 0.1, 1 and 5 Gyr, as well as the light-weighted stellar age and metallicity. The bulge luminosity is dominated by old stellar populations whereas more than 50\% of the emission in the disk comes from stars younger than 5 Gyr.
} 
\label{fig:sfh}
\end{figure}
\end{center}      
% :::::::: FIGURE 5 ::::::::::::::::::::::::::::::::::::

The differential growth of bulge and disk is better illustrated in Fig.~\ref{ry2} where we show the difference between the observed $g$-band SBP relative to the SBPs computed for the eight aforementioned \tcut\ values. The horizontal line corresponds to a dimming of 0 mag, and black symbols show \isan\ determinations for \tcut=0~Gyr, that is, after subtraction of nebular line emission only.
Suppression of young ($<$30 Myr) ionizing stellar populations effects a dimming of $\delta\mu\sim$0.2 mag, with the outer disk becoming fainter by $\sim$1.5 mag at a \tcut=9 Gyr, that is, within the range of values previously estimated from evolutionary synthesis models in Sect.~\ref{ev-dio}. The bar, originally traceable as a weak surface brightness enhancement at \rr/\rbulge$\ga$1 in the observed SBP (black; see also Fig.~2 of \tref{BP18} for SBPs from higher-resolution SDSS data with an FWHM=1\farcs3 as compared to an FWHM$\sim$2\farcs7 for CALIFA) is now better visible out to \rr/\rbulge$\sim$2. Moreover, a noticeable feature from Fig.~\ref{ry2} is a slight steepening of the disk $\delta\mu$ (gray lines depicting linear fits for \rr/\rbulge$>$2) for an age $>$1 Gyr, which might indicate inside-out galaxy growth.

That the build-up timescale of stellar populations is inversely related to their present-day \sstar\ (a trend referred to as local or 
subgalactic downsizing by \tref{Rosales-Ortega et al. 2012} and \tref{BP18} on the basis of spatially resolved IFS studies of stellar metallicity and age, respectively; see also \tref{Ganda et al. 2007}) is illustrated in Fig.~\ref{ry3}. 
Consistently with the evidence from Fig.~\ref{ry2}, the upper panel shows that \dmb\ increases by $>$1 mag from the low-\sstar\ disk periphery to the dense galactic center.
As opposed to the stagnation of SF activity within \rbulge, the continued growth of the outer disk is also reflected in its 
mass fraction of stars younger than 1 Gyr (lower panel), which is higher by a factor $\ga$4.

From these considerations it follows that only after removal of stellar populations of age $\la$9 Gyr does the excess emission of the outer disk (aka the \dio\ effect) in \object{NGC 0776} become sufficiently suppressed and its residual older stellar component become roughly compatible with the bulge from the evolutionary point of view (i.e., with regard to its SED and light-weighted age). Only then would the traditional bulge-disk decomposition concept be logically sound and ensure an unbiased determination of the luminosity and color of the bulge (and bar). Clearly, the exploration of an optimal age threshold \tcut\ that could permit such an age homogenization of stellar populations in the bulge and disk is a nontrivial task that requires further investigation.

Nevertheless, already the cursorily discussion here shows that subtraction of an exponential model to the \dio-corrected outer disk of an LTG can lead to a significant increase in the luminosity of the bulge, in addition to allowing a better recovery of a possibly present bar \tref{(Breda et al., in prep.)}. For instance, fitting an exponential to the SBP of \object{NGC 0776} for \tcut=9 Gyr (dark red curve in Fig.~\ref{ry1}\tref{d}), that is, the \tcut\ coming closest to the light-weighted age of the bulge 
($\sim$9.7 Gyr; cf. lower panel of Fig.~\ref{fig:sfh}), yields a total magnitude for the disk that is fainter by 1.5~mag 
than that determined from the observed (\tcut=0 Gyr) SBP (14.47 mag and 12.95 mag, respectively). 
This in turn implies a luminosity for the excess emission above the disk (i.e., bulge + bar) 
that is higher by a factor $\sim$3 and a rise in the B/T ratio of the galaxy from 0.15 to 0.43.
Clearly, a positive trend between the B/T ratio and the minimum present-day age (\tcut) of stellar populations extracted from an IFS cube is to be expected in an inside-out galaxy growth (and \SFQ) scenario.
% ============================================================================================================
\subsection{Effect of nebular emission on \dio\ in local LTGs \label{sect-ry-neb}}
% ============================================================================================================
The effect of nebular emission on broadband magnitudes and colors has been examined in several previous studies \citep[][among others]{Huchra77,Krueger95,Izotov97,P98,AFvA03,SdB09,Atek11,PapOst12} showing that it becomes important when the emission-line EWs rise above a few percent of the effective width of broadband filters.
Because the \ewha\ in the disks of centrally SF-quenched local LTGs (typically, \brem{iB} and \brem{iC} systems with log(\mstar/\msun)$>$10.3) is generally low \citep[$<$40 \AA; e.g.,][]{Kennicutt89,CatT17,Belfiore18,BP18,Kalinova21}, nebular emission is practically negligible in the context of \dio.
% ::: Fig. 6 (EW Ha vs dimming) :::::::::::::::::::::::::::::::::::::::::::::::::::::::::
\begin{center}
\begin{figure}
\begin{picture}(99,130)
\put(0,50){\includegraphics[width=8.6cm]{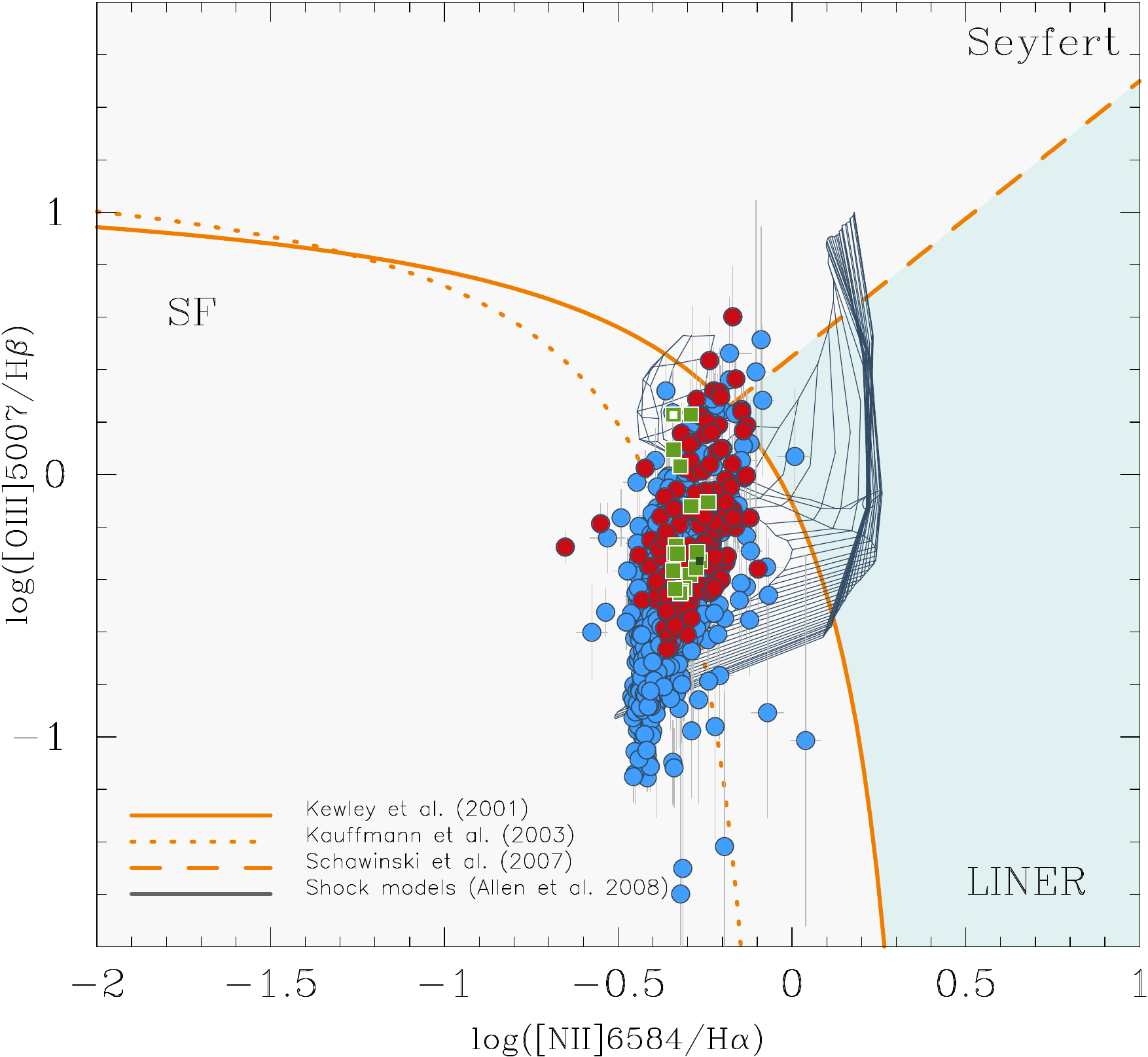}}
\put(0,0){\includegraphics[width=8.8cm]{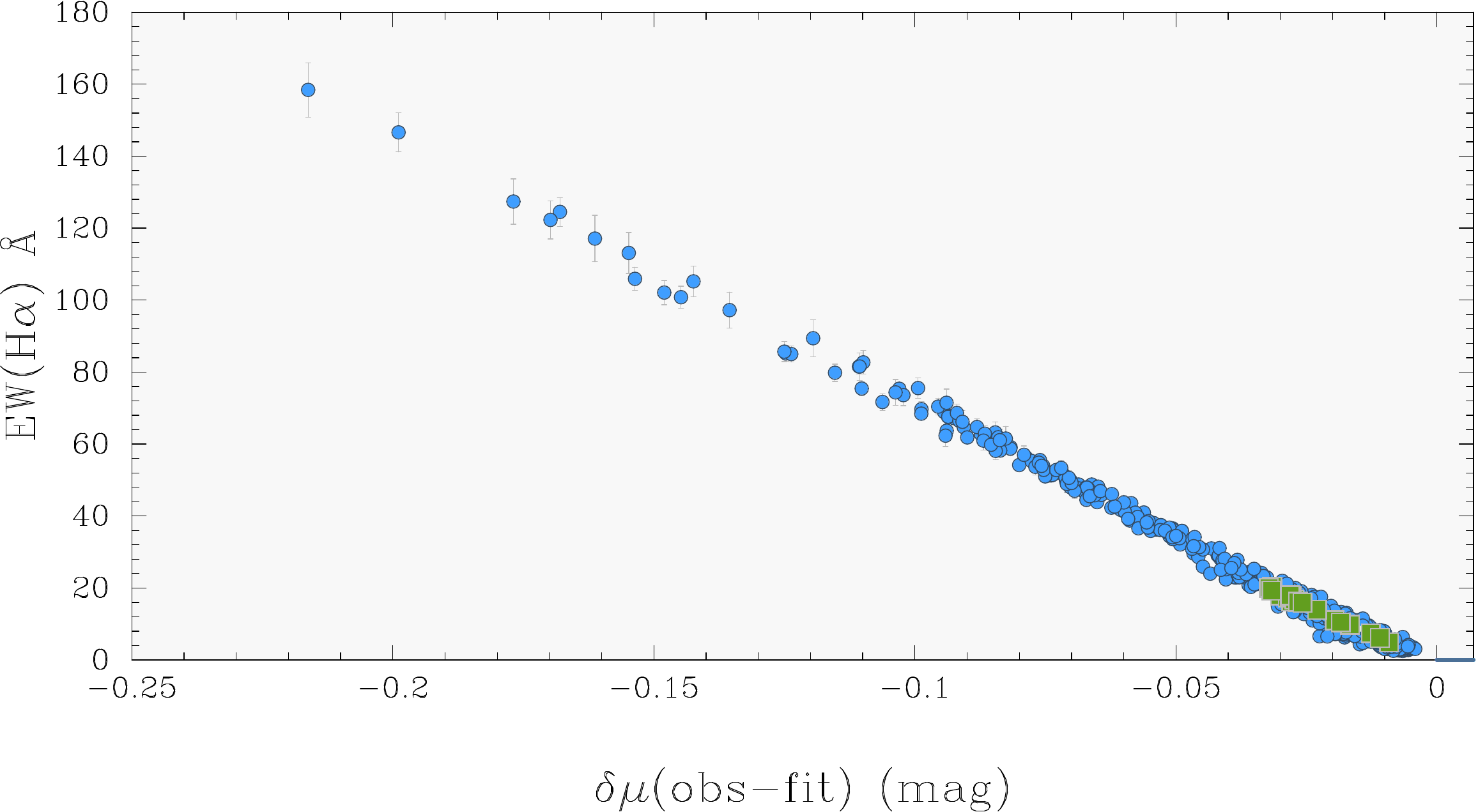}}
\end{picture}
\caption{\brem{a)} \tln2ha\ vs \tlo3hb\ diagram for \object{NGC 0776}. Single-spaxel determinations within \rbulge\ 
and in the outer disk (red and blue, respectively) are shown with dots, whereas squares depict average values within \isan. 
The locus that is characteristic of AGN and LINERs and the locus corresponding to photoionization by OB stars in SF regions are indicated with demarcation lines from \citet[][dotted curve]{Kauffmann03}, \citet[][solid curve]{Kewley01} and \citet[][dashed line]{Schawinski2007}. The grid of thin gray lines depicts the parameter space that can be accounted for by pure shock excitation, as predicted by \cite{Allen2008} for a magnetic field of 1~$\mu$G, and a range of shock velocities between 100 and 1000 \kmsec, for gas densities between 0.1 and 100 cm$^{-3}$. It is apparent that the diagnostic line ratios both for the bulge and outer disk fall close to the empirical envelope characterizing SF regions and in the locus of 'composite' sources (i.e. between the curves by \textcolor{myblue4}{Kauffmann et al.}
and \textcolor{myblue4}{Kewley et al.}.
\brem{b)} \object{NGC 0776}: Difference between the $r$-band magnitude of the observed spectrum and that of the stellar fit to it vs. \ewha\ (\AA) for individual spaxels (dots) and \isan\ (squares). A linear fit (line) indicates that nebular line emission in this 'composite'  system enhances the $r$-band emission by $\sim$0.7 mag per $10^3$ \AA\ in \ewha.}
\label{ry-neb}
\end{figure}
\end{center} 

 This can be illustrated again using \object{NGC 0776} as an example. BPT line ratios place both the bulge and the disk of this galaxy on the locus of SF or Composite sources (Fig.~\ref{ry-neb}\tref{a}). It may therefore be regarded as an example for a higher-mass LTG with an intermediate level of contamination ($\sim$40\%) of the \ha+[N{\sc ii}]$_{6548,6584}$ blend by nitrogen lines. Dots correspond to single-spaxel determinations within \rbulge\ (red) and in the disk (blue), and squares show mean values within \isan.
 
Panel~\tref{b} shows the observed \ewha\ versus the enhancement $\delta\mu$(obs-fit) of the $r$ magnitude due to nebular line emission (i.e., the difference between the magnitude of the observed spectrum and the stellar fit to it, as obtained by applying \RY\ for a \tcut=0). The low average \ewha\ within individual \isan\ ($<$20 \AA) translates into a negligible $\delta\mu$, with the effect of nebular line contamination becoming appreciable ($>$0.1 mag) only locally for a few individual spaxels in H{\sc ii} regions in which the \ewha\ rises above $10^2$ \AA. 
A linear fit yields the relation $\delta\mu \approx$0.74 $r$ mag/k\AA\ implying that nebular line emission does not notably 
enhance the surface brightness enhancement of the disk.

The situation is most certainly different in high-$z$ proto-LTGs where rest-frame EWs in the disk likely exceed several hundred \AA\ and nebular emission may even dominate optical broadband magnitudes.
If the early phase of bulge growth is driven by inward migration and coalescence of massive ($10^{8-9}$ \msun) star-forming clumps emerging out of violent disk instabilities \citep{Noguchi99,Bournaud07,Elm08,Mandelker14,Mandelker17} then a strong \dio\ in these systems could develop early on, within the theoretically estimated clump migration timescale \tmig\ 
of a few $10^8$ yr. In addition to negative age and \ml\ gradients (\tref{BP18}), a plausible expectation from clump migration is a bulge-to-disk EW contrast that develops as early as within the first one Gyr of galaxy evolution.

For example, an SF clump with \zsun/20 forming at a radius $\sim$7 kpc and reaching the center of the galaxy after a \tmig$\sim$700 Myr with a mean radial velocity of $\sim$9.8 \kmsec, will experience a decrease in its \ewha\ from initially $\sim 3\times 10^3$ \AA\ to 60 (2) \AA\ for an exponentially declining SFR with an e-folding time of 300 (100) Myr. The \ewha\ excess of the disk relative to the bulge translates by the empirical relation from Fig.~\ref{ry-neb}\tref{b} into a surface brightness enhancement $\delta\mu_0$ ($\sim$\dio) of $>$2 $r$ mag. The temporal evolution of the \ewha\ profile will of course depend on various factors, such as the fraction of clumps being massive enough to survive SF-driven feedback and reach \rbulge\ as dynamically bound entities \citep{Tamburello17}, the level of in situ SF in the bulge that is fed by inflowing intraclump gas \citep{Hopkins2012}, the dilution of emission-line EWs by the inwardly increasing stellar continuum, and the stellar metallicity.\\
If, on the other hand, bulge formation starts with a phase of dissipative gas collapse \citep[wet compaction in the notation by][]{DekelBurkert14} then a strong \dio\ should first emerge 
once a high-\sstar\ core surrounded by an outwardly propagating quenching wave \citep{Tacchella15} has developed. The radially evolving high-sSFR ring around the SF-quenched proto-bulge that is inferred from postprocessing of cosmological simulations in \citet{Tacchella16} probably implies an inversion of negative to positive \ewha\ gradients as the galaxy 
completes its wet compaction phase and enters the inside-out \SFQ.

Studies with the JWST, ELT and Euclid could address whether positive EW gradients in high-$z$ galaxies emerge prior to or after the appearance of a high-\sstar\ proto-bulge, thereby offering observational constraints on the relative role of clump migration and wet compaction during the early stage of bulge formation. Further valuable insights might be gained from high-resolution cosmological simulations incorporating a detailed treatment of nebular emission and its expected effect on radial color and EW profiles \citep[e.g.,][]{Hirschmann17}.

A fact deserving special attention in bulge-disk decomposition studies of higher-$z$ LTGs is that strong emission lines lead to a selective surface brightness enhancement of the disk (consequently, an increased \dio) in various redshift intervals, depending on photometric filter. 
For example, the $I$-band surface brightness of the disk is elevated at 
$z\simeq 0.17$ because of contamination by the \ha\ line, with a  second and third peak occurring at $z\sim$0.54 and $\sim$1.18 due to the [O{\sc iii}]$_{4959,5007}$ and [OII]$_{3727,29}$ lines. Similarly, the $H$-band surface brightness of the disk is enhanced by hydrogen Paschen lines at $z\sim 0.7$ and by the \ha\ line at $z\sim 1.4$.

The essential aspects of this issue were discussed in \citet{PapOst12}. These authors have examined the influence of nebular emission on photometric studies of high-sSFR galaxies that consist of a high-surface-brightness stellar core and a more extended nebular envelope\footnote{Whereas these authors used the local blue compact dwarf (BCD) galaxy \object{I Zw 18} to exemplify a high-$z$ protogalaxy where SF-feedback leads to extended nebular emission and the spatial decoupling of ionized gas from ionizing stellar clusters, their considerations also apply to any other source of energy and momentum surrounded by a nebular envelope, for example, quasars with a Ly$\alpha$ halo \citep[e.g.,][]{VM07,Humphrey13,Borisova16,Wisotzki16,Lec20}.}.
They showed that the shift of various strong emission lines to filter transmission curves across $z$ leads to a wide range of  
combinations of colors and a core-to-envelope color contrast $\delta_{\rm ce}$. These, when interpreted in terms of purely stellar SEDs, or when ignoring the fact that the (stellar+nebular) SED of a galaxy varies with galactocentric radius, can prompt a variety of erroneous conclusions on the nature and evolutionary status of a higher-$z$ galaxies. For example, as these authors remark, at $0.15\la z \la 0.3$, the large $\delta_{\rm ce}$ ($\sim$0.8 mag) in $V$-$I$ and  moderately blue colors of the core (0.5 mag) superficially suggest an old disk hosting nuclear SF activity. The opposite conclusion would be drawn from the $\delta_{\rm ce}$ in $B$-$V$ ($\sim$0.5 mag), which could be taken as evidence for a young stellar disk encompassing a slightly older core. Likewise, in other $z$'s, observed-frame colors together with an overall low $\delta_{\rm ce}$ could lead to the classification of a starburst galaxy as quiescent.

From Fig.~15 by \citet{PapOst12} is also apparent that nebular line contamination does not boost \dio\ solely in narrow $z$-intervals that could easily be excluded from automated bulge-disk decomposition studies of large galaxy samples, but affects several broad windows in $z$, a fact requiring a careful treatment of this issue. It is also important that the selective enhancement of \dio\ 
in these $z$-intervals, correspondingly an artificial dimming and reddening of the bulge relative to the disk (cf. Sect. \ref{photometry}), could affect the net (disk-subtracted) SED of the bulge such as to potentially mimic bulge growth in discrete major episodes at various $z$ or might lend superficial support to a duality in the origin of bulges.
% ==================================================================================
\section{Effect of \dio\ on photometric properties of bulges \label{photometry}}
% ==================================================================================
Before turning to potential implications of \dio\ on galaxy scaling relations (Sect. \ref{dis}), 
we briefly discuss here its expected principle effect on determinations of photometric properties of bulges.

% ==========================================================================================================
\subsection{Effect of a centrally quenched disk on luminosity determinations for the bulge \label{profiles}}
% ==========================================================================================================
The range of \dio\ estimated in Sect.~\ref{dio} (up to $\sim$2.5 mag in $B$ and $\sim$0.7 mag in $K$) translates into an overestimation 
of the disk below the bulge by a factor between $\sim$1.7 and $\sim$9. 
The underestimation of the bulge luminosity as result of the oversubtraction of the underlying disk depends on two competing factors, 
namely the \dio\ itself and the luminosity ratio \bid\ of the bulge to the inner disk.
For this reason a faint bulge residing in a disk with a low \dio\ might be more affected than a prominent bulge 
in a fully SF-quenched disk with a high \dio.

\begin{center}
\begin{figure}
\begin{picture}(86,72)
\put(2,0){\includegraphics[width=8.4cm]{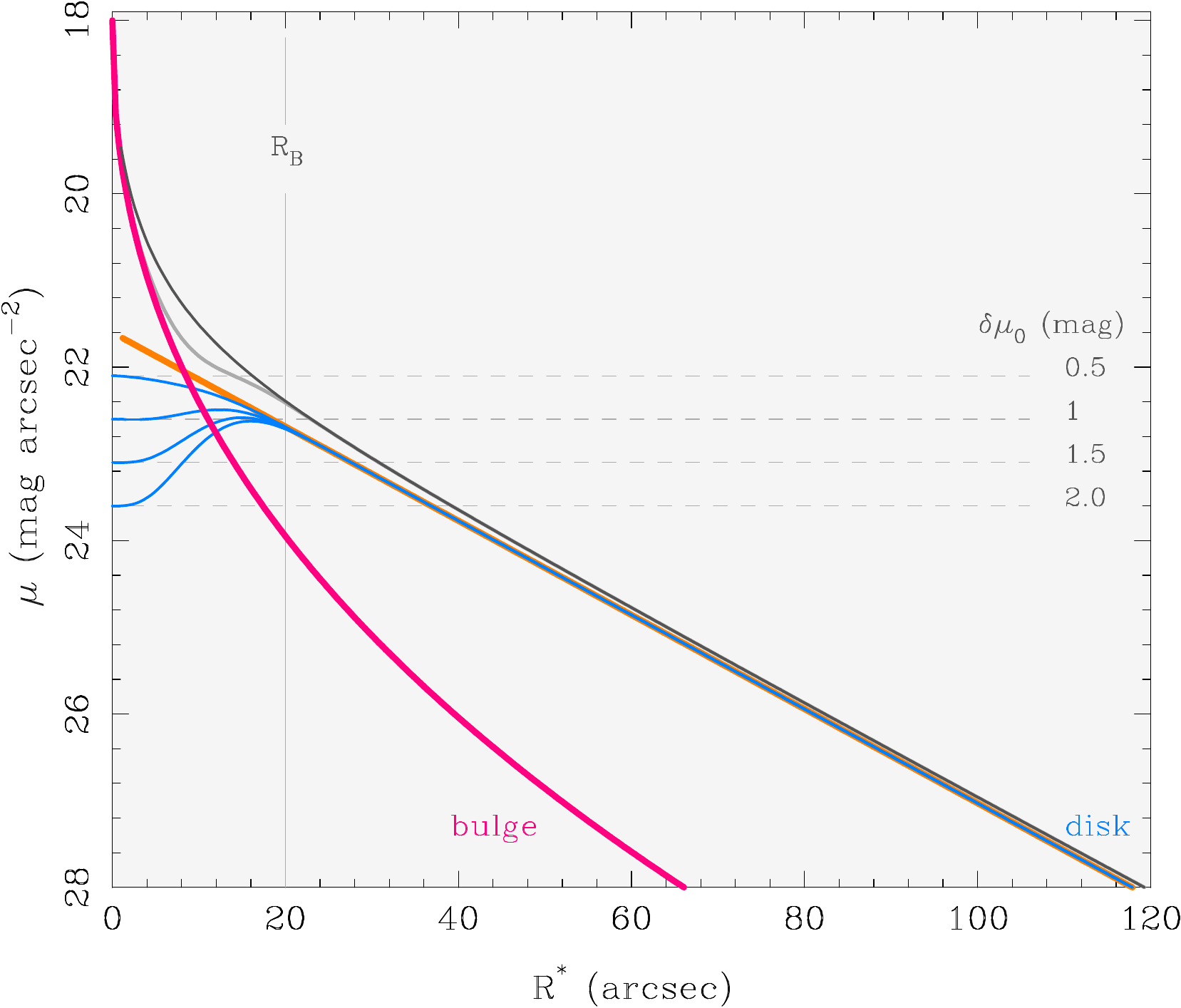}}
\end{picture}
\caption{Synthetic surface brightness profile (SBP) of a galaxy that consists of a bulge (red; approximated by a S\'ersic model) 
and a more extended disk (blue; approximated by the \brem{modexp} function; cf. Eq.~\ref{modexp1}).
Inside the isophotal radius of the bulge at 24 \sbb\ (\rbulge=20\arcsec) star formation quenching leads to a central 
dimming of the disk by 0.5$\leq \delta\mu_0 \leq$2 relative to the exponential fit for \rr$\geq$\rbulge\ (orange).
The composite bulge+disk SBP for a $\delta\mu_0$ of 0.5 mag and 2 mag is shown with the dark and light gray curves, , respectively. 
\label{sb1}} 
\end{figure}
\end{center}

This can be illustrated by the example of a synthetic LTG (Fig.~\ref{sb1}) that consists of an old bulge and a centrally SF-quenched disk. The $B$-band SBP of the bulge is approximated by a S\'ersic profile,
\begin{equation}
\mu(R^{\star})_B = \mu_{\rm 0,B} + 1.086 \cdot (R^{\star}/\beta)^{1/\eta} \label{eq:Sersic}
\end{equation}
with $\mu_{\rm 0,B}$=18 \sbb, $\beta$=0\farcs4 and $\eta$=2.3, and the disk by the modified exponential (\brem{modexp}) function proposed in \citet{P96a}, which allows a depression of an exponential intensity profile  
inside a cutoff radius:
\begin{equation}
I(R^{\star}) = I_{\rm exp} \cdot
\big[1-\epsilon_1\,\exp(-P_3(R^*))\big]
\label{modexp1} 
\end{equation}
where $P_3(R^{\star})$ is defined as
\begin{equation}
P_3(R^*) =
\left(\frac{R^{\star}}{\epsilon_2\,\alpha}\right)^3+\left(\frac{R^{\star}}{\alpha}\,\frac{1-\epsilon_1}{\epsilon_1}\right).
\label{modexp2} 
\end{equation}
The first term in Eq.~\ref{modexp1} stands for the standard exponential law $I_{\rm exp} = I_{\rm 0}\cdot \exp(-R^{\star}/\alpha)$, where $I_0$ and $\alpha$ are the central intensity and angular scale length of an exponential profile, respectively.
The \brem{modexp} function involves two further parameters, namely the central intensity depression 
$\epsilon_1=\Delta I/I_0$ relative to the exponential model and the core radius $R_{\rm c}=\epsilon_2 \cdot \alpha$
within which the flattening occurs\footnote{The \brem{modexp} was developed to approximate the underlying stellar host of local BCDs, for which indirect evidence for a central flattening exists \citep{P96a,Noeske03}, similar to several normal and nucleated dwarf ellipticals \citep[cf., e.g.,][]{BinCam93}. This flexible functional form also permits simulating a centrally depleted disk profile \tref{(B20b)}. 
Details about the connection between $\epsilon_{1,2}$ and \dio\ are provided in the appendix.}.
Here, we adopt for the disk an (extrapolated) central surface brightness $\mu_{\rm 0,D}=21.6$ \sbb, an $\alpha$=20\arcsec, and an $\epsilon_2=1$, meaning that the \brem{modexp} profile deviates from the exponential law inward of one disk scale length. 
For simplicity, we assume that the cutoff radius ($R_{\rm c}\approx$20\arcsec) roughly corresponds to the isophotal radius of the bulge at 24 $B$ \sbb\ (cf. Fig.~\ref{sb1}). 
The depression parameter $\epsilon_1$ can be adjusted such as to simulate profiles with a central disk surface brightness being by $\delta\mu_0$ (mag) fainter than $\mu_{\rm 0,D}$. With the condition $\epsilon_1$=dex($\delta\mu_0$/-2.5)-1, this translates into an $\epsilon_1$ between 0.37 and 0.84 for 0.5$\leq \delta\mu_0 \leq$2. 

The apparent magnitude of the bulge (14.27 mag) and the galaxy (12.8 mag) yield a bulge-to-total (B/T) and bulge-to-disk (B/D) ratio of 0.26 and 0.35, respectively, which is in the range of typical values for local high-mass LTGs with log(\mstar/\msun)$>$11 \citep[e.g.,][]{Men17}. 
% :::::::: FIGURE 7 ::::::::::::::::::::::::::::::::::::::::::::::::::::::::::::::::::::::::::::
%\setlength{\unitlength}{1mm}
% :::::::: FIGURE 7 ::::::::::::::::::::::::::::::::::::::::::::::::::::::::::::::::::::::::::::
The superposition of a S\'ersic with a \brem{modexp} profile yields a good match to the SBPs of typical LTGs \citep[e.g.,][among others]{dJvK94,deJong96a,And94,And95,GA01,DP08,Martinsson13,Erwin15,Erwin21,Barsanti21,Costantin21}. Quite importantly, it is in practice impossible to visually infer the presence of even a strong ($\delta\mu_0$=2) central down-bending in the disk 
from such a composite profile because the bulge dominates the line-of-sight intensity out to $\sim$\rbulge/2 (see also \textcolor{myblue4}{B20b}). 

\begin{center}
\begin{figure}[h] 
\begin{picture}(86,58)
\put(0,0){\includegraphics[width=8.6cm]{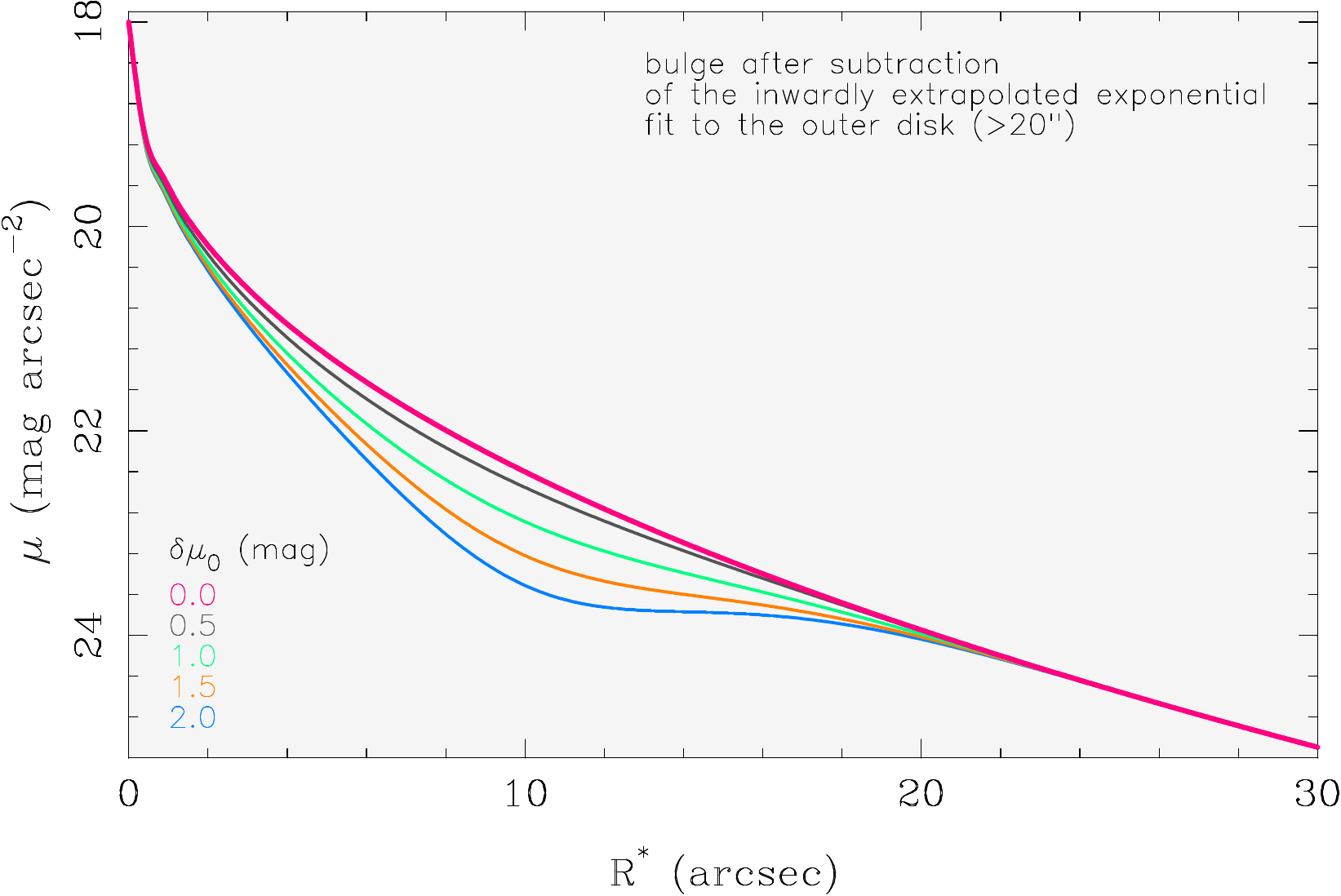}}
\end{picture}
\caption{Net emission of the bulge, as obtained by subtracting the inwardly extrapolated exponential fit for the visible outer disk at \rr$>$20\arcsec\ from the total (S\'ersic+\brem{modexp}) SBP in Fig.~\ref{sb1}.
\label{sb2}} 
\end{figure}
\end{center}

Figure~\ref{sb2} shows the net bulge emission that would be obtained by subtracting an exponential fit to the outer disk ($>$20\arcsec)
from the composite SBP in Fig.~\ref{sb1}. This example, which essentially simulates the standard practice in 1D and 2D decomposition,
illustrates that the deviation of the retrieved profile for the bulge from its true S\'ersic form (red) increases with \dio,
with a tendency for a steepening within the bulge effective radius (\reff$\approx$11\arcsec) and the appearance of a plateau at \reff$\la$\rr$\la$\rbulge.
% :::::::: FIGURE 8 ::::::::::::::::::::::::::::::::::::::::::::::::::::::::::::::::::::::::::::
%\setlength{\unitlength}{1mm}
% :::::::: FIGURE 8 ::::::::::::::::::::::::::::::::::::::::::::::::::::::::::::::::::::::::::::
Depending on the specifics of the fitting procedure (in particular, the error assigned to the inner points and corrections 
for the point spread function, PSF; cf. Sect.~\ref{bd}), the S\'ersic model parameters for the bulge can be biased in various ways
\citep[][for further remarks]{Breda19}. For instance, fitting only its inner part (1\farcs5$\leq$\rr$\leq$\reff) would 
in the case of $\delta\mu_0$=2~mag an $\eta\sim$1.75 yield, thus leading to the classification of the input classical bulge ($\eta$=2.3) as a pseudo-bulge. 
Evidently, the luminosity of the bulge is increasingly underestimated with increasing \dio\ (aka $\delta\mu_0$): 
the bulge apparent magnitude inside \rbulge\ (14.64 mag) decreases to 14.75, 14.94, 15.09 and 15.21 for a $\delta\mu_0$
of 0.5, 1.0, 1.5 and 2.0 mag, respectively, which for the synthetic model under study translates into an error of up to 40\%. 
Additionally, the underestimation of the luminosity of the bulge scales inversely to the bulge luminosity itself, consequently, 
the described bias is aggravated for intrinsically faint bulges (i.e., a low \bid\ ratio)): 
if the simulated bulge in Fig.~\ref{sb1} were fainter by just one mag fainter, then already a $\delta\mu_0$=0.5 (1.0) mag 
would result in the underestimation of its luminosity by 47\% (84\%)\footnote{See Sect.~\ref{app:dimming} for further quantitative estimates as a function of the true magnitude of the bulge and the intensity profile of the disk.}.

Another insight from our foregoing remarks is that, in the presence of \SFQ, standard bulge-disk decomposition 
should lead to discrepant determinations of the bulge S\'ersic $\eta$ in different filters, in the sense of 
a positive correlation between $\eta$ and central filter wavelength.
This is because the effect of the oversubtraction of the star-forming outer disk on the net SBP of the bulge is stronger in the blue
(because \dio\ $\propto \lambda^{-1}$; cf. Fig.~\ref{mlratio}). Together with the evidence from Fig.~\ref{sb2}, this 
implies that the underestimation of S\'ersic $\eta$ decreases with increasing $\lambda$ and becomes minimal in the NIR.
Instead of attempting to eliminate this seeming discrepancy by forcing S\'ersic model parameters to smoothly vary along $\lambda$ according to an empirical functional form \citep[e.g.,][]{H13,Vika14}, we might take advantage of it in order to obtain
indirect constraints on \dio\ and the true radial intensity profile of the disk below the bulge.

% :::::::: FIGURE 9 ::::::::::::::::::::::::::::::::::::::::::::::::::::::::::::::::::::::::::::
%\setlength{\unitlength}{1mm}
\begin{center}
\begin{figure}
\begin{picture}(86,60)
\put(0,0){\includegraphics[width=8.6cm]{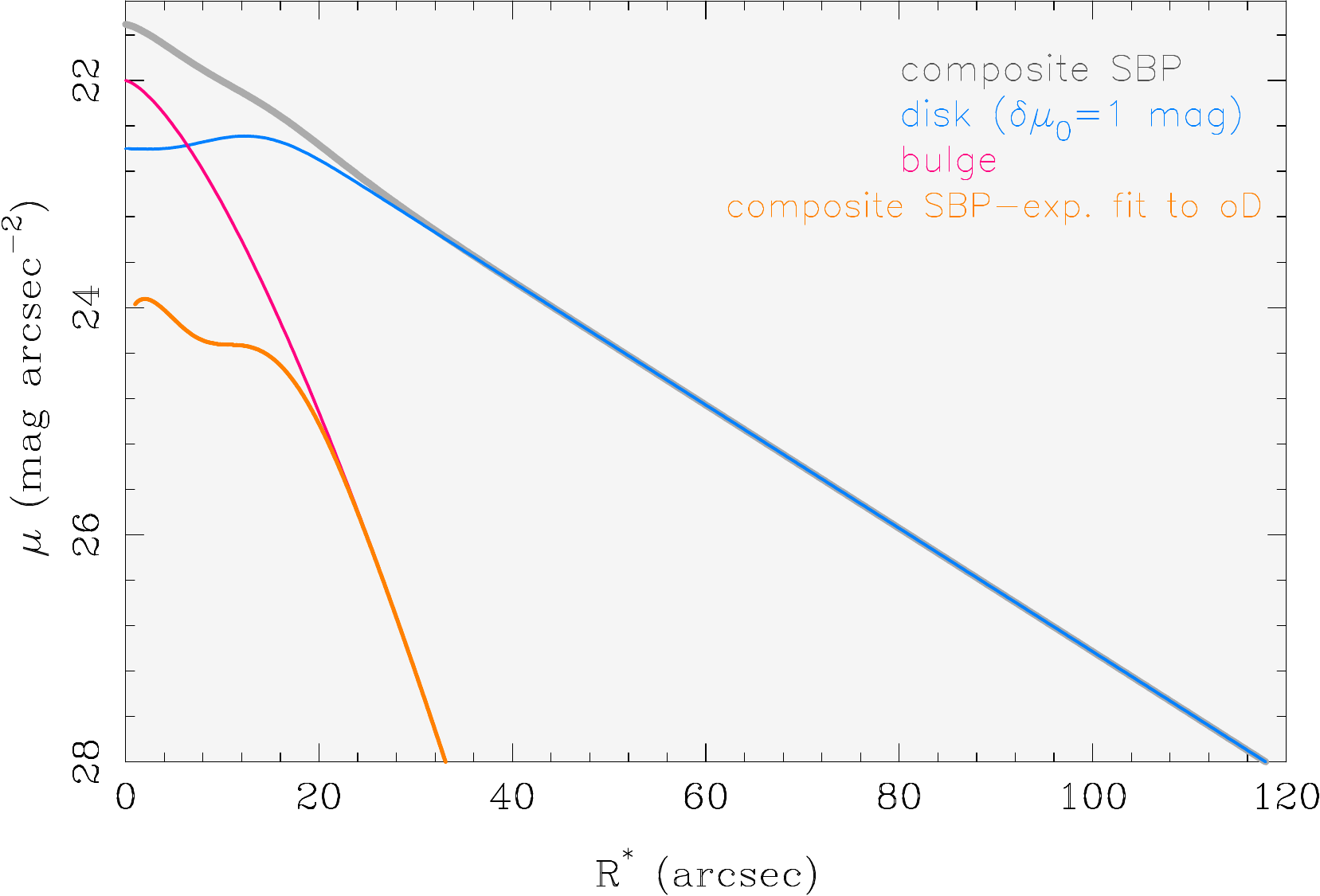}}
\end{picture}
\caption{Simulation of a seemingly bulgeless galaxy that consists of a faint bulge (red; B/T ratio $\sim$0.1) and an  
SF-quenched disk (blue) with $\delta\mu_0$=1 mag. The residual bulge emission after subtraction of an exponential model to the outer
disk (\rr$>$20\arcsec) is shown in orange.
\label{sim2}} 
\end{figure}
\end{center}      
% :::::::: FIGURE 9 ::::::::::::::::::::::::::::::::::::::::::::::::::::::::::::::::::::::::::::
% ::::::::::::::::::::::::::::::::::::::::::::::::::::::::::::::::::::::::::::::
\subsection{Bulgeless galaxies from the perspective of \dio\ \label{bulgeless}}
% ::::::::::::::::::::::::::::::::::::::::::::::::::::::::::::::::::::::::::::::
The existence of bulgeless galaxies, that is, LTGs with a minor, if any at all, central luminosity excess 
above the disk \citep[e.g.,][]{Kormendy10,Coelho13,Bizzocchi14,Grossi18} 
has triggered an intense controversy \citep[e.g.,][]{Governato10,Pilkington11}, 
given that these galaxies are not expected in the $\Lambda$CDM cosmology.

In the light of our previous considerations, it is worth contemplating whether the conclusion that these systems 
lack a bulge of appreciable mass remains compelling. 
Figure~\ref{sim2} simulates an LTG with a B/T ratio of $\sim$0.1 through the superposition 
of the same \brem{modexp} model for the disk as in Fig.~\ref{sb1} (blue) with a shallower S\'ersic profile 
for the bulge (red) that is given by $\mu_{\rm 0,B}$=22 \sbb, $\beta$=10\arcsec\ and $\eta$=0.7. 
The composite SBP (gray curve) is a nearly perfect exponential over 6 mag in surface brightness. 
A fit to the outer disk severely underestimates the luminosity of the bulge ($\sim$16.7 mag, i.e. 1.2 mag fainter 
than its true apparent magnitude of 15.5 mag), which now appears as a low-surface brightness 
($>$24 \sbb) core enclosing just $<$4\% of the total emission of the galaxy.

The main insight from Fig.~\ref{sim2} is that even a modest \dio\ ($\delta\mu_0\sim$1 mag) would comfortably allow a low-luminosity bulge to coexist with a centrally SF-quenched disk in a nearly perfect exponential composite SBP and evade detection with the standard bulge-disk decomposition technique. Although \dio\ is mitigated in the $I$ band and in the NIR, this alone does not warrant that a putative bulge can invariably be recovered in a bulgeless LTG because its detectability also depends on its luminosity (or the \bid) and the core radius $R_{\rm c} = \epsilon_2 \cdot \alpha$ of the SF-quenched central zone of the disk (Sect.~\ref{profiles} and \ref{app:dimming}).

Summarizing, it is in practice impossible to photometrically rule out the presence of a centrally SF-quenched disk beneath the bulge.
Numerous combinations of a \brem{modexp} profile for the disk and a S\'ersic profile for the bulge are possible that adequately  
reproduce typical SBPs of local LTGs over their entire morphological spectrum, from bulge-dominated all the way to prima facie bulgeless systems.   

% ================================================================================================
\subsection{Effect of a centrally quenched disk on color profiles within the bulge \label{colors}}
% ================================================================================================
% :::::::: FIGURE 8 ::::::::::::::::::::::::::::::::::::::::::::::::::::::::::::::::::::::::::::
%\setlength{\unitlength}{1mm}
\begin{center}
\begin{figure}[!ht] 
\begin{picture}(86,104)
\put(0,52){\includegraphics[width=8.6cm]{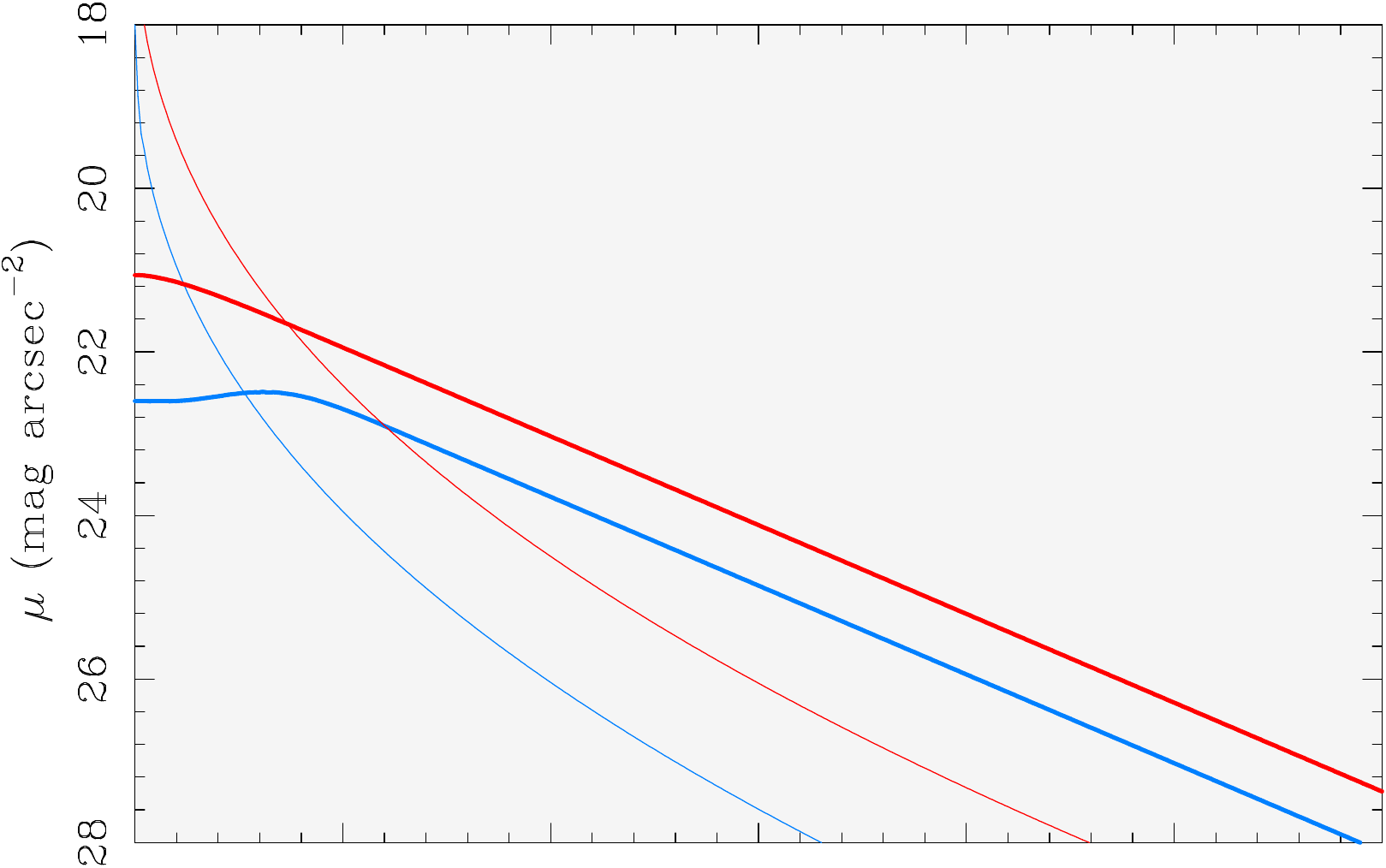}}   
\put(0.5,0){\includegraphics[width=8.75cm]{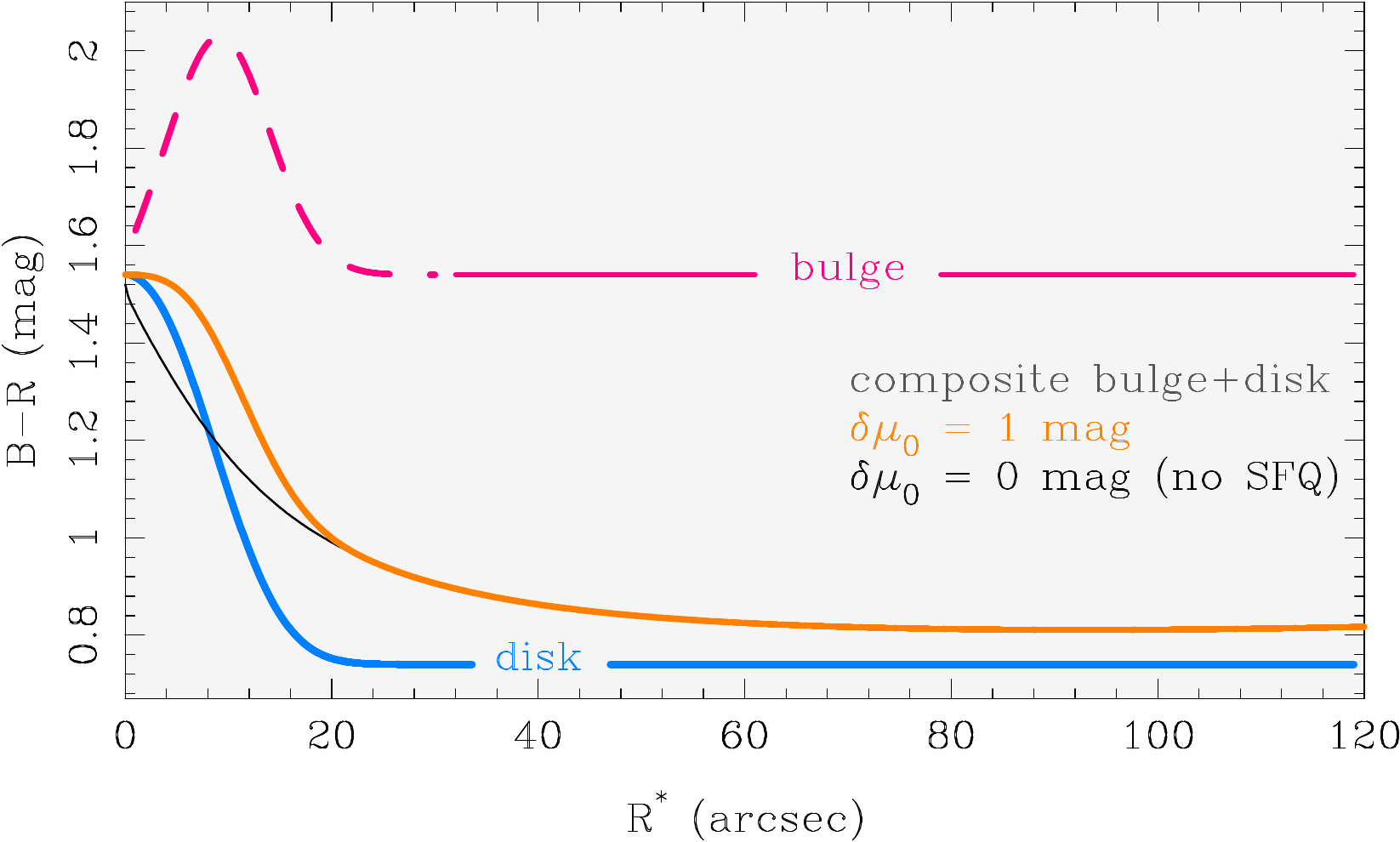}} 
\end{picture}
\caption{\brem{upper panel:} Surface brightness profiles in $B$ and $R$ (blue and red, respectively) for the bulge and disk (thin and thick curves, respectively) 
for a synthetic galaxy with a constant $B$-$R$ color of 1.54 mag in its bulge and 0.74 mag in its outer disk (\rr$>$\rbulge=20\arcsec). 
The disk is modeled by \brem{modexp} profiles with a different central depression parameter $\epsilon_1$ in $B$ and $R$ such as to simulate a central SF quenching with a $\delta\mu_0$=1~mag in $B$ and a color that smoothly increases inward of \rbulge\ from 0.74 mag to 1.54 mag (cf. lower panel).
\brem{lower panel:} True $B$-$R$ color of the bulge (solid red line), true-color profile of the disk (blue), and 
observed color profile resulting from the superposition of the bulge and the disk in the case of \SFQ\ (orange) and no \SFQ\ (black). The dashed red curve shows the color profile that would be obtained for the bulge if the correction for the underlying disk were made by assuming a pure exponential (instead of a \brem{modexp}) profile for it, that is, when \SFQ\ is neglected.
\label{col0}} 
\end{figure}
\end{center}      
% :::::::: FIGURE 8 ::::::::::::::::::::::::::::::::::::::::::::::::::::::::::::::::::::::::::::
Color gradients within \rbulge\ may hold valuable clues to bulge formation and evolution. It is thus worthwhile to examine how they are 
influenced by the line-of-sight intensity contribution of the underlying disk, both in the presence and absence of \SFQ.

For the sake of our discussion, we assume that the bulge in Fig.~\ref{sb1} has a constant $B$(Johnson)-$R$(Cousins) color of 1.54 mag, in agreement with predictions from {\sc P\'egase}~2.0 \citep{FRV97} for \sfhc\ and an age of $\sim$13 Gyr (cf. Fig.~\ref{mlratio}).
The disk is approximated both in $B$ and $R$ by a \brem{modexp} with the same $\alpha$ and cutoff radius (20\arcsec$\equiv$\rbulge), but a slightly different central intensity depression $\epsilon_1$, such as to obtain a $\delta\mu_0$ of 1 mag and 0.2 mag, respectively (upper panel of Fig.~\ref{col0}). 
These two SBPs yield a constant color of 0.74 mag (i.e., the value expected for 
\sfha\ at the present age) outside the bulge that gradually rises inside \rbulge\ to a central value of 1.54 mag, such as to simulate \SFQ\
(blue curve in the lower panel). 

The color profile resulting from superposition of the bulge with the \brem{modexp} profile for the disk is shown in orange, and black corresponds to the color profile expected in the case of no \SFQ, that is, a purely exponential profile for both $B$ and $R$ with an overall constant color of 0.74 mag.
% :::::::: FIGURE 9 ::::::::::::::::::::::::::::::::::::::::::::::::::::::::::::::::::::::::::::
%\setlength{\unitlength}{1mm}
\begin{center}
\begin{figure}
\begin{picture}(86,60)
\put(0,0){\includegraphics[width=8.6cm]{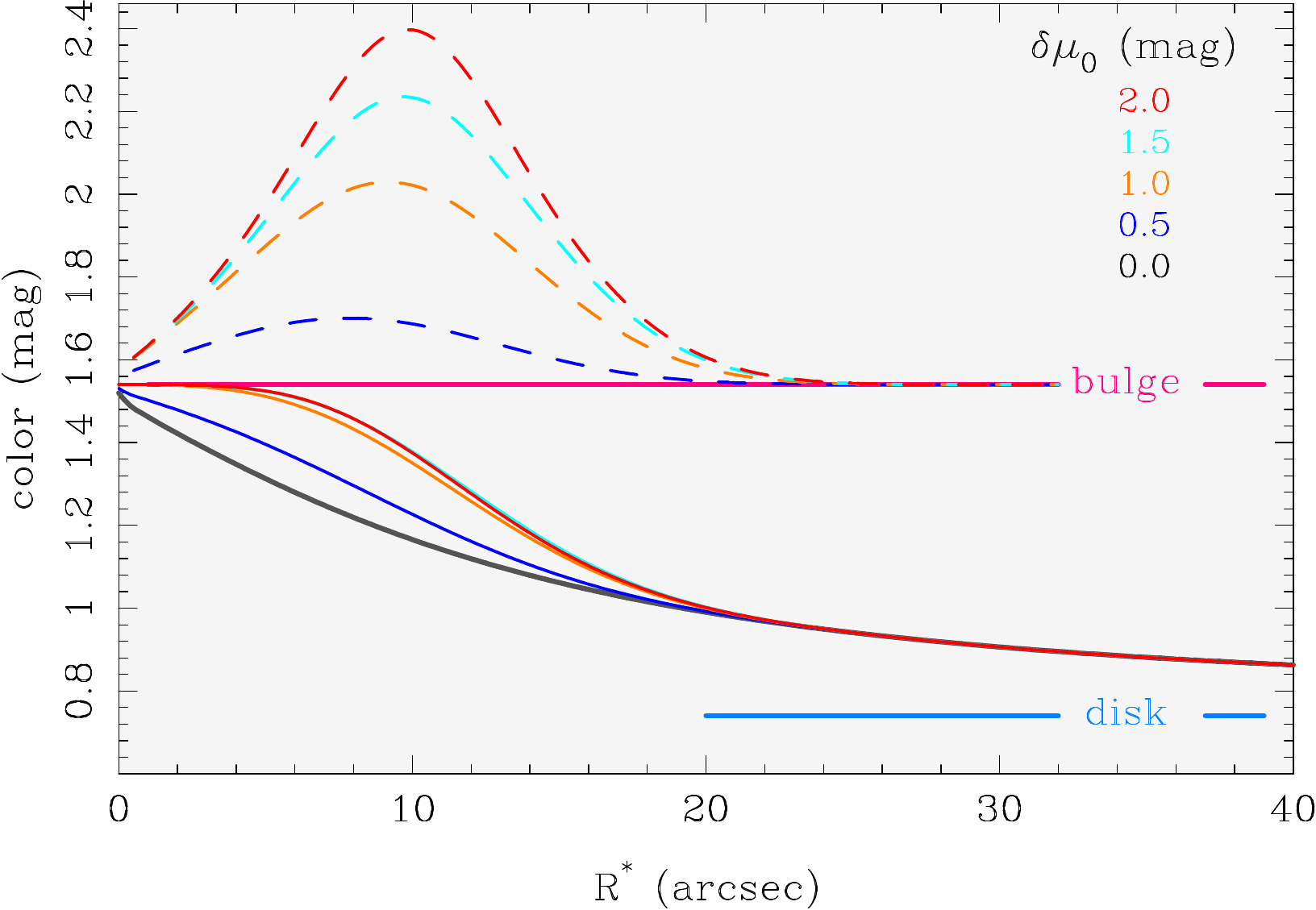}}
\end{picture}
\caption{Central part of $B$-$R$ color profiles for the composite (S\'ersic+\brem{modexp}) galaxy model in Fig.~\ref{col0} for a $\delta\mu_0$ between 0 and 2 mag (solid curves). Dashed curves show disk-corrected color profiles computed on the standard assumption that the disk follows an exponential profile all the way to its center. The true color of the bulge (1.54 mag) and of the  
outer disk (\rr$>$\rbulge; 0.74 mag) are depicted by the horizontal lines.}
\label{col1}
\end{figure}
\end{center}      
% :::::::: FIGURE 9 ::::::::::::::::::::::::::::::::::::::::::::::::::::::::::::::::::::::::::::
Regardless of whether the disk is centrally SF-quenched or not, the observed negative color gradient within \rbulge\ 
($\sim$ --0.5 mag/\rbulge) is purely a projection effect that is driven by the outwardly increasing line-of-sight contribution of the disk.
Using this color gradient to place constraints on bulge evolution and demographics is obviously pointless.
Without prior correction for the underlying disk, negative color gradients within \rbulge\ cannot be interpreted in terms of a 
possible radial decrease in stellar age or metallicity within the bulge, and offer useful constraints on bulge formation and evolution\footnote{More generally, any superposition of evolutionary and spatially distinct stellar components results in (or alters existing) radial color gradients. One similar but reverse example is offered by blue compact dwarf galaxies: In their majority, these systems exhibit intense SF in the central part of an evolved ($B$-$R$ $\sim$1 mag) stellar host. The interplay between the radially decreasing SFR surface density and the increasing line-of-sight contribution of the old underlying stellar component manifests itself in positive color gradients out to approximately two exponential scale lengths of the host. \citet{P02} have shown that correction for the host can reduce the integral color of the starburst component by $\sim$0.5 $B$-$R$ mag and change its radial color gradient by up to 1 mag/kpc. Thus, modeling and subtracting of the underlying host is crucial for a proper age dating of the young stellar component in these systems. Evidently, similar considerations apply to any other radially resolvable luminosity-weighted observable in a 3D geometry (EW, light-weighted age and metallicity, stellar velocity dispersion, Lick indices, and diagnostic emission-line ratios).}. 
Unfortunately, a review of the rich literature about LTGs reveals only very few photometric studies acknowledging the influence of the disk on color gradients within the bulge \citep[e.g.,][]{Head14}. 

Next, we comment on the color profile that would be obtained for the bulge if correction for the underlying disk were attempted on the standard assumption that the intensity of the latter follows a purely exponential instead of a \brem{modexp} profile. 
The color profile obtained in this way (dashed red curve in the lower panel of Fig.~\ref{col0}) is overestimated throughout within the bulge, reaching at \reff$\simeq$\rbulge/2 (11\arcsec) a maximum deviation of +0.5 mag from the true color. 
More generally, correction of bulge color profiles (or color maps) for the underlying disk without taking \SFQ\ into account 
(i.e., assuming a purely exponential model for the disk) leads to an increasingly overestimated color with increasing \dio. 
This is better illustrated in Fig.~\ref{col1} where we plot the observed and disk-corrected color profile (solid and dashed curves, respectively) for a $\delta\mu_0$ between 0 mag (no \SFQ) and 2 mag. As $\delta\mu_0$ increases the color gradient at \rbulge/2$\la$\rr$\la$\rbulge\ steepens from --0.35 mag/\arcmin\rbulge\ for $\delta\mu_0$=0 mag (black) to --0.76 mag/\rbulge\ for $\delta\mu_0$=2 mag (red). The disk-corrected color of the bulge is overestimated for any $\delta\mu_0$>0 mag, exhibiting around \reff\ a broad excess by 0.2 mag ($\delta\mu_0$=0.5) and up to 0.9 mag ($\delta\mu_0$=2).

Summarizing, negative color gradients within galaxy bulges can be entirely due to the outwardly increasing line-of-sight contribution of the underlying disk, both in the absence and presence of \SFQ. In the case of a centrally SF-quenched LTG, correction for the disk using an inadequate model for its radial intensity (i.e., a purely exponential profile with constant color) leads to a local and global 
overestimation of the color of the bulge, in particular to an artificially red circumnuclear zone that could be taken as evidence for enhanced dust obscuration.
Additionally, the oversubtraction of the blue SED of the disk can lead to an abnormally red SED for the bulge and drive spectrophotometric bulge-disk decomposition studies toward an exceedingly high age and metallicity (B20b). This is particularly true for LTGs at redshift intervals where \dio\ is enhanced by nebular line emission in the disk (cf. Sect.~\ref{sect-ry-neb}).

% %%%%%%%%%%%%%%%%%%%%%%%%%%%%%%
\section{Discussion \label{dis}}
% %%%%%%%%%%%%%%%%%%%%%%%%%%%%%%
The goal of this study is to draw attention to the implications of \SFQ\ for structural studies of galaxies and motivate an exploration of this hitherto uncharted territory. Extending our discussion in Sect.~\ref{photometry}, in the following we point out that the neglect of \dio\ can introduce complex biases with considerable relevance to our understanding of the formation history and demographics of galaxy bulges (Sect.~\ref{bd}) and that these biases are not simple to overcome because they impact structural studies differentially, depending on galaxy mass and redshift (Sect.~\ref{downsizing}). In particular, the neglect of \dio\ might appreciably affect the scatter and slope of galaxy scaling relations (Sect.~\ref{smbh}), which calls for a closer observational and theoretical examination of this effect and for the development of empirical approaches for its correction in a statistical sense. Possible routes toward this goal are discussed in Sect.~\ref{outlook}.
% ::::::::::::::::::::::::::::::::::::::::::::::::::::::::::::::::::::::::::::::
\subsection{Principle photometric biases due to the neglect of \dio\ \label{bd}}
% ::::::::::::::::::::::::::::::::::::::::::::::::::::::::::::::::::::::::::::::
As discussed in Sect.~\ref{photometry}, the neglect of \dio\ in dedicated bulge-disk decomposition studies can result in biased determinations of the photometric properties of galaxy bulges. Expected effects include \brem{a)} a systematic underestimation of the luminosity of the bulge in a manner that is inversely related to its prominence (specifically, the \bid\ ratio) and proportional to the degree ($\delta\mu_0$) and radial extent ($R_{\rm c}=\epsilon_2\cdot\alpha$) of \SFQ, and \brem{b)} a potential underestimation of the S\'ersic exponent $\eta$. The latter entails an overestimation of the model-dependent effective radius of the bulge \citep[because \reff\ is by definition inversely related to $\eta$; e.g.,][]{Trujillo01}, consequently also an underestimation of the mean stellar velocity dispersion $\sigma_{\star}$ therein\footnote{In practice, the adopted value for $\sigma_{\star}$ will depend on aperture corrections \citep[e.g.,][]{Davies87,J95,ZB97,Jablonka07} which in turn require assumptions on the SBP of the bulge.}. The underestimation of $\eta$ likely results in a tendency for the classification of moderately luminous classical bulges as pseudo-bulges on the basis of the commonly adopted empirical cutoff at $\eta \sim 2$.
Finally, as previously shown, the neglect of \dio\ can result in the erroneous classification of low-B/D LTGs as bulgeless disks.

Whereas it is plausible that any galaxy decomposition scheme that lacks a relevant structural component or adopts an inadequate functional form for it can entail a bias, quantitative predictions on the concrete implications of this for a multicomponent bulge-(bar)-disk fit that involves multiple nonlinearly coupled free parameters is not straightforward. 
This is particularly true when photometric uncertainties, hence the statistical weights that largely guide the best-fitting solution, happen to correlate with evolutionary patterns in galaxies
(colors, sSFR and emission-line EWs). This is precisely the case in LTGs: because the S/N scales with the surface brightness (or \sstar), the highest weight is given to the SF-quenched core (\rr$\la$\rbulge), that is, the radial zone showing the strongest negative color and age gradients simultaneously with the steepest positive sSFR and \ewha\ gradients \citep[cf., e.g.,][among others]{BP94,GA01,CatT17,Belfiore18,Breda20a}. 
A further complication stems from the fact that this core is most affected by PSF smearing, 
which can propagate to further uncertainties in the S\'ersic fit for the bulge \citep[see the discussion in][]{Breda19}.

How the interplay between these factors is imprinted on the best-fitting solution is not immediately clear. 
For instance, the scaling of photometric errors inversely to surface brightness can result in an artificial compaction of the disk (underestimation of its $\alpha$ and overestimation of its $\mu_0$), which in turn translates into an underestimation of \bid\ (the disk luminosity fraction inside \rbulge) and could bias the S\'ersic model parameters for the bulge \citep{P96a}. 
The neglect of a possibly present bar poses a further complication \citep[cf., e.g.,][\tref{BP18}]{Men08}; this is especially true for higher-mass LTGs because the bar prominence increases with LTG luminosity \citep{Bittner17,Gadotti20}. Moreover, small-scale features 
such as nuclear rings \citep{ButaCombes96,Comeron10} or -disks \citep{Bittner20}, or double-bars \citep{dLC19} may have an influence given their high surface brightness (and weight in the fit).

It is thus conceivable that the neglect of \dio\ is largely responsible for the fact that bulge-disk decomposition solutions in different bands are frequently mutually inconsistent or untenable from the evolutionary point of view.
For instance, \citet{dosReis20} report several cases of formally (in terms of $\chi^2$) irreproachable fits with GALFIT \citep{Peng2010} that translate into unphysical colors for the bulge and disk when the best-fitting solution is extrapolated beyond \rr$\ga$\reff, however.
As remarked in Sect.~\ref{profiles}, forcing the S\'ersic exponent $\eta$ to vary smoothly across $\lambda$ according to an ad hoc functional form might offer a remedy, but no solution to the problem.

An important step forward would be to empirically examine in synthetic multiband galaxy images how the neglect of \SFQ\ is imprinted on determinations of structural properties of bulges 
($\eta$, effective radius \reff\ and effective surface brightness \mueff, total magnitude \mmb). To this end, standard bulge-disk decomposition into a S\'ersic and a purely exponential 
model would need be applied to synthetic galaxies whose disk is given by a \brem{modexp} or other functional forms \citep[e.g., 
a variant of the five-parameter core-S\'ersic profile;][]{Graham03,Trujillo04,Bonfini14}, possibly with the addition of a \citet{Ferrers1877} component approximating a bar. 
In addition to a grid of models that densely cover possible configurations of a central flattening or downbending of the disk inside the bulge (i.e., the $\epsilon_{1,2}$ parameters 
of Eq.~\ref{modexp2}), this task should ideally include the radially varying contribution of nebular emission to broadband SBPs in different evolutionary stages of an LTG (Sect.~\ref{sect-ry-neb}).

% ::::::::::::::::::::::::::::::::::::::::::::::::::::::::::::::::::::::::::::::::::::
\subsection{Galaxy downsizing and the differential nature of \dio\ \label{downsizing}}
% ::::::::::::::::::::::::::::::::::::::::::::::::::::::::::::::::::::::::::::::::::::
The factor \flb\ by which the luminosity of the bulge is underestimated due to the neglect of \dio\ depends on two competing but  interrelated factors, 
namely the \dio\ itself and the luminosity fraction \bid\ of the bulge inside \rbulge. For the sake of simplicity we assume that \bid\ roughly scales with B/T 
(however, see Sect.~\ref{cd}). 

Some qualitative predictions can be made from empirical insights and plausibility arguments.
A reasonable expectation is that \dio\ affects determinations of photometric and structural properties of bulges differentially, that is, in a manner that at a given age scales  inversely with galaxy mass \mstar. 

From the perspective of galaxy downsizing \citep{Cowie96} the formation timescale of galaxies ia anticorrelated with \mstar, with the dominant phase of their build-up being delayed to a lower $z$ as \mstar\ decreases \citep[staged galaxy formation;][]{Noeske07b}. Manifestations of this phenomenon include the inverse relation between \mstar\ and burst parameter or \ewha\ in BCDs \citep{Salzer1989b,Krueger95}, and the anticorrelation between \mstar\ (or \sstar) and various proxies to the ongoing and average past SF activity, such as \ewha\ or SFR/<SFR> \citep{Bri04}, mass doubling time \citep{Noeske07b} and luminosity-weighted age \citep{Gal05}.
The advent of wide-field IFS helped to better recognize that an analogy to downsizing on galactic scales exists on subgalactic scales 
(subgalactic downsizing, in the notation by \tref{BP18}) 
in the sense that the formation timescale of stellar populations scales inversely to with their mean \sstar: As documented by several recent studies, the dense galaxy centers assemble 
first and followed by a prolonged, still ongoing build up of their periphery \citep[e.g.,][]{Perez13,Fang13,Gon14,Gomes16b}, as also shown from UV observations in, e.g., \citet{MM07}, \citet{SalimRich10}, \citet{Salim12} and \citet{GdP07}.
In the specific context of LTGs this trend is echoed by the increasing bulge-to-disk age contrast with increasing \mstar, for example, or by the radially increasing luminosity contribution 
of stellar populations of intermediate-to-young age (Sect.~\ref{ifs-dio}).
% :::::::: FIGURE 13 ::::::::::::::::::::::::::::::::::::::::::::::::::::::::::::::::::::::::::::
\begin{center}
\begin{figure}[t]
\begin{picture}(86,98)
\put(0,47){\includegraphics[width=8.6cm]{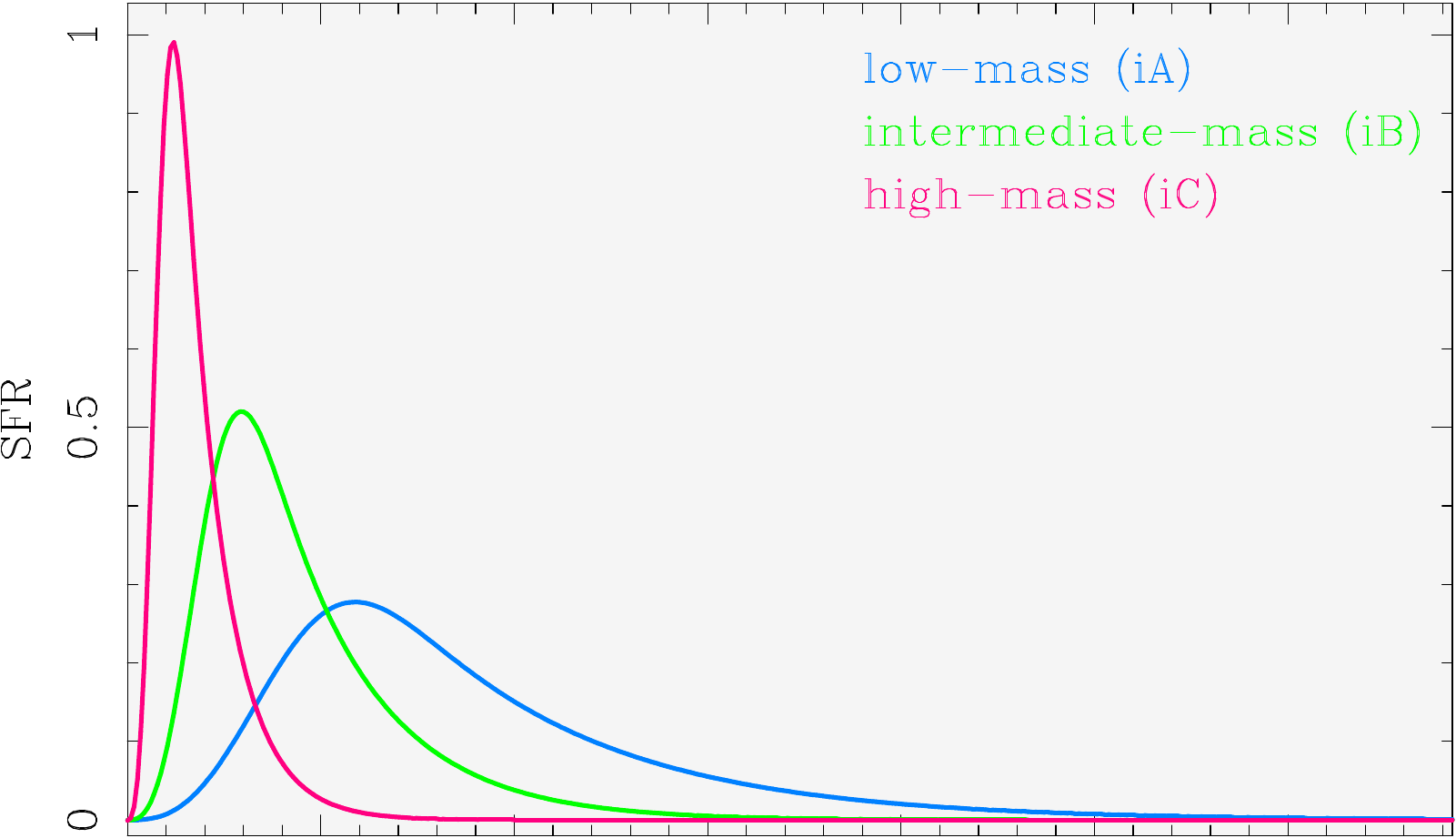}}
\put(0,0){\includegraphics[width=8.6cm]{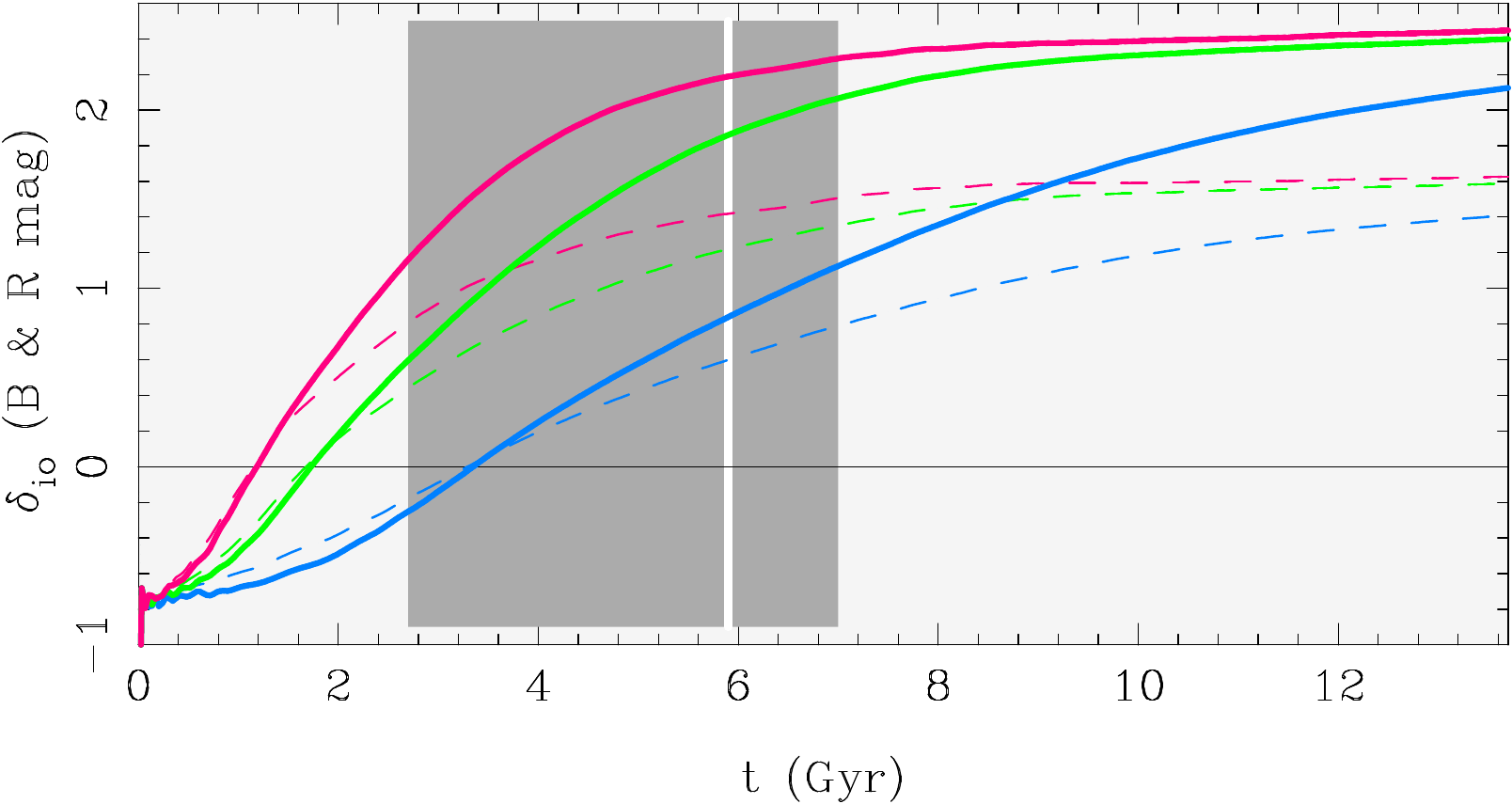}}
\end{picture}
\caption{\brem{top:} SFR parameterizations simulating a downsizing trend for galaxy bulges by assuming that their age at the peak of their SFR and their SF e-folding timescale inversely with \mstar. High-, intermediate- and low-mass bulges (\brem{iC}, \brem{iB} and \brem{iA}, respectively, in the classification by \tref{BP18}) reach their maximum SFR at 0.47, 1.2 and 2.4 Gyr.
\brem{bottom:} Evolution of \dio\ (approximated by $2.5\cdot \log(\psi)$, as in Fig.~\ref{mlratio}) in the $B$ and $R$ band (solid and dashed curves, respectively) for the three SFH scenarios for the bulge (upper panel) when a constant SFR for the outer disk is assumed. 
The shaded gray area depicts the redshift interval 0.76$\leq z \leq$2.3 that will be covered both by the MOONS spectrograph at VLT (0.65-1.8 $\mu$m) and the $B$-band filter ($\sim$0.37-0.54 $\mu$m). 
The amplitude of \dio\ at $z\sim1$ (vertical white line) depends on galaxy mass. It is moderate ($\sim$0.85 $B$ mag) for low-mass ($<$L$^{\star}$) galaxies and reaches $\sim$2 mag for massive LTGs in advanced stages of SF quenching.
\label{ipeak}} 
\end{figure}
\end{center}      
% :::::::: FIGURE 13 ::::::::::::::::::::::::::::::::::::::::::::::::::::::::::::::::::::::::::::

This (sub)galactic downsizing may be simulated with delayed SFH scenarios (Fig.~\ref{ipeak}) that yield a zero-order approximation to a \mstar-dependent formation history for bulges in the three LTG classes tentatively defined in \tref{BP18}: Similar to Fig.~\ref{mlratio}, the disk outside the bulge is assumed to form continuously with a constant SFR, whereas the SFR of \brem{iC}, \brem{iB} and \brem{iA} bulges is approximated by models involving an SFR that reaches its maximum at 0.47, 1.2 and 2.4 Gyr, respectively, and 
then exponentially declines with an e-folding time that scales inversely with \mstar. As in Fig.~\ref{mlratio}, nebular emission is taken into account and a constant metallicity of \zsun\ (\zsun/5) is assumed for the bulge (disk).

The lower panel shows the evolution of \dio\ in the $B$ and $R$ band (solid and dashed curve, respectively) for the three scenarios. 
\dio\ becomes positive after 1-3 Gyr, as SF and nebular contamination gradually cease in the bulge and its \ml\ 
rises above that of the outer disk, and then smoothly increases to its present value of $\ga$2 $B$ mag and 1.5 $R$ mag. 

An insight from this simplified parameterization of galaxy downsizing 
is that \dio\ increases faster for massive LTGs and vice versa. Taking the evidence from Fig.~\ref{ipeak} at face value, at $z\sim1$ \citep[vertical line; roughly at the onset of the decline of the cosmic SFR density; cf.][]{MD14}, massive galaxies in an advanced stage of \SFQ\ have developed a \dio\ that is higher by $>$1 mag than low-mass galaxies that still sustain a significant level of SF in their bulges. The gap between the two gradually shrinks to less than 0.7 $B$ mag since only $\sim$4 Gyr ago ($z\sim 0.36$).

The key implication is that (the neglect of) \dio\ does not just add a scatter to any galaxy scaling relation that involves bulge luminosities, but affects its slope to a degree that depends both on galaxy mass and $z$.
This calls for caution when (magnitude- or mass-selected) galaxy samples spanning a range in redshift are jointly analyzed 
with a focus on the structural and photometric evolution of their bulge.
For example, in the spectral range covered by the multi-object spectrograph MOONS at VLT \citep[0.65-1.8 $\mu$m;][]{Cirasuolo20} 
single-fiber spectroscopy can be supplemented by bulge-disk decomposition in the restframe $B$-band at 0.76 $\leq z \leq$ 2.3 (shaded gray area). 
Within this interval, \dio\ shows roughly the same increase for all LTGs ($\sim$1.3 mag), however, its value at $z=1$ is nearly three times higher for massive 
(\brem{iC}) galaxies (2.2 mag) than for low-mass (\brem{iA}) galaxies (0.85 mag). 
Therefore, photometric corrections adopting an average value for \dio\ are inadequate for compensating for the effect of \SFQ\
\footnote{Star formation quenching obviously has a broader range of implications for fixed-aperture spectroscopic surveys
\citep[we refer to several previous studies addressing aperture effects from various viewpoints, e.g.,][and references therein]{Hopkins03,Bri04,IP13,IP16}. In particular, \citet{Gomes16b} have shown that for a model of inside-out galaxy growth (or \SFQ) that reproduces the size of local early-type galaxies, the 3\arcsec\ aperture of the SDSS misses a progressively large portion of the outer star-forming zone for $z<1$.
This leads to a severe underestimation of their total SFR and their incorrect classification as "retired" on the basis of the spectroscopic 
classification by \citet{sta08}.
With regard to the upcoming MOONS, the spectroscopic fiber ($\diameter$=1\arcsec) will map a projected area of $\sim$8 kpc at $z\sim1$, 
therefore encompassing at least parts of the star-forming outer disk both for high-mass centrally SF-quenched and lower-mass star-forming galaxies. 
One implication of this is that dilution of AGN-specific emission-line diagnostics by the star-forming disk will be stronger in massive quenched galaxies because nebular emission within their bulges is weaker. This effectively entails a detection bias against jet-mode AGN because these are predominantly hosted by SF-quenched, virtually gas-devoid stellar spheroids \citep{KH13,HeckmanBest14} 
with a high Lyman-continuum photon escape fraction, hence weak nebular emission \citep{P13,GP16-ETGs}.}.

Quantitative inferences based on the fractional underestimation of the bulge luminosity \flb\ have to await 
detailed studies of \dio\ and \bid\ (Sect.~\ref{outlook}). Nevertheless, it is reasonable to assume that \flb\ scales inversely with \mstar, being low in massive \brem{iC} LTGs (log(\mstar/\msun)$>$10.7) and comparatively high in intermediate-\mstar\ (log(\mstar/\msun)$\sim$10.5) \brem{iB} LTGs.
This is because the former, although having developed the highest \dio\ (Fig.~\ref{ipeak}), 
typically host prominent bulges, which implies a high \bid. 
These two factors likely counterbalance each other, effectively reducing \flb.
Intermediate-mass LTGs, on the other hand, at the same time combine a significant \dio\ with a modest \bid, which likely acts toward raising their \flb.
Finally, low-mass LTGs (\brem{iA}) show a comparable level of SF in their centers and disk periphery, thus a generally low \dio, as is also apparent from their low bulge-to-disk age contrast (cf. \tref{BP18}) and flat age gradients within \rbulge\ \tref{(B20a)}.
This is also consistent with their BPT classification in the locus of SF regions, which suggests the absence of strong AGN activity 
as a possible agent of inside-out \SFQ. 
Notwithstanding this fact, because of the faintness of their bulge (i.e., very low \bid), even a small \dio\ could 
raise \flb\ and even prevent detection of their bulge (Fig.~\ref{sim2}). 
Based on these considerations, we expect \flb\ in the low-mass end of the LTG sequence to show a large dispersion around a generally low median value.
% =======================================================================================================
\subsection{Effect of \dio\ on the scatter and slope of fundamental scaling relations \label{smbh}}
% =======================================================================================================
A closer investigation of \dio\ is of considerable interest also because correction for this effect in galaxy decomposition studies 
will likely influence the scatter and slope of some galaxy scaling relations, eventually opening new prospects for improving our understanding of bulge evolution and demographics.
Because this correction will increase the bulge luminosity and probably also the S\'ersic $\eta$, it will move some previously categorized pseudo-bulges into the locus of classical bulges, possibly leading to sharper observational photometric discriminators for bulge classification. 
For instance, it is imaginable that the failure to establish $\eta$ as a robust bulge classification means, despite several efforts 
over the past three decades \citep[e.g.,][for a recent review]{Neumann17}, partly stems from the neglect of \dio\ (Sect.~\ref{profiles}).
Furthermore, the expected increase in the parametric (S\'ersic model-dependent) effective surface brightness, effective radius and $\sigma_{\star}$ therein could be relevant to for example the \citet{Kormendy77} relation (see also \tref{Kormendy \& Djorgovski 1989} and \tref{Ziegler et al. 1999}) and for the \citet{FJ76} relation for bulges, as well as for any relation contrasting structural characteristics of bulges (e.g., $\eta$, \reff, \mueff, \mmb) with, for example, the $\alpha_4$ parameter \citep{Fabricius12}, ellipticity, concentration index, velocity anisotropy, and nonthermal radio power. Furthermore, because disk-dominated LTGs are more affected by \dio\ than bulge-dominated early-type galaxies, it might naturally be expected that bulges in these two galaxy classes describe slightly different slopes in various scaling relations.

The principle effect of accounting for \dio\ might be illustrated with the example of the \mbh-bulge relation \citep[e.g.,][]{Magorrian98,FerMer00} in which   
the bulge mass is commonly approximated by its absolute $K$-band magnitude or mean $\sigma_{\star}$ within \reff.
This relation is fundamental to our understanding of the regulatory role of AGN in galaxy evolution, as also reflected in the fact that deviations from  the average ratio \mbh/\mstar$\approx$0.002 \citep{HeckmanBest14} are regarded as signatures of distinct evolutionary routes in the coevolution of SMBHs and galaxy spheroids.
For example, \citet{KBC11} studied a sample of CBs (high-luminosity, pressure-supported bulges), PBs (intermediate-luminosity bulges with significant rotational support) 
and bulgeless disks, all having direct kinematical determinations of their \mbh. 
These authors find that similar to ellipticals, CBs obey the bulge versus \mbh\ relation, whereas PBs
and bulgeless galaxies show a large scatter or systematic deviations from it \citep[however, see][]{Bennert21}. 
Five of the 11 PBs included in this study are indeed overluminous by 1-2 $K$ mag for their \mbh. 

\begin{center}
\begin{figure}
\begin{picture}(86,74)
\put(6,0){\includegraphics[clip, width=7.6cm]{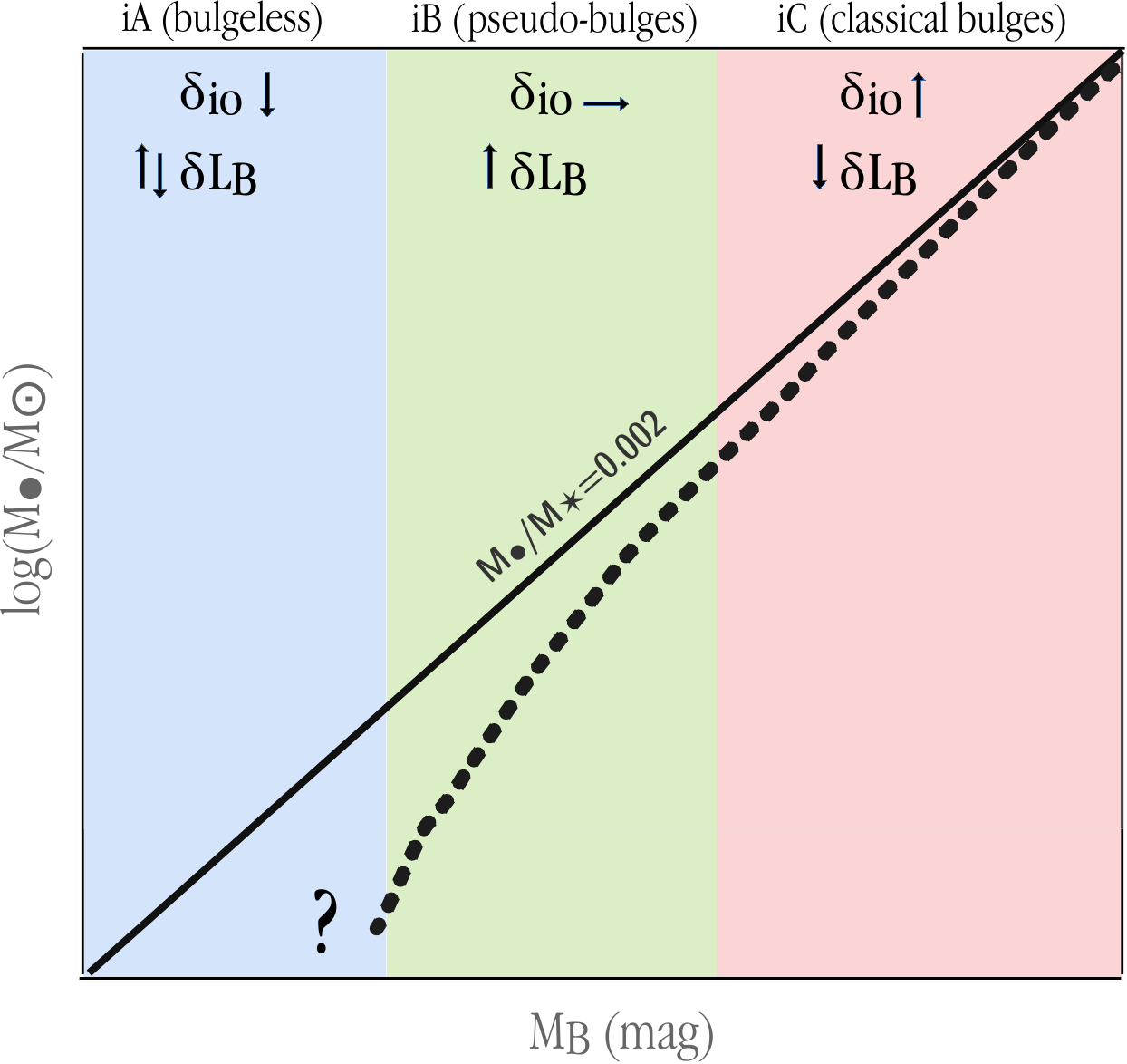}}
\end{picture}
\caption{Schematic sketch of the relation between SMBH mass \mbh\ and absolute magnitude of the bulge M$_{\rm B}$, with the solid diagonal line depicting the commonly adopted mean ratio \mbh/\mstar$\approx$0.002 \citep[cf., e.g.,][]{HeckmanBest14} and the dashed line its expected change after correction for the \dio\ effect.
Arrows in the upper part point to the amplitude of \dio\ and the expected fractional under-estimation of the bulge luminosity \flb\ for the three LTG classes proposed in 
\tref{BP18}. Bulgeless, pseudo-bulge and classical-bulge galaxies may loosely be associated with type \brem{iA}, \brem{iB} and \brem{iC} LTGs in the mass intervals 
log(\mstar/\msun)$<$10.3, $\sim$10.5 and $>$10.7, respectively. 
Whereas \dio\ increases with \mstar, its effect on \flb\ is likely comparatively small for high-mass LTGs because of the prominence of their bulges, while it is expected to increase for intermediate-mass LTGs that simultaneously show a significant \dio\ and a modest B/D ratio. We expect \flb\ for low-mass LTGs to show a large scatter around a low median value because of their very low \dio\ and B/D ratio.}
\label{fig:SMBH} 
\end{figure}
\end{center}

A correction for \dio\ could enhance the luminosity of bulges by a factor $\propto$\mstar$^{-1}$ (cf. Sect.~\ref{downsizing}), which would have a greater effect for PBs than for CBs and would lead to a down-bending of the bulge-\mbh\ relation below log(\mstar/\msun)$\sim$10.7, as sketched in Fig.~\ref{fig:SMBH}. From the perspective of the standard scenario \citep[][for a review]{KorKen04}, this down-bending
could be taken as supportive evidence for the prevailing view that CBs and PBs 
are evolutionary distinct, with the former showing an affinity to massive stellar spheroids that form early on and experience a synchronous growth with their SMBH, whereas the latter primarily resulting from rearranged disk material that piles up into a central luminosity excess in the course of the secular evolution of disk-dominated galaxies.

Alternatively, a down-bending of the bulge-\mbh\ relation for sub-L$^{\star}$ galaxies could strengthen the conclusion by \citet[][their Fig.~9b]{HeckmanBest14} 
that \mbh\ is not a fixed fraction of \mstar\ but decreases from a mean \mbh/\mstar\ ratio of $10^{-2.5}$ at \mstar$\sim 10^{11.5}$ \msun\ to $\sim 10^{-4}$ at \mstar$\sim 10^{10}$ \msun.
Furthermore, a dependence of the \mbh/\mstar\ ratio on \mstar\ would be consistent with the scarcity or weakness of accretion-powered nuclear activity in low-mass LTGs \citep[][see also \tref{BP18}]{KH13}. 

Summarizing, in the light of our considerations above, a correction of bulge magnitudes for \dio\ appears to be fundamental for sharpening  
observational constraints on the intrinsic scatter of the bulge-\mbh\ relation \citep[e.g.,][]{GRB09} and the possible dependence of its slope 
on \mstar\ (Fig.~\ref{fig:SMBH}) or galaxy morphology \citep[e.g.,][]{Sahu20}.

% :::::::: FIGURE 1 ::::::::::::::::::::::::::::::::::::
%\setlength{\unitlength}{1mm}
%MBMBH.png

% ::::::::::::::::::::::::::::::::::::::::::::::::::::::::::::::::::::::::::::::::::::::::::::
\subsection{Combined effect of \dio\ with a possible central depletion of the disk \label{cd}}
% ::::::::::::::::::::::::::::::::::::::::::::::::::::::::::::::::::::::::::::::::::::::::::::
Our foregoing considerations are based on the conservative assumption that the stellar surface density \sstar(\rr) of the disk is exponential all the way to its center and the central flattening or down-bending of its SBP is solely due to a higher \ml\ inside the SF-quenched bulge.
However, the exponentiality of \sstar\ is not for granted. Theoretical work points to a central depletion of the disk 
\citep{KD95,Obreja13,WidDub05,Du20} although little is quantitatively known on its possible connection to \mstar\ and other galaxy properties 
(e.g., angular momentum, presence of an AGN). As an example, \citet{Obreja13} find from hydrodynamical simulations a steep decrease in the disk's \sstar\ 
by $\sim$2.5 dex within \rr$\sim$1 kpc ($\sim$1/2 \rbulge).

Circumstantial observational support for a central depletion in the disk comes from recent work by \tref{B20b}.
These authors showed that the stellar SED of the outer disk, if scaled to the light fraction implied within \rbulge\ by inward extrapolation 
of a purely exponential model and then subtracted from the integrated spectrum therein, yields an unphysical (negative) or implausible net stellar SED for the bulge. 
The latter was deemed to be the case when the disk-corrected SED for the bulge is abnormally red, which would force spectral synthesis models to converge 
to the maximum possible age and metallicity allowed by the construction of the SSP library used, or yield for the net-bulge a \mstar\ that exceeds the value obtained 
by fitting the integrated SED within \rbulge. These authors also tested the effect that a \brem{modexp} model for the disk would have on the net SED of the bulge. 
They showed that models with a central flattening reduce the percentage of implausible solutions to 15\% 
whereas a down-bending of the disk to a central intensity of 0 (i.e. a \brem{modexp} with $\epsilon_1=1$, cf. Eq.~\ref{modexp1}) 
in all cases yields an acceptable spectral fit. 
We recall that the first \brem{modexp} model (centrally flat exponential profile) can be accounted for by a modest \dio\ 
whereas the second model ($\epsilon_1$=1) cannot just be due to a variation in the \ml\ ratio, but requires a central evacuation of the disk. Because the pilot study by \tref{B20b} has considered only these two \brem{modexp} profiles, it was unable to pinpoint the form of the inner disk below the bulge or place constraints on the relative importance of \dio\ and a possible central decrease in \sstar.
For instance, it is conceivable that a \dio\ at the upper range of estimated values ($\sim$2.5 $B$ mag) could alleviate the need for a strong 
central reduction of \sstar. It is also important to bear in mind that \brem{modexp} is a functional form developed in another context (dwarf galaxies) 
and may not be optimal for galactic disks.

Star formation quenching and a central depletion of the disk are presumably nonmutually exclusive phenomena, and both imply an upward revision of the luminosity of the bulge. A step forward toward their better understanding might be possible through spectrophotometric and kinematical decomposition of IFS data (e.g., by extending the method by \tref{B20b} by a dense grid of \brem{modexp} models for $0\leq \epsilon_1 \leq 1$, or other functional forms that allow for a centrally depressed exponential disk). 
A combined analysis of observational constraints gained from this task with galaxy simulations that incorporate detailed prescriptions for \SFQ, as well as Schwarzschild orbital superposition modeling \citep[e.g.,][]{Zhu18a,Zhu18b,Du20,Jagvaral21} would be of considerable interest.
% :::::::::::::::::::::::::::::::::::::::::::::::::::::::::::::::::::::::::::::::::::::::::
\subsection{Approaches toward an empirical correction of \dio\ \label{outlook}}
% :::::::::::::::::::::::::::::::::::::::::::::::::::::::::::::::::::::::::::::::::::::::::
Because the \dio\ effect is a direct and causal consequence of star formation quenching, understanding it is almost 
tautologous to the understanding of spiral galaxy evolution, including the physical drivers, timescales and spatial characteristics of inside-out 
cessation of SF. Elucidating this subject in its broad complexity is clearly a long-term endeavor. 
For this reason, it appears worthwhile to explore empirical recipes that could permit an approximate short-term a posteriori correction of the large existing 
body of photometric quantities for bulges from previous galaxy decomposition studies. 
Ideally, such corrective formulae should involve easily accessible observables (e.g., EWs, colors) or quantities inferred from spectral modeling (bulge-to-disk age contrast, \dmb).
This task can presumably only be achieved in a statistical sense: The rich variety of substructures in the central parts of LTGs \citep[nuclear disks and rings, single- and double-bars, barlenses; cf., e.g.,][]{Bittner17,dLC19,Laurikainen18,Gadotti20} together with the significant scatter ($\sim$1.7 Gyr/\rbulge) of stellar age gradient determinations therein \tref{(B20a)} document a diversity in the biography of individual systems.

The development of such a rectification concept will most likely require determination and suppression of \dio\ in a large, representative sample of local LTGs in conjunction with multiband photometric analyses of simulated galaxies in various stages of \SFQ\ with the goal of elaborating empirical constraints on the impact of \dio\ on bulge-disk decomposition (Sect.~\ref{bd}).

The first task could exploit image processing (e.g., unsharp masking), and SED fitting of spatially resolved data in combination with an age-slicing tool like \RY.\\
\brem{a)} One possibility is to mask or bidimensionally subtract spiral features prior to bulge-disk decomposition. This has been done in detailed studies of small galaxy samples in order to stabilize the initial guess on ellipticity and center \citep[cf. the iterative technique by][]{Men08}. 
While feasible, such an approach requires careful interactive work and thus appears impractical for automated studies of large galaxy samples. Machine-learning-supported suppression of nonaxissymmetric features might be an alternative but would require training on local galaxy samples, 
which could turn into a disadvantage for studies of irregular clumpy galaxies at a higher $z$.
An additional concern is that high-contrast grand-design spiral features are generally prominent in massive LTGs, whereas lower-mass spirals tend to be flocculent and to show low-contrast spiral patterns \citep[e.g.,][]{Bittner17}. This could result in a bias against lower-mass LTGs, especially when studying 
higher-$z$ samples with PSF-degraded imaging data. Another limitation of this approach is that it is hardly suitable for suppression of diffuse ionized gas (DIG) and 
intraspiral arm emission from moderately evolved ($\sim$200 Myr) stellar populations that experience significant evolution in their \ml\ over a galaxy rotation period.
The DIG, although faint (\ewha$<$6 \AA), can contribute up to $\sim$50\% of the total \ha\ luminosity of star-forming galaxies \citep{Ferguson96,Weilbacher18,denBrok19}.

\brem{b)} NIR photometry allows for the reduction of \dio\ by a factor $\sim$3 (cf. Sect.~\ref{ev-dio}), but it is prone to contamination by the nebular continuum \citep[e.g.,][]{Krueger95}. On the other hand, thanks to the rapidly increasing amount of NIR data with an adequate resolution (e.g., with VISTA and soon Euclid), it offers important prospects for the suppression of \dio\ in low-$z$ galaxy samples.

\brem{c)} Another possible approach could take advantage of \RY\ (Sect.~\ref{ifs-dio}) to suppress the excess emission from 
the star-forming outer disk and obtain an estimate on the size and intensity profile of the SF-quenched core (i.e., the $\epsilon_{1,2}$ parameters of Eq.~\ref{modexp2}).  
Limitations of this approach stem from uncertainties in population spectral synthesis \citep[][for a discussion]{GP16-RY,BP18}, its high 
computational expense and the requirement for an objective definition of the optimal \tcut\ for a galaxy.

\brem{d)} A complement to the aforementioned methods could involve bulge-(bar)-disk decomposition of stellar surface density (instead of surface brightness) maps
\citep[][see also \tref{Wuyts et al. 2012}]{Lang14}. In this regard, the availability of spatially resolved, low-resolution ($R\sim 60$) SEDs 
from J-PAS \citep[Javalambre Physics of the Accelerating Universe Astrophysical Survey;][]{Molles11,Bonoli21} opens a promising avenue.

Despite their advantages and disadvantages, a combination of the approaches above could yield statistical estimates of \dio\ as a function of \mstar\ and B/D ratio, and might support the refinement of various spectrophotometric decomposition concepts proposed recently \citep[e.g.,][]{Johnston12,Johnston17,Tabor17,Tabor19,Men19,Breda20b,Barsanti21}. 
A correction for \dio\ could also help provide accurate 2D \ml\ and \sstar\ maps needed for advanced Schwarzschild orbital decomposition and foster synergy between theory and observations toward a better understanding of the star formation quenching phenomenon.
% ===============================================
\section{Summary and conclusions \label{summary}}
% ===============================================
The aim of this study was to draw attention to the implications of star formation quenching (\SFQ) for bulge-disk decomposition studies of 
late-type galaxies (LTGs) and motivate further observational and theoretical work in this field.

The standard practice of bulge-disk decomposition is to fit and subtract an exponential model to the disk (more specifically, to its visible periphery 
outside the bulge) in order to isolate the net luminosity of the bulge. Regardless of whether the modeling of disk and bulge is done sequentially or simultaneously, or in 1D or 2D, 
the key assumption in all cases is that the inwardly extrapolated fit to the outer disk faithfully reproduces the radial intensity profile in its observationally inaccessible 
inner zone, below the bulge.

However, this postulate would only be valid if the star formation history (SFH) and specific SFR of the disk were spatially invariant, which is irreconcilable with the well established fact of \SFQ\ inside the bulge radius \rbulge\ of massive ($\ga$L$^{\star}$) LTGs. 
A central depression of star formation (SF) implies a higher \ml\ ratio for the inner SF-quenched disk (\rr$<$\rbulge) 
than for the SF-enhanced outer disk (\rr$>$\rbulge). For this reason, the inwardly extrapolated exponential model must lead to an 
overestimation (and oversubtraction) of the disk inside \rbulge, consequently to a systematic underestimation of the luminosity of the bulge.
We refer to the difference between the true magnitude of the disk inside \rbulge\ and the value implied by the exponential 
fit to the outer disk as \dio\ ($\geq$0 mag). This quantity can be approximated by the difference $\delta\mu_0$ between the true and extrapolated central surface brightness of the disk, or as 2.5$\cdot$log($\psi$) where $\psi$ denotes the ratio of the mean \ml\ 
of the disk inside and outside \rbulge.

\brem{i)} We showed that if the bulge and the inner disk share roughly the same SFH, then \dio\ in present-day LTGs can reach 2.5 mag in $B$ and 0.7 mag in $K$. This was demonstrated with evolutionary synthesis models involving continuous SF in the outer disk and an exponentially decreasing SF with an e-folding time of 0.5 and 1 Gyr within \rbulge. These estimates remain valid for SFH scenarios that simulate a downsizing trend, with bulges in massive galaxies experiencing the dominant phase of their 
formation early on and vice versa.
Additionally, observational support for a \dio\ well in excess of 1 $g$-band mag is provided through post-processing of spatially resolved spectral synthesis models 
for local LTGs with the age-slicing tool {\sc RemoveYoung} \citep{GP16-RY}, extending previous work in \citet{BP18}.

\brem{ii)} In the presence of \SFQ, the exponential surface brightness profile (SBP) of the disk should show a central flattening or down-bending inside the radius of the bulge. 
A suitable functional form for such an SBP is given by the modified exponential (\brem{modexp}) distribution proposed in \citet{P96a}.

\brem{iii)} It is pointed out that the neglect of \dio\ in dedicated bulge-disk decomposition studies can result in biased determinations 
of the photometric and structural properties of galaxy bulges. Expected effects include a) a systematic underestimation of the luminosity 
of the bulge in a manner that is inversely related to its prominence and proportional to the degree of SFQ (aka \dio), 
and b) a potential underestimation of the S\'ersic exponent $\eta$. The latter entails an overestimation of the
model-dependent effective radius of the bulge (because \reff\ and $\eta$ are inversely related to each other), and consequently also an 
underestimation of the mean stellar velocity dispersion $\sigma_{\star}$ therein. These biases likely lead to a tendency for the classification of moderately luminous classical bulges as pseudo-bulges on the basis of the commonly adopted empirical cutoff at $\eta\sim2$. 
Additionally, we showed that the neglect of \dio\ can readily result in the erroneous classification of low-B/D LTGs as bulgeless disks.

\brem{iv)} Negative color gradients within the bulge can entirely be due to the outwardly increasing line-of-sight intensity contribution of the blue star-forming disk. 
This applies both to centrally star-forming and SF-quenched disks. For this reason, correction for the underlying disk (which, in turn, requires an understanding of \dio) 
is crucial for a meaningful study of color gradients in LTG bulges and the evolutionary clues these may hold.
If such a correction is attempted assuming a purely exponential SBP for the disk then the net color of the bulge is invariably overestimated. 
This chromatic bias is aggravated by red rims that can be taken as evidence for dusty circumnuclear SF rings.

\brem{v)} From a synopsis of insights from this study and existing observational and theoretical evidence it is pointed out that massive 
LTGs develop a high \dio\ early on because their centers quench first, whereas lower-mass LTGs experience a retarded evolution that leads 
to a slower rise of \dio\ with time. This differential evolution of \dio\ across galaxy mass (\mstar) can lead to complex biases when 
galaxy samples spanning a range in redshift ($z$) or \mstar\ are structurally analyzed and compared with each other, and propagate into errors in the 
scatter and slope of any galaxy scaling relation that involves photometric quantities for bulges.
As an example, we argued that correction for \dio\ might lead to a down-bending of the bulge versus supermassive black hole mass 
\mbh\ relation below log(\mstar/\msun)$\la$10.7.
A decrease of the \mbh/\mstar\ ratio with decreasing \mstar\ would offer an element toward understanding the virtual absence of accretion-powered nuclear activity in low-mass spiral galaxies.

\brem{vi)} The above conclusions are drawn on the conservative assumption that the radial stellar surface density \sstar\ of the disk is purely exponential, and the central down-bending of its SBP results purely from an increase in its \ml\ ratio within the bulge.
However, a possible central depletion of the disk \citep{Breda20b} would further enhance photometric biases due to \dio.

\brem{vii)} A significant \dio\ ($\ga$2 mag in restframe $r$ band) is expected in young high-$z$ galaxies due to aging of SF clumps as they inwardly migrate from the disk. 
An important contribution to the early rise of \dio\ likely comes from the outwardly increasing contamination of broadband fluxes by nebular emission,
both due to the lower age of SF clumps and the lower dilution of their emission-line equivalent widths (EWs) by the local continuum. Spatially resolved studies of high-$z$ protogalaxies with the JWST, ELT and Euclid could test whether positive EW gradients emerge prior to or after the appearance of a dense, SF-quenched bulge, and might in this way place observational constraints on the relative role of clump migration and dissipative gas collapse during the early phase of bulge formation. 

\brem{viii)} Possible approaches toward a statistical estimation of \dio\ and its effect on the large existing body of photometric and structural determinations for bulges were discussed.

\begin{acknowledgements}
The authors thank the anonymous referee for valuable comments and suggestions, and colleagues Catarina Lobo, Jos\'e Afonso and Jarle Brinchmann (Instituto de Astrofísica e Ciências do Espaço; IA) for inspiring discussions. P.P. thanks Funda\c{c}\~{a}o para a Ci\^{e}ncia e a Tecnologia (FCT) for managing research funds graciously provided to Portugal by the EU. This work was supported through FCT grants UID/FIS/04434/2019, UIDB/04434/2020, UIDP/04434/2020 and the project "Identifying the Earliest Supermassive Black Holes with ALMA (IdEaS with ALMA)" (PTDC/FIS-AST/29245/2017). 
I.B. was supported by IA through the research grants CIAAUP-30/2019-BID and IA2020-05-BIDP and by the FCT PhD::SPACE Doctoral Network (PD/00040/2012) through the fellowship PD/BD/52707/2014. 
A.H. acknowledges support through contract CIAAUP-37/2018-CTTI.
J.M.G. is supported by the DL 57/2016/CP1364/CT0003 contract and acknowledges the previous support by the fellowships CIAAUP-04/2016-BPD in the context of the FCT project UID/-FIS/04434/2013 and POCI-01-0145-FEDER-007672, and SFRH/BPD/66958/2009 funded by FCT and POPH/FSE (EC). 
C.P. acknowledges support from DL 57/2016 (P2460) from the `Departamento de F\'{i}sica, Faculdade de Ci\^{e}ncias da Universidade de Lisboa'.
We acknowledge support from the FCT-CAPES Transnational Cooperation Project "Parceria Estrat\'egica em Astrof\'isica Portugal-Brasil". This study uses data provided by the Calar Alto Legacy Integral Field Area (CALIFA) survey (http://califa.caha.es), 
funded by the Spanish Ministry of Science under grant ICTS-2009-10, and the Centro Astron\'omico Hispano-Alem\'an.
It is based on observations collected at the Centro Astron\'omico Hispano Alem\'an (CAHA) at Calar Alto, operated jointly 
by the Max-Planck-Institut f\"ur Astronomie and the Instituto de Astrofísica de Andalucía (CSIC).
This research has made use of the NASA/IPAC Extragalactic Database (NED) which is operated by the Jet Propulsion Laboratory, 
California Institute of Technology, under contract with the National Aeronautics and Space Administration.
\end{acknowledgements}

% ::::::::::::::::::::

% ::::::::::::::::::::
\begin{appendix}
% :::::::::::::::::::::::::::::::::::::::::::::::::::::::::::::::::::::::::::::::::::::::::::::
\section{Supplementary notes to Sect.~\ref{photometry} \label{app:phot1}}
\subsection{\dio\ vs. central intensity depression and cutoff radius of a \brem{modexp} profile \label{app1}}
% :::::::::::::::::::::::::::::::::::::::::::::::::::::::::::::::::::::::::::::::::::::::::::::

In this section we supplement the discussion in Sect.~\ref{profiles} by a quantitative determination of \dio\ for different \brem{modexp} profiles for the disk. We recall that \brem{modexp} involves two further free parameters in addition to the central surface brightness $\mu_0$ (\sbb) and scale length $\alpha$ (\arcsec) defining a purely exponential model.
The first one $\epsilon_1=\Delta I/I_0$ constraints the depression $\Delta I$ of the disk relative to the central intensity $I_0$
of a purely exponential profile. Correspondingly, the dimming $\delta\mu_0$ (mag) of the disk at \rr=0\arcsec\
relative to the central surface brightness of an exponential model is $-2.5\,\log(1-\epsilon_1)$. 
In Sect.~\ref{profiles} we considered cases of a modest central dimming ($0.5\leq \delta\mu_0\,{\rm(mag)}\leq 2$), based on estimates with \pegase\ 
on the \ml\ contrast between the star-forming disk and the SF-quenched bulge.
This range of $\delta\mu_0$ appears adequate when assuming that the stellar surface density \sstar\ profile of the disk is purely exponential 
all the way to \rr=0\arcsec, thus \SFQ\ is the only cause for the surface brightness depression of the disk in its central part.
If, additionally, the disk is centrally depleted (cf. Sect.~\ref{cd}), then a higher $\delta\mu_0$ will need be invoked 
(e.g., $\sim$5 mag in the case of almost complete central evacuation with an $\epsilon_1\approx 0.99$).

The second free parameter $\epsilon_2$ quantifies the radius $R_{\rm c}$ (in units of $\alpha$) inward of which \brem{modexp} deviates from 
the exponential model, it can therefore be regarded as the galactocentric distance out to which the physical mechanisms behind \SFQ\ appreciably influence the SFH (and radial \ml\ profile) of a galaxy. While, to the best of our knowledge, theoretical or observational inferences do not exist for this radius, it is reasonable to assume that it depends on the nature and timescale of the dominant SF quenching mechanism. For example, in the context of 'morphological quenching' \citep{Martig09} one may expect \SFQ\ to be spatially confined to within  
\rbulge\ (i.e., $\epsilon_2=1$ for the synthetic galaxy in Fig.~\ref{sb1} where \rbulge=$\alpha$) while extending well beyond \rbulge\ in the case of AGN-driven \SFQ\ \citep{Silk97}. As for other proposed mechanisms, such as inhibition of cold gas inflow from the cosmic web due to virial shocks in high-mass galaxies \citep{Dekel09}, little can be quantitatively said on $\epsilon_2$ and the 
expected age and \ml\ gradients therein (see \tref{B20a} for a discussion). Clearly, since $\epsilon_2$ might offer a discriminator between different models for \SFQ, it would be interesting to explore techniques for its determination. 

\begin{center}
\begin{figure}
\begin{picture}(86,120)
\put(0,60){\includegraphics[clip, height=5.5cm]{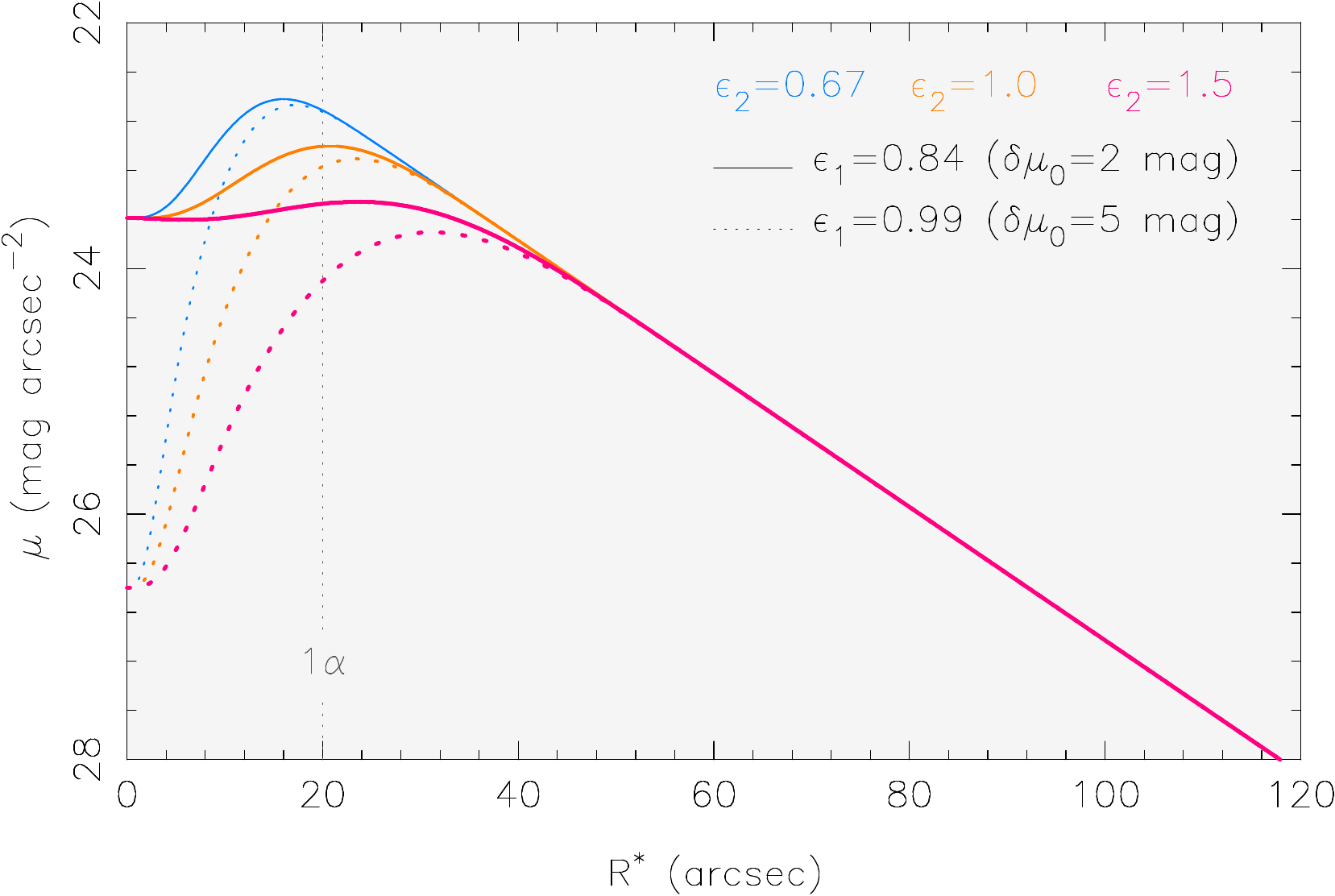}}
\put(0,0){\includegraphics[clip, height=5.5cm]{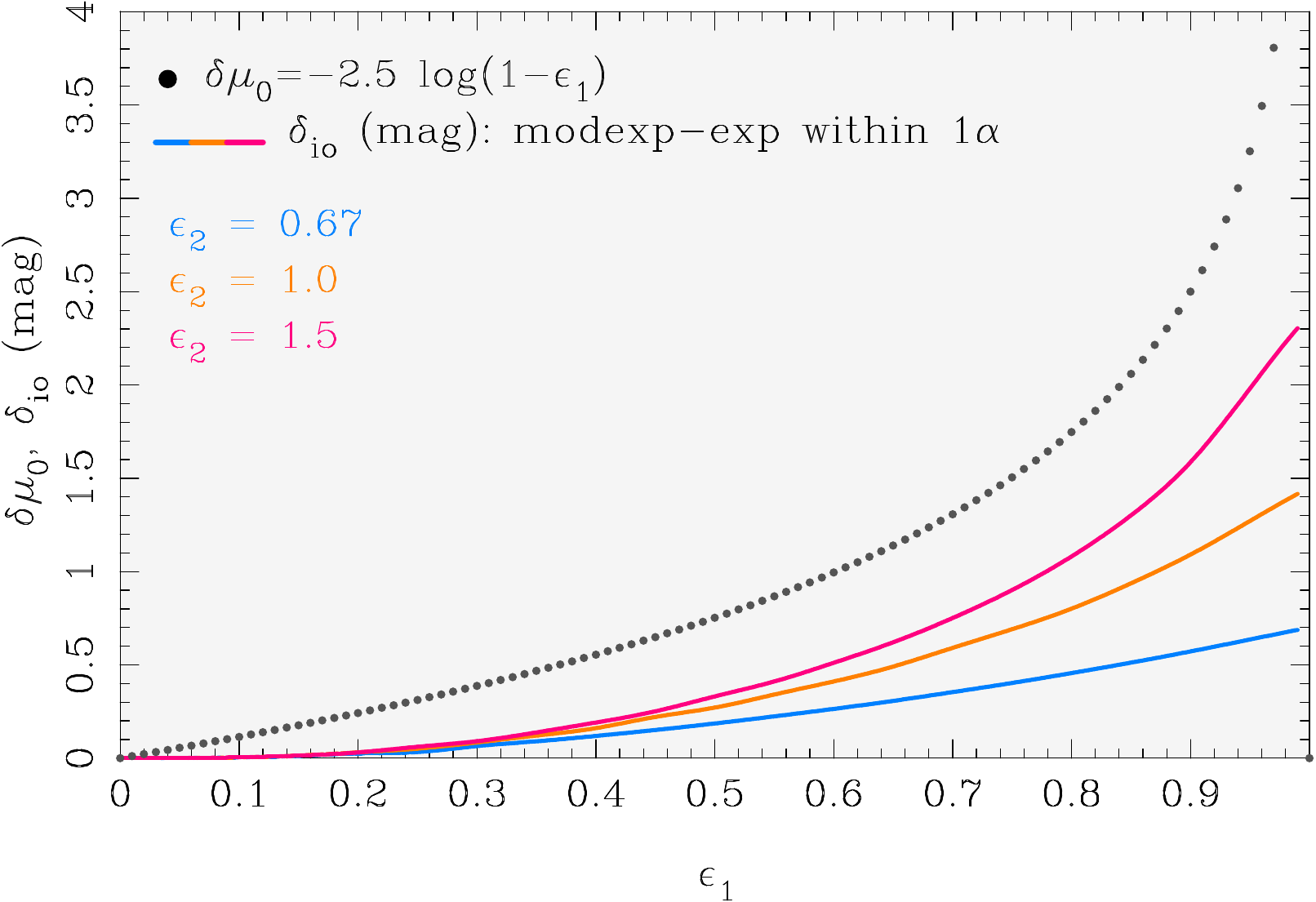}}
\end{picture}
\caption{\brem{upper panel:} Illustration of modified exponential (\brem{modexp}) profiles (Eq.~\ref{modexp1}) with a central intensity depression $\epsilon_1$ of 0.84 and 0.99 (solid and dotted curves, respectively) and a cutoff radius $\epsilon_2$ of 0.67, 1.0 and 1.5 $\alpha$ (blue, orange and red, respectively), with $\alpha$ (20\arcsec$\equiv$\rbulge\ for the synthetic SBP in Fig.~\ref{sb1}) denoting the exponential scale length of the disk. \brem{bottom panel:} $\epsilon_1$ vs. depression $\delta\mu_0$ of the central surface brightness of a \brem{modexp} profile relative to a purely exponential profile (dots). Curves show the variation across $\epsilon_1$ of the difference \dio\ (mag) between the integrated magnitude within \rbulge\ of \brem{modexp} profiles and the value corresponding to a pure exponential profile.
It can be seen that for an $\epsilon_2\geq 1$ the disk suffers a dimming between 0.9 mag ($\epsilon_1=0.84$) and 2.5 mag ($\epsilon_1=0.99$). }
\label{emag} 
\end{figure}
\end{center}      

Figure~\ref{emag} (upper panel) illustrates the variation of \brem{modexp} profiles for a disk model with the same $\mu_0$ and $\alpha$ as in Fig.~\ref{sb1} (21.6 \sbb\ and 20\arcsec, respectively) for three $\epsilon_2$ (0.67, 1.0 and 1.5; blue, orange and red, respectively) and two values for $\epsilon_1$, the first one (0.84) characterizing the upper range of values that one might expect for a purely exponential \sstar\ profile, and the second one (0.99) approximating a centrally depleted exponential profile. These $\epsilon_1$ values correspond to a $\delta\mu_0$ of, respectively, 2 and 5 mag. 
The lower panel shows as a function of $\epsilon_1$ the variation of $\delta\mu_0$ (dots) and dimming \dio\ (mag) of the integrated magnitude of the disk within \rr=$\alpha$.
%The ratio $\delta\mu_0$/\dio\ varies with $\epsilon_2$ becoming larger as $\epsilon_2$ decreases.
The computed \dio\ for $\epsilon_1=0.84$ ranges between 0.9 and 1.25 mag for an $\epsilon_2$ of 1.0 and 1.5, respectively, reaching values of 1.4 and 2.5 mag for a depleted disk with $\epsilon_1=0.99$.

Polynomial fits for 0.1$\leq\epsilon_1\leq 0.99$ yield the relations
$$\delta_{\rm io}\,{\rm (mag)} = -0.0181 + 0.0903\,\epsilon_1 + 0.6291\,\epsilon_1^2\quad {\rm (for\,\, \epsilon_2=0.67)} $$
$$\delta_{\rm io}\,{\rm (mag)} = 0.0693 - 0.5582\,\epsilon_1 + 1.887\,\epsilon_1^2\qquad\quad {\rm (for\,\, \epsilon_2=1.0)} $$
$$\delta_{\rm io}\,{\rm (mag)} = -0.188 + 1.748\,\epsilon_1 -3.78\,\epsilon_1^2 + 4.53\,\epsilon_1^3 \quad {\rm (for\,\, \epsilon_2=1.5)} $$
%Following the example in Fig.~\ref{sb1}, we define \rbulge\ at 20\arcsec\ ($\equiv$ one exp. scale length $\alpha$)
%and study how variation of the depression parameter $\epsilon_1$  and the cutoff radius $\epsilon_2$, expressed in $\alpha$ changes \dio.

% ::::::::::::::::::::::::::::::::::::::::::::::::::::::::::::::::::::::::::::::::::::::::::::
\subsection{Retrieved vs. true bulge magnitude when neglecting \dio\ \label{app:dimming}}
% ::::::::::::::::::::::::::::::::::::::::::::::::::::::::::::::::::::::::::::::::::::::::::::
\begin{center}
\begin{figure*}
\begin{picture}(210,52)
\put(0,0){\includegraphics[clip, height=5.5cm]{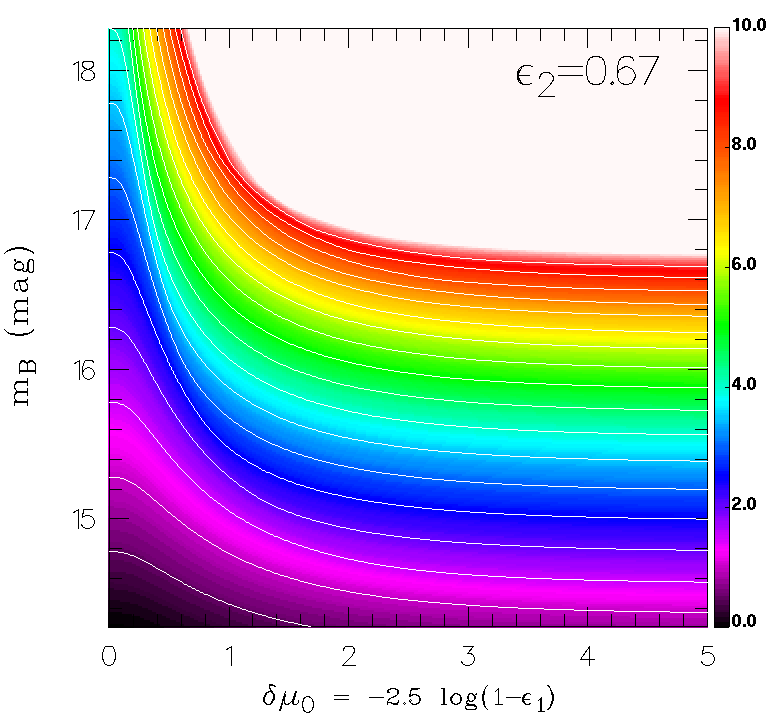}}
\put(60,0){\includegraphics[clip, height=5.5cm]{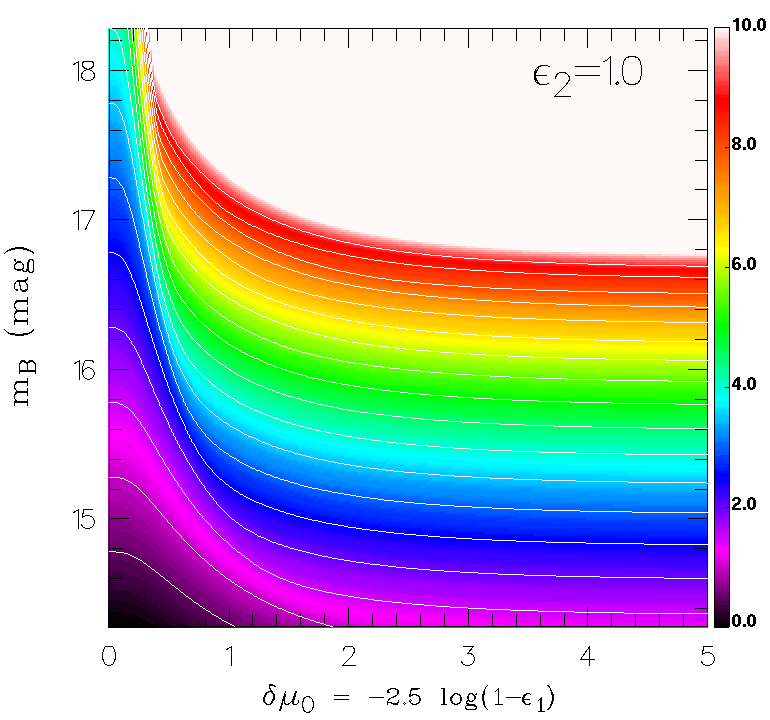}}
\put(120,0){\includegraphics[clip, height=5.5cm]{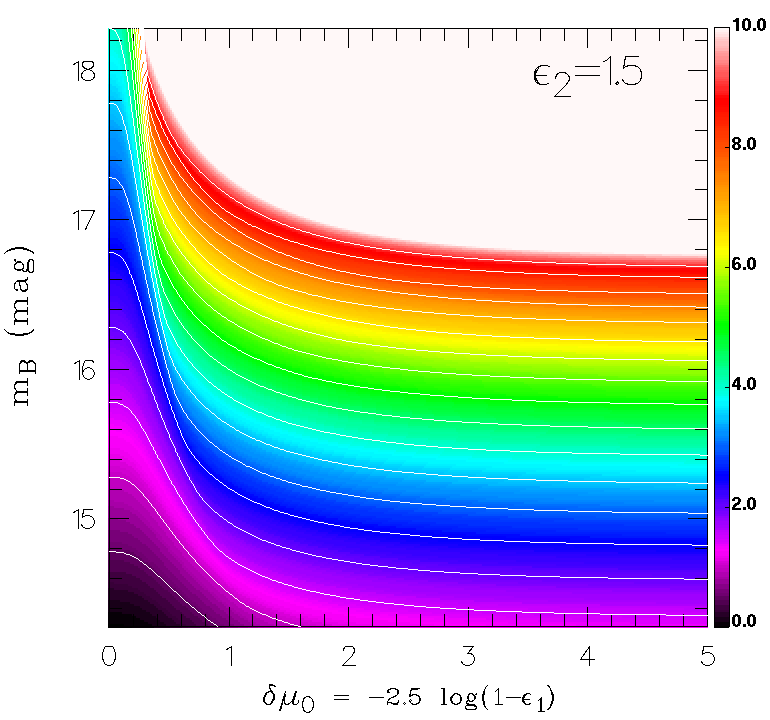}}
\end{picture}
\caption{Simulation of the underestimation (in mag) of the apparent magnitude of the bulge inside \rbulge\ (vertical bar) 
%inside 20\arcsec\ (one disk's exponential scale length $\alpha$) through standard
when applying standard bulge-disk decomposition (into a S\'ersic model for the bulge and a pure exponential model for the disk) to a galaxy 
with a centrally SF-quenched disk. The latter is approximated by a \brem{modexp} profile with cutoff radii $\epsilon_2$ of 0.67, 1.0 and 1.5 (l.h.s., central and r.h.s. panel, respectively; cf. Fig.~\ref{emag}) and a central surface depression $\delta\mu_0 = -2.5\,\log(1-\epsilon_1)$ between 0 and 5 mag. The (extrapolated) central surface brightness $\mu_0$ (21.6 \sbb) and exponential scale length $\alpha$ (20\arcsec) of the disk are identical to those adopted in Fig.~\ref{sb1}. The true apparent magnitude $m_{\rm B}$ of the bulge (abscissa) varies between 14.27 mag (the value for the S\'ersic profile in Fig.~\ref{sb1}) and $\sim$18 mag. Contours go from 0.5 mag to 9 mag in increments of 0.5 mag.}
\label{e1vsdm} 
\end{figure*}
\end{center}    
In this section we supplement our remarks on Fig.~\ref{sb2} with further quantitative inferences on the under-estimation 
of the luminosity of the bulge when bulge-disk decomposition of a centrally SF-quenched LTG is made assuming 
that the exponential model to the outer, SF-elevated zone of the disk is valid all the way to the center of the galaxy.
In particular, we show that the fainter the bulge is (i.e., the lower the \bid) the higher is the fractional 
under-estimation of its luminosity (Sect.~\ref{profiles}\&\ref{downsizing}). 
This means that a low-luminosity bulge in a modestly SF-quenched disk can be more affected by \dio\ 
than an intrinsically bright bulge in a fully SF-quenched disk.
Clearly, the estimates below only address a narrow aspect of the influence of inside-out \SFQ\ for galaxy decomposition.
As pointed out in Sect.~\ref{bd} a comprehensive investigation of this subject will require simulations of bulge-disk decomposition 
for synthetic multi-band galaxy images. These should integrate realistic prescriptions for \SFQ\ as a function of age and galaxy mass, 
and reproduce the observational imprints (e.g., radial age, \ml\ and \ewha\ gradients) of this process both for barless
and barred galaxies. Additionally, observationally and theoretically motivated alternatives to \brem{modexp} should be explored.

Similar to Fig.~\ref{sb2}, the estimates next are obtained by decomposing synthetic SBPs that consist of a S\'ersic model for the bulge and a \brem{modexp} for the disk into, respectively, a S\'ersic and a pure exponential profile. The net (disk-subtracted) profiles for the bulge were in turn used to infer how the neglect of \dio\ affects the retrieved apparent magnitude of the bulge inside \rbulge.
These simulations refer to a disk with an (extrapolated) central surface brightness of 21.6 \sbb\ and an exponential scale length $\alpha=20$\arcsec\ (values identical to those in Fig.~\ref{sb1}) and cover a range $0\leq \epsilon_1 \leq 0.99$ (correspondingly, a $\delta\mu_0$ between 0 and 5 mag) for three \SFQ\ radii $\epsilon_2$ (0.67, 1 and 1.5 $\alpha$). 
Additionally, they address how at a given $\epsilon_{1,2}$ the fractional under-estimation of the 
luminosity of the bulge increases as its intrinsic luminosity decreases; for this, we simulate a decrease of the magnitude of the bulge from its original value (m$_B$=14.27 mag; cf. Fig.~\ref{sb1}) to $\sim$18 mag.

The main insight from Fig.~\ref{e1vsdm} is that the under-estimation of the bulge magnitude (vertical bar at the r.h.s. of each diagram) depends both on m$_B$ itself and the shape of the SF-quenched disk (i.e., on $\epsilon_{1,2}$). 
For example, a relatively bright bulge with the properties assumed in Fig,~\ref{sb1} is relatively immune to \dio, as its apparent magnitude within \rbulge\ would be under-estimated by merely $\approx$0.5 (1) mag for a $\delta\mu_0=1$ (2) mag, when $\epsilon_2=1$. 
However, a bulge with the same compactness (i.e., $\beta$ and $\eta$) yet lower luminosity will suffer an artificial dimming by 1.4 mag if its m$_B$ were one mag fainter (15.27 mag), and by 3.7 mag for a m$_B$=16.27 mag. The detection bias against intrinsically faint bulges will be further aggravated for a larger cutoff radius. For instance, an $\epsilon_2=1.5$ will rise the underestimation of the bulge magnitude to 2.6 mag and 5.2 mag, respectively.
%However, a bulge with the same compactness (i.e., $\beta$ and $\eta$) yet lower luminosity would suffer an artificial 
%dimming of 1.4 (2.6) mag if its m$_B$ were one mag fainter (15.27 mag), and of 3.7 (5.2) mag for a m$_B$=16.27 mag.
The evidence from Fig.~\ref{e1vsdm} underscores our remarks in Sect.~\ref{smbh} that studies of intrinsically bright bulges in \brem{iC} LTGs are only moderately affected  by \dio\ whereas a stronger effect is expected for lower-luminosity bulges in \brem{iB} ($\sim$L$^{\star}$) LTGs.
\end{appendix}
\end{document}